%% file: main.tex
\pdfoutput=1
\documentclass[12pt,a4paper]{article}

\usepackage{ifthen} 
\newboolean{pdflatex}
\setboolean{pdflatex}{true} 

\newboolean{articletitles}
\setboolean{articletitles}{true} 

\newboolean{uprightparticles}
\setboolean{uprightparticles}{false} 

\def\paperauthors{LHCb collaboration} 
\def\paperasciititle{Analysis of Lb -> pK mu+ mu- decays} 
\def\papertitle{Analysis of $\it{\Lambda}^\mathrm{0}_b \rightarrow pK^-\mu^+\mu^-$ decays} 
\def\paperkeywords{{High Energy Physics}, {LHCb}} 
\def\papercopyright{\the\year\ CERN for the benefit of the LHCb collaboration} 
\def\paperlicence{CC BY 4.0 licence}
\def\paperlicenceurl{https://creativecommons.org/licenses/by/4.0/}

\input{preamble}
\usepackage{longtable} 
\usepackage{multicol}
\usepackage{multirow}
\usepackage{booktabs}

\input{additional-symbols}

\usepackage{diagbox} 

\begin{document}

\renewcommand{\thefootnote}{\fnsymbol{footnote}}
\setcounter{footnote}{1}

\input{title-LHCb-PAPER}


\renewcommand{\thefootnote}{\arabic{footnote}}
\setcounter{footnote}{0}

\cleardoublepage


\pagestyle{plain} 
\setcounter{page}{1}
\pagenumbering{arabic}


\input{body}

\input{acknowledgements}

\input{appendix}


\addcontentsline{toc}{section}{References}
\bibliographystyle{LHCb}
\bibliography{main,standard,LHCb-PAPER,LHCb-DP,LHCb-TDR}

\newpage
\input{Authorship_LHCb-PAPER-2024-024}

\end{document}

%% file: preamble.tex
\usepackage[top=1in, bottom=1.25in, left=1in, right=1in]{geometry}

\usepackage{microtype}
\usepackage{lineno}  
\usepackage{xspace} 
\usepackage{caption} 

\usepackage{graphicx}  
\usepackage{color}
\usepackage{colortbl}
\graphicspath{{./figs/}} 

\usepackage{amsmath} 
\usepackage{amssymb}
\usepackage{amsfonts}
\usepackage{upgreek} 

\newcommand*\patchAmsMathEnvironmentForLineno[1]{%
\expandafter\let\csname old#1\expandafter\endcsname\csname #1\endcsname
\expandafter\let\csname oldend#1\expandafter\endcsname\csname
end#1\endcsname
 \renewenvironment{#1}%
   {\linenomath\csname old#1\endcsname}%
   {\csname oldend#1\endcsname\endlinenomath}%
}
\newcommand*\patchBothAmsMathEnvironmentsForLineno[1]{%
  \patchAmsMathEnvironmentForLineno{#1}%
  \patchAmsMathEnvironmentForLineno{#1*}%
}
\AtBeginDocument{%
\patchBothAmsMathEnvironmentsForLineno{equation}%
\patchBothAmsMathEnvironmentsForLineno{align}%
\patchBothAmsMathEnvironmentsForLineno{flalign}%
\patchBothAmsMathEnvironmentsForLineno{alignat}%
\patchBothAmsMathEnvironmentsForLineno{gather}%
\patchBothAmsMathEnvironmentsForLineno{multline}%
\patchBothAmsMathEnvironmentsForLineno{eqnarray}%
}

\usepackage{hyperxmp}

\usepackage[pdftex,
            pdfauthor={\paperauthors},
            pdftitle={\paperasciititle},
            pdfkeywords={\paperkeywords},
            pdfcopyright={Copyright (C) \papercopyright},
            pdflicenseurl={\paperlicenceurl}]{hyperref}

\usepackage[colorinlistoftodos,textsize=scriptsize]{todonotes}

\usepackage[bottom,flushmargin,hang,multiple]{footmisc}

\usepackage[all]{hypcap} 

\input{lhcb-symbols-def} 

\usepackage{cite} 
\usepackage{mciteplus}

%% file: lhcb-symbols-def.tex
\usepackage{xspace} 
\usepackage{upgreek}

\def\lhcb   {\mbox{LHCb}\xspace}

\def\MagUp {\mbox{\em Mag\kern -0.05em Up}\xspace}

\ifthenelse{\boolean{uprightparticles}}%
{
 
 \def\Pgamma      {\ensuremath{\upgamma}\xspace}

 \def\Pmu         {\ensuremath{\upmu}\xspace}

 \def\Ppi         {\ensuremath{\uppi}\xspace}

 \def\Ppsi        {\ensuremath{\uppsi}\xspace}

 \def\PDelta      {\ensuremath{\Delta}\xspace}                 
 \def\PXi         {\ensuremath{\Xi}\xspace}                 
 \def\PLambda     {\ensuremath{\Lambda}\xspace}                 
 \def\PSigma      {\ensuremath{\Sigma}\xspace}                 
 \def\POmega      {\ensuremath{\Omega}\xspace}                 
 \def\PUpsilon    {\ensuremath{\Upsilon}\xspace}
 \let\oldPi\Pi
 \def\PPi         {\ensuremath{\oldPi}\xspace}

 \def\PB      {\ensuremath{\mathrm{B}}\xspace}                 
                  
 \def\PD      {\ensuremath{\mathrm{D}}\xspace}

 \def\PJ      {\ensuremath{\mathrm{J}}\xspace}                 
 \def\PK      {\ensuremath{\mathrm{K}}\xspace}

 \def\Pb      {\ensuremath{\mathrm{b}}\xspace}                 
 \def\Pc      {\ensuremath{\mathrm{c}}\xspace}

 \def\Pi      {\ensuremath{\mathrm{i}}\xspace}

 \def\Pp      {\ensuremath{\mathrm{p}}\xspace}

 \def\Ps      {\ensuremath{\mathrm{s}}\xspace}

 \def\thebaroffset{0.0em}
}
{
 
 \def\Pgamma      {\ensuremath{\gamma}\xspace}

 \def\Pmu         {\ensuremath{\mu}\xspace}

 \def\Ppi         {\ensuremath{\pi}\xspace}

 \def\Ppsi        {\ensuremath{\psi}\xspace}                 
                  
 \mathchardef\PDelta="7101
 \mathchardef\PXi="7104
 \mathchardef\PLambda="7103
 \mathchardef\PSigma="7106
 \mathchardef\POmega="710A
 \mathchardef\PUpsilon="7107
 \mathchardef\PPi="7105
                  
 \def\PB      {\ensuremath{B}\xspace}                 
                  
 \def\PD      {\ensuremath{D}\xspace}

 \def\PJ      {\ensuremath{J}\xspace}                 
 \def\PK      {\ensuremath{K}\xspace}

 \def\Pb      {\ensuremath{b}\xspace}                 
 \def\Pc      {\ensuremath{c}\xspace}

 \def\Pi      {\ensuremath{i}\xspace}

 \def\Pp      {\ensuremath{p}\xspace}

 \def\Ps      {\ensuremath{s}\xspace}

 \def\thebaroffset{0.18em}
}
\newcommand{\offsetoverline}[2][\thebaroffset]{\kern #1\overline{\kern -#1 #2}}%

\makeatletter
\ifcase \@ptsize \relax
  \newcommand{\miniscule}{\@setfontsize\miniscule{4}{5}}
\or
  \newcommand{\miniscule}{\@setfontsize\miniscule{5}{6}}
\or
  \newcommand{\miniscule}{\@setfontsize\miniscule{5}{6}}
\fi
\makeatother

\DeclareRobustCommand{\optbar}[1]{\shortstack{{\miniscule (\rule[.5ex]{1.25em}{.18mm})}
  \\ [-.7ex] $#1$}}

\def\mup        {{\ensuremath{\Pmu^+}}\xspace}
\def\mun        {{\ensuremath{\Pmu^-}}\xspace} 

\def\mumu       {{\ensuremath{\Pmu^+\Pmu^-}}\xspace}

\def\g      {{\ensuremath{\Pgamma}}\xspace}

\def\squark    {{\ensuremath{\Ps}}\xspace}

\def\cquark    {{\ensuremath{\Pc}}\xspace}

\def\bquark    {{\ensuremath{\Pb}}\xspace}

\def\pion   {{\ensuremath{\Ppi}}\xspace}

\def\pip    {{\ensuremath{\pion^+}}\xspace}
\def\pim    {{\ensuremath{\pion^-}}\xspace}

\def\kaon    {{\ensuremath{\PK}}\xspace}

\def\KorKbar {\kern \thebaroffset\optbar{\kern -\thebaroffset \PK}{}\xspace}

\def\Kp      {{\ensuremath{\kaon^+}}\xspace}
\def\Km      {{\ensuremath{\kaon^-}}\xspace}

\def\Kstarz  {{\ensuremath{\kaon^{*0}}}\xspace}

\def\Kstar   {{\ensuremath{\kaon^*}}\xspace}

\def\D       {{\ensuremath{\PD}}\xspace}

\def\DorDbar {\kern \thebaroffset\optbar{\kern -\thebaroffset \PD}\xspace}

\def\Dp      {{\ensuremath{\D^+}}\xspace}
\def\Dm      {{\ensuremath{\D^-}}\xspace}

\def\DpDm    {\ensuremath{\Dp {\kern -0.16em \Dm}}\xspace}

\def\Dsm     {{\ensuremath{\D^-_\squark}}\xspace}

\def\B       {{\ensuremath{\PB}}\xspace}

\def\BorBbar {\kern \thebaroffset\optbar{\kern -\thebaroffset \PB}\xspace}
\def\Bz      {{\ensuremath{\B^0}}\xspace}

\def\Bd      {{\ensuremath{\B^0}}\xspace}

\def\BdorBdbar {\kern \thebaroffset\optbar{\kern -\thebaroffset \Bd}\xspace}
\def\Bu      {{\ensuremath{\B^+}}\xspace}

\def\Bp      {{\ensuremath{\Bu}}\xspace}

\def\Bs      {{\ensuremath{\B^0_\squark}}\xspace}

\def\BsorBsbar {\kern \thebaroffset\optbar{\kern -\thebaroffset \Bs}\xspace}

\def\jpsi     {{\ensuremath{{\PJ\mskip -3mu/\mskip -2mu\Ppsi}}}\xspace}
\def\psitwos  {{\ensuremath{\Ppsi{(2S)}}}\xspace}

\def\Y#1S{\ensuremath{\PUpsilon{(#1S)}}\xspace}

\def\proton      {{\ensuremath{\Pp}}\xspace}

\def\Lz          {{\ensuremath{\PLambda}}\xspace}
\def\Lbar        {{\ensuremath{\offsetoverline{\PLambda}}}\xspace}
\def\LorLbar     {\kern \thebaroffset\optbar{\kern -\thebaroffset \PLambda}\xspace}

\def\Lc          {{\ensuremath{\Lz^+_\cquark}}\xspace}

\def\Lb           {{\ensuremath{\Lz^0_\bquark}}\xspace}
\def\Lbbar        {{\ensuremath{\Lbar{}^0_\bquark}}\xspace}

\def\BF         {{\ensuremath{\mathcal{B}}}\xspace}
\def\BR         {\BF}

\newcommand{\decay}[2]{\ensuremath{#1\!\to #2}\xspace} 

\def\to                 {\ensuremath{\rightarrow}\xspace}

\def\qsq       {{\ensuremath{q^2}}\xspace}

\def\AT#1     {\ensuremath{A_{\mathrm{T}}^{#1}}\xspace}           

\def\C#1      {\ensuremath{\mathcal{C}_{#1}}\xspace}                       
\def\Cp#1     {\ensuremath{\mathcal{C}_{#1}^{'}}\xspace}                    
\def\Ceff#1   {\ensuremath{\mathcal{C}_{#1}^{\mathrm{(eff)}}}\xspace}        
\def\Cpeff#1  {\ensuremath{\mathcal{C}_{#1}^{'\mathrm{(eff)}}}\xspace}       
\def\Ope#1    {\ensuremath{\mathcal{O}_{#1}}\xspace}                       
\def\Opep#1   {\ensuremath{\mathcal{O}_{#1}^{'}}\xspace}                    

\newcommand{\aunit}[1]{\ensuremath{\text{\,#1}}}       

\newcommand{\tev}{\aunit{Te\kern -0.1em V}\xspace}
\newcommand{\gev}{\aunit{Ge\kern -0.1em V}\xspace}
\newcommand{\mev}{\aunit{Me\kern -0.1em V}\xspace}
\newcommand{\kev}{\aunit{ke\kern -0.1em V}\xspace}
\newcommand{\ev}{\aunit{e\kern -0.1em V}\xspace}
 
\newcommand{\mevc}{\ensuremath{\aunit{Me\kern -0.1em V\!/}c}\xspace}
\newcommand{\gevc}{\ensuremath{\aunit{Ge\kern -0.1em V\!/}c}\xspace}
\newcommand{\mevcc}{\ensuremath{\aunit{Me\kern -0.1em V\!/}c^2}\xspace}
\newcommand{\gevcc}{\ensuremath{\aunit{Ge\kern -0.1em V\!/}c^2}\xspace}
\newcommand{\gevgevcccc}{\ensuremath{\gev^2\!/c^4}\xspace} 

\def\fb   {\ensuremath{\aunit{fb}}\xspace}
\def\invfb   {\ensuremath{\fb^{-1}}\xspace}

\def\deriv {\ensuremath{\mathrm{d}}}

\def\gsim{{~\raise.15em\hbox{$>$}\kern-.85em
          \lower.35em\hbox{$\sim$}~}\xspace}
\def\lsim{{~\raise.15em\hbox{$<$}\kern-.85em
          \lower.35em\hbox{$\sim$}~}\xspace}

\def\sPlot{\mbox{\em sPlot}\xspace}

\def\pt         {\ensuremath{p_{\mathrm{T}}}\xspace}

\def\ptot       {\ensuremath{p}\xspace}

\def\evtgen     {\mbox{\textsc{EvtGen}}\xspace}

\def\geant      {\mbox{\textsc{Geant4}}\xspace}

\def\photos     {\mbox{\textsc{Photos}}\xspace}

\def\pythia     {\mbox{\textsc{Pythia}}\xspace}

\def\tell1  {TELL1\xspace}
\def\ukl1   {UKL1\xspace}

\newcommand{\eg}{\mbox{\itshape e.g.}\xspace}

\newcommand{\lhcborcid}[1]{\href{https://orcid.org/#1}{\hspace*{0.1em}\raisebox{-0.45ex}{\includegraphics[width=1em]{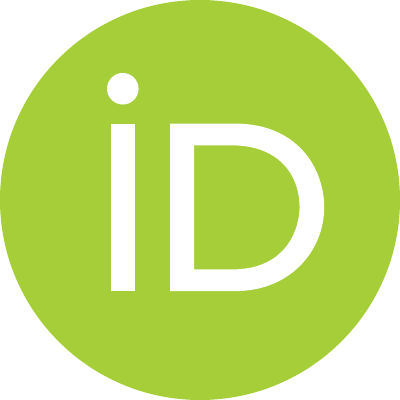}}}}


%% file: additional-symbols.tex
\newcommand{\Lbpkmm}{\ensuremath{\Lb\to p\Km\mumu}\xspace}
\newcommand{\Lbpkjpsi}{\ensuremath{\Lb\to \jpsi\proton\Km}\xspace}
\newcommand{\Lbpkg}{\ensuremath{\Lb\to\proton\Km\g}\xspace}
\newcommand{\mpksq}{\ensuremath{m^{2}_{pK}}\xspace}
\newcommand{\mpk}{\ensuremath{m_{pK}}\xspace}
\newcommand{\mpkmm}{\ensuremath{m_{pK\mu\mu}}\xspace}
\newcommand{\Kobs}[1]{\ensuremath{\offsetoverline{K}_{#1}}}
\def\zfit  {\mbox{\textsc{zfit}}\xspace}

\renewcommand{\vec}[1]{\boldsymbol{#1}}

%% file: title-LHCb-PAPER.tex

\begin{titlepage}
\pagenumbering{roman}

\vspace*{-1.5cm}
\centerline{\large EUROPEAN ORGANIZATION FOR NUCLEAR RESEARCH (CERN)}
\vspace*{1.5cm}
\noindent
\begin{tabular*}{\linewidth}{lc@{\extracolsep{\fill}}r@{\extracolsep{0pt}}}
\ifthenelse{\boolean{pdflatex}}
{\vspace*{-1.5cm}\mbox{\!\!\!\includegraphics[width=.14\textwidth]{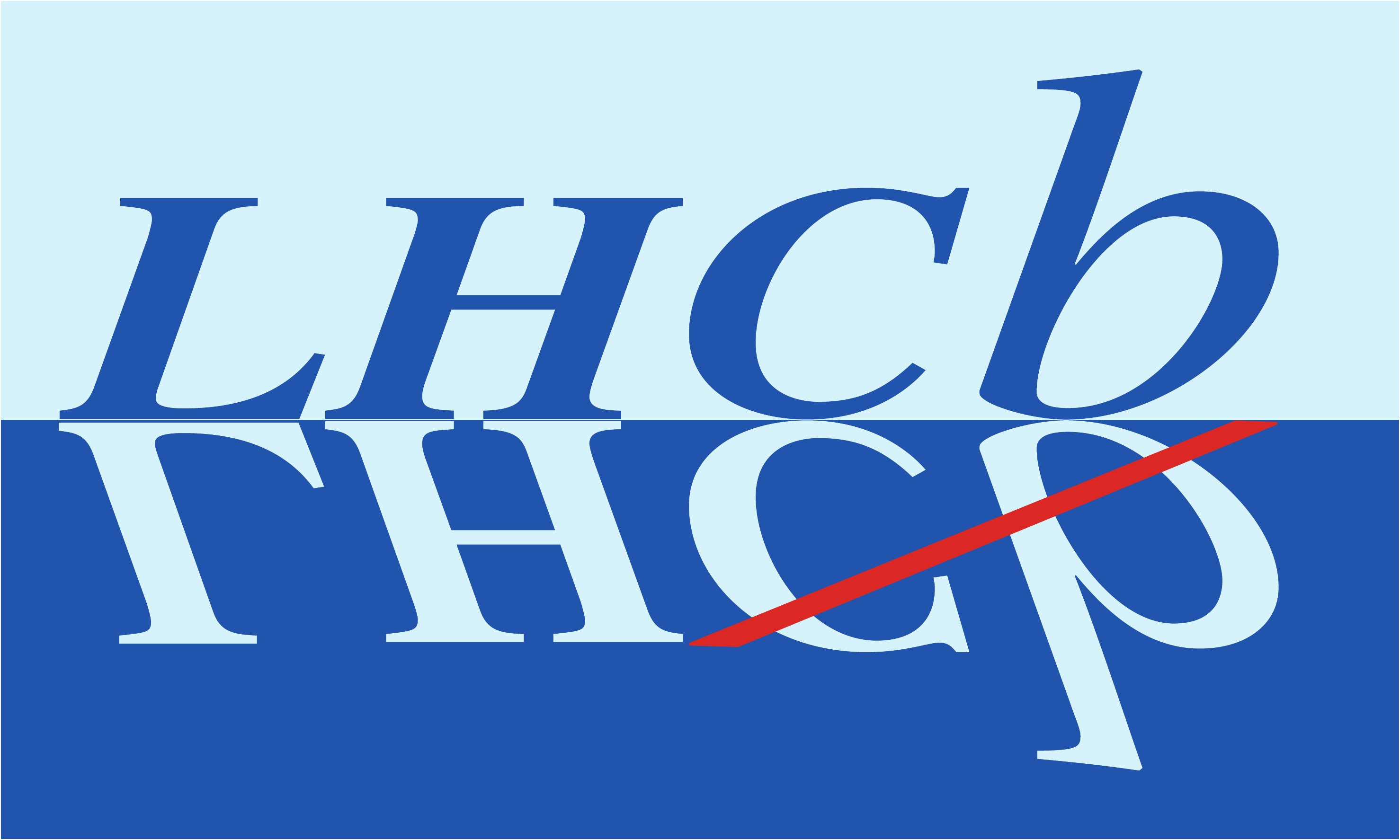}} & &}%
{\vspace*{-1.2cm}\mbox{\!\!\!\includegraphics[width=.12\textwidth]{figs/lhcb-logo.eps}} & &}%
\\
 & & CERN-EP-2024-212 \\  
 & & LHCb-PAPER-2024-024 \\  
 & & 19 December 2024 \\ 
 & & \\
\end{tabular*}

\vspace*{4.0cm}

{\normalfont\bfseries\boldmath\huge
\begin{center}
  \papertitle 
\end{center}
}

\vspace*{2.0cm}

\begin{center}
\paperauthors\footnote{Authors are listed at the end of this paper.}
\end{center}

\vspace{\fill}

\begin{abstract}
  \noindent
  The differential branching fraction and angular coefficients of \Lbpkmm decays are measured in bins of the dimuon mass squared and dihadron mass.
  The analysis is performed using a data set corresponding to 9\invfb of integrated luminosity collected with the \lhcb detector between 2011 and 2018. 
  The data are consistent with receiving contributions from a mixture of \Lz resonances with different spin-parity quantum numbers.
  The angular coefficients show a pattern of vector--axial vector interference that is a characteristic of the type of flavour-changing neutral-current transition relevant for these decays. 
\end{abstract}

\vspace*{2.0cm}

\begin{center}
  Published in
  JHEP 12 (2024) 147
\end{center}

\vspace{\fill}

{\footnotesize 
\centerline{\copyright~\papercopyright. \href{\paperlicenceurl}{\paperlicence}.}}
\vspace*{2mm}

\end{titlepage}


\newpage
\setcounter{page}{2}
\mbox{~}

%% file: body.tex
\section{Introduction}
\label{sec:introduction}

The decay of a \Lb baryon to a \Lz resonance and a pair of oppositely charged muons is mediated by a \bquark- to \squark-quark flavour-changing neutral-current (FCNC) transition. 
In the Standard Model (SM) of particle physics, such transitions are suppressed as they proceed via loop-order Feynman diagrams.
In extensions of the SM, the decay rate and angular distribution can be significantly modified~\cite{Beck:2022spd}.
The rate and angular distribution of \decay{\Lb}{\Lz(1116)\mumu} decays, where the $\Lz(1116)$ baryon is the weakly decaying ground-state, have previously been studied by the LHCb and CDF collaborations~\cite{LHCb-PAPER-2015-009,LHCb-PAPER-2018-029,CDF:2011buy}.\footnote{
The inclusion of charge-conjugate processes is implied throughout this paper unless explicitly stated.
}
While the properties of \decay{\Lb}{\Lz(1116)\mumu} decays are consistent with SM expectations at the current level of experimental precision~\cite{Blake:2019guk}, tensions are seen in measurements of \bquark- to \squark-quark FCNC transitions involving \Bp, \Bz and \Bs mesons. 
The decay rates of these processes are found to be systematically below SM predictions~\cite{LHCb-PAPER-2016-012,LHCb-PAPER-2021-014,CMS:2018qih,BELLE:2019xld,BaBar:2012mrf}. 
Angular observables in \decay{\Bz}{\Kstar(892)^{0}\mumu} and \decay{\Bp}{\Kstar(892)^{+}\mumu} decays are also found to differ from expectations when the dimuon mass squared, \qsq, is less than $8\gevgevcccc$~\cite{LHCb-PAPER-2014-006,LHCb-PAPER-2020-002,LHCb-PAPER-2020-041,LHCb-PAPER-2021-022,LHCb-PAPER-2024-011,CMS:2020oqb,CMS:2017rzx,ATLAS:2018gqc}. 
There is currently no conclusion on whether these discrepancies provide evidence for a genuine breakdown of the SM or are reflective of the challenges of calculating the rates of exclusive processes in a nonperturbative regime of QCD, see for example Refs.~\cite{Ciuchini:2017mik,Capdevila:2017bsm,Gubernari:2022hxn,Hurth:2023jwr,Bordone:2024hui}.
It is therefore important to search for similar discrepancies in other hadronic systems, which require different theoretical treatments. 

In this paper, \Lbpkmm decays with intermediate \Lz resonances decaying to $\proton\Km$ are considered.
Rare FCNC decays of a \Lb baryon to the $p\Km\mumu$ final state were first observed by the LHCb collaboration in Ref.~\cite{LHCb-PAPER-2016-059}.
The differential branching fraction of the decay was measured in the range $0.1 < \qsq < 6.0\gevgevcccc$ in Ref.~\cite{LHCb-PAPER-2019-040}. 
Subsequently, the differential branching fraction to the narrowest strongly decaying state, the $\Lz(1520)$ resonance, was measured in Ref.~\cite{LHCb-PAPER-2022-050} in bins of \qsq. 
The full dihadron spectrum comprises a relatively large number of \Lz states with different $J^P$ quantum numbers. 
There are no accurate predictions of the differential decay rates of many of these states. 
Standard Model predictions, with reliable uncertainties, are only available for \decay{\Lb}{\Lz(1520)\mumu} decays, based on Lattice QCD~\cite{Meinel:2020owd,Meinel:2021mdj} or dispersive bounds~\cite{Amhis:2022vcd}. 
The hadronic form factors for transitions from \Lb baryons to other strongly decaying $\Lz$ resonances have only been estimated using a simplified quark model~\cite{Mott:2011cx}. 
This paper presents a measurement of the branching fraction, and a first measurement of the angular distribution, of \decay{\Lb}{p\Km\mumu} decays.
Coefficients of the angular distribution are determined using the formalism described in  Ref.~\cite{Beck:2022spd}. 

The branching fraction and angular coefficients are measured in bins of \qsq and dihadron mass, \mpk. 
In \qsq, the data are binned in the ranges 0.10--0.98, 1.1--2.0, 2.0--4.0, 4.0--6.0, 6.0--8.0, 11.0--12.5 and 15.0--17.5\gevgevcccc. 
In \mpk, the data are binned in the ranges 1.4359--1.5900, 1.59--1.75, 1.75--2.20 and 2.20--5.41\gevcc. 
The first bin in \mpk isolates the narrow $\Lz(1520)$ state, which is a prominent feature of the dihadron spectrum. 
The last bin in \mpk contains several broad \Lz resonances whose properties are poorly known. 
At large \qsq, the 1.4359--1.5900 and 1.59--1.75\gevcc bins are combined and the others are kinematically inaccessible. 
The \qsq regions between 0.98--1.10 and 12.5--15.0\gevgevcccc are removed from the data sample as they contain contributions from $\phi$ and \psitwos meson decays. 
Candidates with $8.0 < \qsq < 11.0\gevgevcccc$ are dominated by \Lbpkjpsi decays and are retained as a control sample.  

At the LHC, \Lb baryons are observed to be produced with a net polarisation below 1\%~\cite{LHCb-PAPER-2020-005}. 
In this paper, they are treated as having zero polarisation in order to simplify the number of angular coefficients that need to be considered. 
In this case, the angular distribution of the decay can be described by three angles: 
$\theta_\mu$, the angle between the direction of the \mup (\mun) in the \mumu rest frame and the direction of the \mumu pair in the \Lb (\Lbbar) rest frame; 
$\theta_p$, the angle between the direction of the proton (antiproton) in the dihadron rest frame and the direction of the dihadron pair in the \Lb (\Lbbar) rest frame; 
$\phi$, the angle between the \mumu and dihadron decay planes in the \Lb rest frame. 
The resulting angular distribution is complex due to the contribution from a multitude of different \Lz states, with different $J^{P}$ quantum numbers, that decay to the same $p\Km$ final state. 
For \Lz states with $J \leq \tfrac{5}{2}$, the differential decay rate can be written in terms of a set of angular coefficients, $K_i(\qsq,\mpksq)$, as 
\begin{align}
    \frac{\deriv^5\Gamma}{\deriv\vec{\Phi}} = \frac{3}{8\pi} \sum\limits_{i=1}^{46} K_i(\qsq,\mpksq) f_i(\vec{\Omega})~,
\end{align}
where $\vec{\Phi} = (\qsq,\mpksq,\cos\theta_\mu,\cos\theta_p,\phi)$, $\vec{\Omega} = (\cos\theta_\mu,\cos\theta_p,\phi)$ and with the angular dependencies, $f_{i}(\vec{\Omega})$, given in Table~\ref{tab:legendre:basis}. 
The rate-averaged angular coefficients \Kobs{i} in bins of \qsq and \mpksq are given by
\begin{align}
\Kobs{i} = \int\limits_{{\rm bin}}K_{i}(\qsq,\mpksq) \deriv\qsq\deriv\mpksq \Bigg/ \int\limits_{{\rm bin}}\frac{\deriv^{2}\Gamma}{\deriv\qsq\deriv\mpksq} \deriv\qsq\deriv\mpksq~.
\end{align} 
Due to the large number of observables involved, the coefficients are determined using the method of moments (see \eg Ref.~\cite{Beaujean:2015xea}) rather than by fitting the angular distribution of the data. 
Nonzero production polarisation of the \Lb baryons significantly increases the number of possible angular observables that can be measured but, due to the angular structure, does not affect the determination of the coefficients measured in this paper.

\begin{table}[!htb]
\caption{
Orthogonal basis functions for the angular terms $f_1(\vec{\Omega})$--$f_{46}(\vec{\Omega})$ that arise for unpolarised \Lb baryons decaying to \Lz resonances with $J \leq \tfrac{5}{2}$. 
Here, $P_l^m(\cos\theta)$ are associated Legendre polynomials.
The basis follows Ref.~\cite{Beck:2022spd}. 
}
\centering
\begin{tabular}{cr|cr}
\toprule
$i$  &  $f_i(\vec{\Omega})$  &  $i$  & $f_i(\vec{\Omega})$  \\ 
\midrule
   1 & $\tfrac{1}{\sqrt{3}}P_0^0(\cos\theta_p)P_0^0(\cos\theta_\mu)\hphantom{\cos \phi}$
& 24 & $\tfrac{1}{2} \sqrt{\tfrac{7}{3}} P_3^1(\cos\theta_p) P_1^1(\cos\theta_\mu) \cos \phi\hphantom{2}$ \\
   2 & $P_0^0(\cos\theta_p)P_1^0(\cos\theta_\mu)\hphantom{\cos \phi}$
& 25 & $\tfrac{1}{2} P_4^1(\cos\theta_p) P_2^1(\cos\theta_\mu) \cos \phi\hphantom{2}$ \\
   3 & $\sqrt{\tfrac{5}{3}} P_0^0(\cos\theta_p)P_2^0(\cos\theta_\mu)\hphantom{\cos \phi}$
& 26 & $\tfrac{3}{2 \sqrt{5}} P_4^1(\cos\theta_p) P_1^1(\cos\theta_\mu) \cos \phi\hphantom{2}$ \\
   4 & $P_1^0(\cos\theta_p)P_0^0(\cos\theta_\mu)\hphantom{\cos \phi}$
& 27 & $\tfrac{1}{3} \sqrt{\tfrac{11}{6}} P_5^1(\cos\theta_p) P_2^1(\cos\theta_\mu) \cos \phi\hphantom{2}$ \\
   5 & $\sqrt{3} P_1^0(\cos\theta_p) P_1^0(\cos\theta_\mu)\hphantom{\cos \phi}$
& 28 & $\sqrt{\tfrac{11}{30}} P_5^1(\cos\theta_p) P_1^1(\cos\theta_\mu) \cos \phi\hphantom{2}$ \\
   6 & $\sqrt{5} P_1^0(\cos\theta_p) P_2^0(\cos\theta_\mu)\hphantom{\cos \phi}$
& 29 & $\sqrt{\tfrac{5}{6}} P_1^1(\cos\theta_p) P_2^1(\cos\theta_\mu) \sin \phi\hphantom{2}$ \\
   7 & $\sqrt{\tfrac{5}{3}} P_2^0(\cos\theta_p)P_0^0(\cos\theta_\mu)\hphantom{\cos \phi}$
& 30 & $\sqrt{\tfrac{3}{2}} P_1^1(\cos\theta_p) P_1^1(\cos\theta_\mu) \sin \phi\hphantom{2}$ \\
   8 & $\sqrt{5} P_2^0(\cos\theta_p) P_1^0(\cos\theta_\mu)\hphantom{\cos \phi}$
& 31 & $\tfrac{5}{3 \sqrt{6}} P_2^1(\cos\theta_p) P_2^1(\cos\theta_\mu) \sin \phi\hphantom{2}$ \\
   9 & $\tfrac{5}{\sqrt{3}} P_2^0(\cos\theta_p) P_2^0(\cos\theta_\mu)\hphantom{\cos \phi}$
& 32 & $\sqrt{\tfrac{5}{6}} P_2^1(\cos\theta_p) P_1^1(\cos\theta_\mu) \sin \phi\hphantom{2}$ \\
  10 & $\sqrt{\tfrac{7}{3}} P_3^0(\cos\theta_p)P_0^0(\cos\theta_\mu)\hphantom{\cos \phi}$
& 33 & $\tfrac{1}{6} \sqrt{\tfrac{35}{3}} P_3^1(\cos\theta_p) P_2^1(\cos\theta_\mu) \sin \phi\hphantom{2}$ \\
  11 & $\sqrt{7} P_3^0(\cos\theta_p) P_1^0(\cos\theta_\mu)\hphantom{\cos \phi}$
& 34 & $\tfrac{1}{2} \sqrt{\tfrac{7}{3}} P_3^1(\cos\theta_p) P_1^1(\cos\theta_\mu) \sin \phi\hphantom{2}$ \\
  12 & $\sqrt{\tfrac{35}{3}} P_3^0(\cos\theta_p) P_2^0(\cos\theta_\mu)\hphantom{\cos \phi}$
& 35 & $\tfrac{1}{2} P_4^1(\cos\theta_p) P_2^1(\cos\theta_\mu) \sin \phi\hphantom{2}$ \\
  13 & $\sqrt{3} P_4^0(\cos\theta_p)P_0^0(\cos\theta_\mu)\hphantom{\cos \phi}$
& 36 & $\tfrac{3}{2 \sqrt{5}} P_4^1(\cos\theta_p) P_1^1(\cos\theta_\mu) \sin \phi\hphantom{2}$ \\
  14 & $3 P_4^0(\cos\theta_p) P_1^0(\cos\theta_\mu)\hphantom{\cos \phi}$
& 37 & $\tfrac{1}{3} \sqrt{\tfrac{11}{6}} P_5^1(\cos\theta_p) P_2^1(\cos\theta_\mu) \sin \phi\hphantom{2}$ \\
  15 & $\sqrt{15} P_4^0(\cos\theta_p) P_2^0(\cos\theta_\mu)\hphantom{\cos \phi}$
& 38 & $\sqrt{\tfrac{11}{30}} P_5^1(\cos\theta_p) P_1^1(\cos\theta_\mu) \sin \phi\hphantom{2}$ \\
  16 & $\sqrt{\tfrac{11}{3}} P_5^0(\cos\theta_p)P_0^0(\cos\theta_\mu)\hphantom{\cos \phi}$
& 39 & $\tfrac{5}{12 \sqrt{6}} P_2^2(\cos\theta_p) P_2^2(\cos\theta_\mu) \cos 2\phi$ \\
  17 & $\sqrt{11} P_5^0(\cos\theta_p) P_1^0(\cos\theta_\mu)\hphantom{\cos \phi}$
& 40 & $\tfrac{1}{12} \sqrt{\tfrac{7}{6}} P_3^2(\cos\theta_p) P_2^2(\cos\theta_\mu) \cos 2\phi$ \\
  18 & $\sqrt{\tfrac{55}{3}} P_5^0(\cos\theta_p) P_2^0(\cos\theta_\mu)\hphantom{\cos \phi}$
& 41 & $\tfrac{1}{12 \sqrt{2}} P_4^2(\cos\theta_p) P_2^2(\cos\theta_\mu) \cos 2\phi$ \\
  19 & $\sqrt{\tfrac{5}{6}} P_1^1(\cos\theta_p) P_2^1(\cos\theta_\mu) \cos \phi$
& 42 & $\tfrac{1}{12} \sqrt{\tfrac{11}{42}} P_5^2(\cos\theta_p) P_2^2(\cos\theta_\mu) \cos 2\phi$ \\
  20 & $\sqrt{\tfrac{3}{2}} P_1^1(\cos\theta_p) P_1^1(\cos\theta_\mu) \cos \phi$
& 43 & $\tfrac{5}{12 \sqrt{6}} P_2^2(\cos\theta_p) P_2^2(\cos\theta_\mu) \sin 2\phi$ \\
  21 & $\tfrac{5}{3 \sqrt{6}} P_2^1(\cos\theta_p) P_2^1(\cos\theta_\mu) \cos \phi$
& 44 & $\tfrac{1}{12} \sqrt{\tfrac{7}{6}} P_3^2(\cos\theta_p) P_2^2(\cos\theta_\mu) \sin 2\phi$ \\
  22 & $\sqrt{\tfrac{5}{6}} P_2^1(\cos\theta_p) P_1^1(\cos\theta_\mu) \cos \phi$
& 45 & $\tfrac{1}{12 \sqrt{2}}P_4^2(\cos\theta_p) P_2^2(\cos\theta_\mu) \sin 2\phi$ \\
  23 & $\tfrac{1}{6} \sqrt{\tfrac{35}{3}} P_3^1(\cos\theta_p) P_2^1(\cos\theta_\mu) \cos \phi$
& 46 & $\tfrac{1}{12} \sqrt{\tfrac{11}{42}} P_5^2(\cos\theta_p) P_2^2(\cos\theta_\mu) \sin 2\phi$ \\
\bottomrule
\end{tabular}
\label{tab:legendre:basis}
\end{table}

The data set used in this paper corresponds to 9\invfb of integrated luminosity of $pp$ collision data collected with the LHCb experiment between 2011 and 2018. 
This paper is organised as follows. 
Section~\ref{sec:method} introduces the method used to determine the differential branching fraction and the \Kobs{i} angular coefficients. 
Section~\ref{sec:detector} provides a description of the LHCb detector and its simulation. 
Section~\ref{sec:selection} describes the selection of candidates from the LHCb data set. 
Section~\ref{sec:mass} discusses the use of the $p\Km\mumu$ mass distribution, \mpkmm, to separate signal from background in the data set. 
Section~\ref{sec:efficiency} describes correcting weights that are needed to account for the nonuniform response of the detector in $\vec{\Phi}$. 
Sources of systematic uncertainty on the measurement of the branching fraction and angular coefficients are discussed in Sec.~\ref{sec:systematics}.
Results are presented in Sec.~\ref{sec:results} and summarised in Sec.~\ref{sec:summary}.

\section{Methodology}
\label{sec:method} 

The branching fraction of the \Lbpkmm decay is measured relative to that of the \Lbpkjpsi decay, where the \jpsi meson subsequently decays to two oppositely charged muons. 
The differential branching fraction as a function of \qsq and \mpksq is
\begin{align}
    \frac{\deriv^2\BR(\Lbpkmm)}{\deriv\qsq \deriv\mpksq} 
    = \frac{N_{\Lbpkmm}}{N_{\Lbpkjpsi}}\frac{\BR(\Lbpkjpsi)\BR(\decay{\jpsi}{\mumu})
}{\Delta(\qsq,\mpksq)} \ ,
\end{align}
where $N$ are the efficiency-corrected and background-subtracted yields of the decays in the relevant \qsq and \mpk ranges, and $\Delta(\qsq,\mpksq)$ is the area of the \qsq and \mpksq bin.
For the \Lbpkjpsi decay, no restriction is made on the \mpk range. 
The branching fractions of the \mbox{\Lbpkjpsi} and \decay{\jpsi}{\mumu}decays are
$(3.17\,^{+0.57}_{-0.45})\times 10^{-2}$~\cite{LHCb-PAPER-2015-032} and 
$(5.961\pm 0.033)\times 10^{-2}$~\cite{PDG2024}, respectively.ulated by summing over the candidates in the dataset with weight $w(\vec{\Phi})$, 
\begin{equation}\label{eq:yield:value}
    N = \sum\limits_{{\rm event}\,n} w(\vec{\Phi}_n)~.
\end{equation}
The weights are given by the ratio $s_n/\varepsilon(\vec{\Phi}_n)$, where: 
$\varepsilon(\vec{\Phi}_n)$ is the candidate efficiency, which depends on the candidate's position in the five-dimensional phase space; 
while the $s_n$ coefficients are calculated using the \sPlot technique~\cite{Pivk:2004ty} from a fit to the four-body mass of the candidate, and are used to statistically subtract the background in the data set.
For the \Lbpkjpsi decay, the sum runs over all candidates without restriction to a particular \mpk bin.
The variance of $N$ is given by
\begin{align}
\label{eq:yield:uncertainty}
\text{Var}\left(N\right) = \sum\limits_{{\rm event}\,n} (w(\vec{\Phi}_n))^2~.
\end{align}

The angular observables can be determined by calculating the moments of the angular distribution.
The rate-averaged angular coefficient across a bin is given by 
\begin{equation}\label{eq:moment:value}
    \Kobs{i} = \frac{1}{N} \sum\limits_{{\rm event}\,n} w(\vec{\Phi}_n) f_i(\vec{\Omega}_n) \ ,
\end{equation}
as described in Ref.~\cite{Beck:2022spd}. 
These angular observables are extracted with respect to the angular basis given in Table~\ref{tab:legendre:basis}.
Finally, the variance on the angular observables is given by 
\begin{align}
\label{eq:moment:uncertainty}
\begin{split}
\text{Var}\left(\Kobs{i}\right) = 
    \frac{1}{N^2}\sum\limits_{{\rm event}\,n}\left( w(\vec{\Phi}_n) \left(f_i(\vec{\Omega}_n) - \Kobs{i}\right) \right)^2 ~.
\end{split}
\end{align}

\section{Detector and simulation} 
\label{sec:detector} 

The LHCb detector~\cite{LHCb-DP-2008-001,LHCb-DP-2014-002} is a
single-arm forward spectrometer covering the pseudorapidity range $2 < \eta < 5$, designed for the study of \bquark- or \cquark-hadron decays. 
The detector includes a high-precision tracking system
consisting of a silicon-strip vertex detector surrounding the $pp$
interaction region, a large-area silicon-strip detector located
upstream of a dipole magnet with a bending power of about
$4{\mathrm{\,T\,m}}$, and three stations of silicon-strip detectors and straw
drift tubes placed downstream of the magnet.
The tracking system provides a measurement of the momentum, \ptot, of charged particles with
a relative uncertainty better than 1\% for $\ptot < 200\gevc$. 
Different types of charged hadrons are distinguished using information from two ring-imaging Cherenkov detectors. 
Muons are identified by a system comprising alternating layers of iron and multiwire proportional chambers.
The online event selection is performed by a trigger, which comprises a hardware stage followed by two software stages.

Samples of simulated events are used to determine the efficiency, $\varepsilon(\vec{\Phi})$, and to study sources of specific backgrounds. 
In the simulation, $pp$ collisions are generated using \pythia~\cite{Sjostrand:2007gs,*Sjostrand:2006za} with a specific \lhcb configuration~\cite{LHCb-PROC-2010-056}.
Decays of unstable particles are described by \evtgen~\cite{Lange:2001uf}, with final-state radiation generated using \photos~\cite{davidson2015photos}. 
The \evtgen generator does not provide models for \Lbpkmm and \Lbpkjpsi decays that are able to describe the full decay structure. 
Instead, the \Lb baryons are decayed according to phase-space availability.
The \Lbpkjpsi decays are weighted to reproduce the amplitude structure determined in Ref.~\cite{LHCb-PAPER-2015-029}.
The interaction of the generated particles with the detector, and its response, are implemented using the \geant toolkit~\cite{Allison:2006ve, *Agostinelli:2002hh} as described in Ref.~\cite{LHCb-PROC-2011-006}. 
The simulated samples are corrected for known differences between data and simulation in the \Lb production kinematics. 
Percent-level corrections are also applied to the efficiency of the hardware trigger and to the tracking and muon identification efficiencies.
These corrections are derived from control samples of \decay{\Bu}{\jpsi\Kp} decays in the data~\cite{LHCb-DP-2013-002}. 
The particle identification (PID) information for hadrons is corrected by replacing the simulated response with values from cleanly selected samples of protons and kaons with similar kinematic properties to the data of interest~\cite{Poluektov:2014rxa,LHCb-DP-2018-001}.

\section{Selection}
\label{sec:selection} 

The analysis uses data triggered in the hardware stage by either a single high transverse momentum muon or by a pair of muons with a large transverse momentum product. 
The first software stage requires an event to contain at least one good-quality track with significant displacement from every $pp$ collision vertex (PV) and large transverse momentum. 
In the second stage, this track is combined with one or more other tracks and filtered according to topological criteria~\cite{BBDT,LHCb-PROC-2015-018}.
Candidates are formed by combining two muons of opposite charge with a proton and a kaon. 
The muons and hadrons are required to have good-quality tracks, have a significant impact parameter (IP) with respect to every PV and form a common vertex with a good vertex-fit quality. 
The \Lb candidate must have a significant transverse momentum and be significantly displaced from every PV.
The candidate is assumed to originate from the PV with which it has the smallest IP. 
The candidates are required to have an IP consistent with zero and a momentum vector aligned with the direction between the origin and decay vertices.

A lower threshold on the angle between the directions of any pair of charged particles is applied to remove cases where a single charged particle results in multiple reconstructed tracks.
Additional requirements are applied to remove specific sources of background. 
The most dangerous sources of background arise from decays with similar topologies, where one or more particles are incorrectly identified.  
For example, \decay{\Bs}{\Kp\Km\mumu} decays, where the \Kp is mistakenly identified as a proton. 
Such decays are suppressed by requiring additional particle identification criteria if the dihadron and four-body mass combination are consistent with originating from a different decay with a similar topology. 
The kaon to proton misidentification probability after particle identification requirements is typically at the percent level.
Such requirements also remove background contributions from \Lbpkmm decays where the proton is mistakenly identified as a kaon and vice versa. 
Four-body hadronic \bquark-hadron decays form a negligible source of background. 

There is also a potentially large background source from semileptonic \decay{\Lb}{\Lc\mun\bar{\nu}_{\mu}} decays, where the \Lc baryon decays to $p\Km\pip$ and the \pip is mistakenly identified as a muon. 
Such decays are suppressed by vetoing candidates where the $p\Km\pip$ mass is consistent with the \Lc baryon mass, after assigning the \pip mass hypothesis to the \mup candidate. 
Processes involving two semileptonic decays, such as \decay{\Lb}{\Lc\mun\bar{\nu}_{\mu}} with \mbox{\decay{\Lc}{p\Km\mup\nu_{\mu}}}, populate the low $p\Km\mumu$ mass region and can be safely ignored in the analysis.  
Semileptonic decays of \Bs mesons, such as \decay{\Bs}{\Dsm\mup\nu_{\mu}} with \mbox{\decay{\Dsm}{\Kp\Km\mun\bar\nu_{\mu}}}, are suppressed by requiring that the candidate $\Kp\Km\mumu$ mass, formed by assigning the kaon mass hypothesis to the proton candidate, is larger than $5\gevcc$. 

After suppressing the major sources of specific backgrounds, the remaining background contribution is dominated by random combinations of protons, kaons, and two oppositely charged muons.
A significant fraction of the background is found to have a dihadron system consistent with coming from \decay{\phi}{\Kp\Km} decays, where a kaon is mistakenly identified as a proton. 
Additional particle identification criteria are applied to suppress this background.  
The remaining background is further suppressed using a multivariate classifier, a boosted decision tree~\cite{xgboost}, trained to separate simulated signal decays from candidates selected from an upper mass sideband of the data. 
The classifier uses features of the data with low correlations to the phase-space variables and the $p\Km\mumu$ mass: the fit quality of the \Lb decay vertex, the angle between the \Lb momentum and its reconstructed flight direction, the \pt of the \Lb candidate, and the consistency of the candidate with originating from a PV.
The response of the classifier is validated on the \Lbpkjpsi data and simulation samples.
The classifier working point is chosen to maximise the expected signal significance in the data.
The chosen working point removes 98\% of the combinatorial background in the upper mass sideband and retains 74\% of simulated signal decays. 

\section{Mass distribution}
\label{sec:mass} 

The \sPlot procedure~\cite{Pivk:2004ty} is used to statistically disentangle the signal and remaining background.
An unbinned maximum-likelihood fit is performed to the \mpkmm distribution in the different bins using the \zfit software package~\cite{ESCHLE2020100508}.
The signal component is described by a Gaussian distribution with power law tails on both sides of the peak~\cite{Skwarnicki:1986xj}, and the combinatorial background is modelled by an exponential distribution.
The tail parameters of the signal shape are determined using the simulated samples.
The peak position and width parameters of the signal are also obtained from simulation and corrected based on fits to the \Lbpkjpsi control sample. 
All bins share the peak position in simulation, as well as the correction factors for the peak position and width in data.
For the \Lbpkjpsi sample, the other signal parameters are independent between the different \mpk bins.
For the other \qsq bins, the other signal parameters are shared between bins with the same \qsq but different \mpk values.
Due to the larger size of the \Lbpkjpsi sample, the models are modified to better describe the shapes seen in the data:   
the signal shape is modified to include sigmoid activation functions on the left- and right-hand side of the distribution to account for the effect of the finite dimuon mass window on \mpkmm;
and the background shape is modified by multiplying it by
\begin{align}
\Theta(\mpkmm - m_{{\rm thr}}) (\mpkmm - m_{{\rm thr}})^{\delta}~,
\end{align} 
where $\Theta$ is the Heaviside step function, $m_{{\rm thr}}$ is the smallest allowed $p\Km\mumu$ mass and is given by the sum of the known \jpsi mass~\cite{PDG2024} and \mpk, and $\delta$ is a parameter that controls the turn-on of the shape from the threshold.

\begin{figure}[htbp!]
    \centering
    \includegraphics[width=\textwidth]{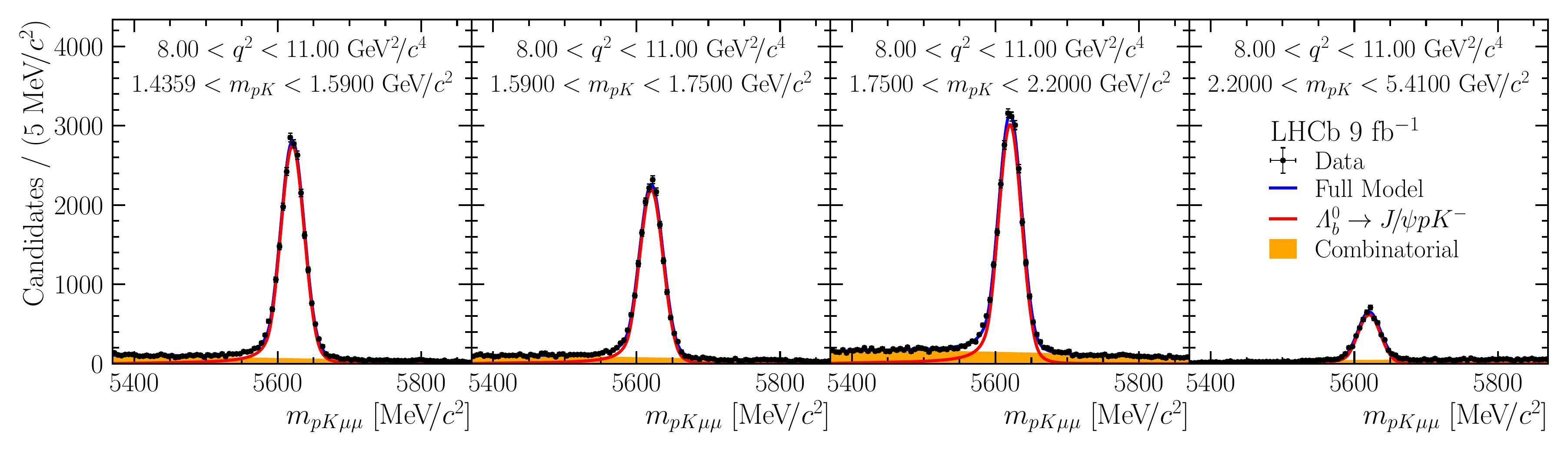}
    \caption{
    Mass distribution of selected \Lbpkjpsi candidates in bins of \mpk. 
    The data are overlaid with the result of the fit described in the text.   
    }
    \label{fig:massfit:jpsi}
\end{figure}

\begin{figure}[htbp!]
    \centering
    \includegraphics[width=\textwidth]{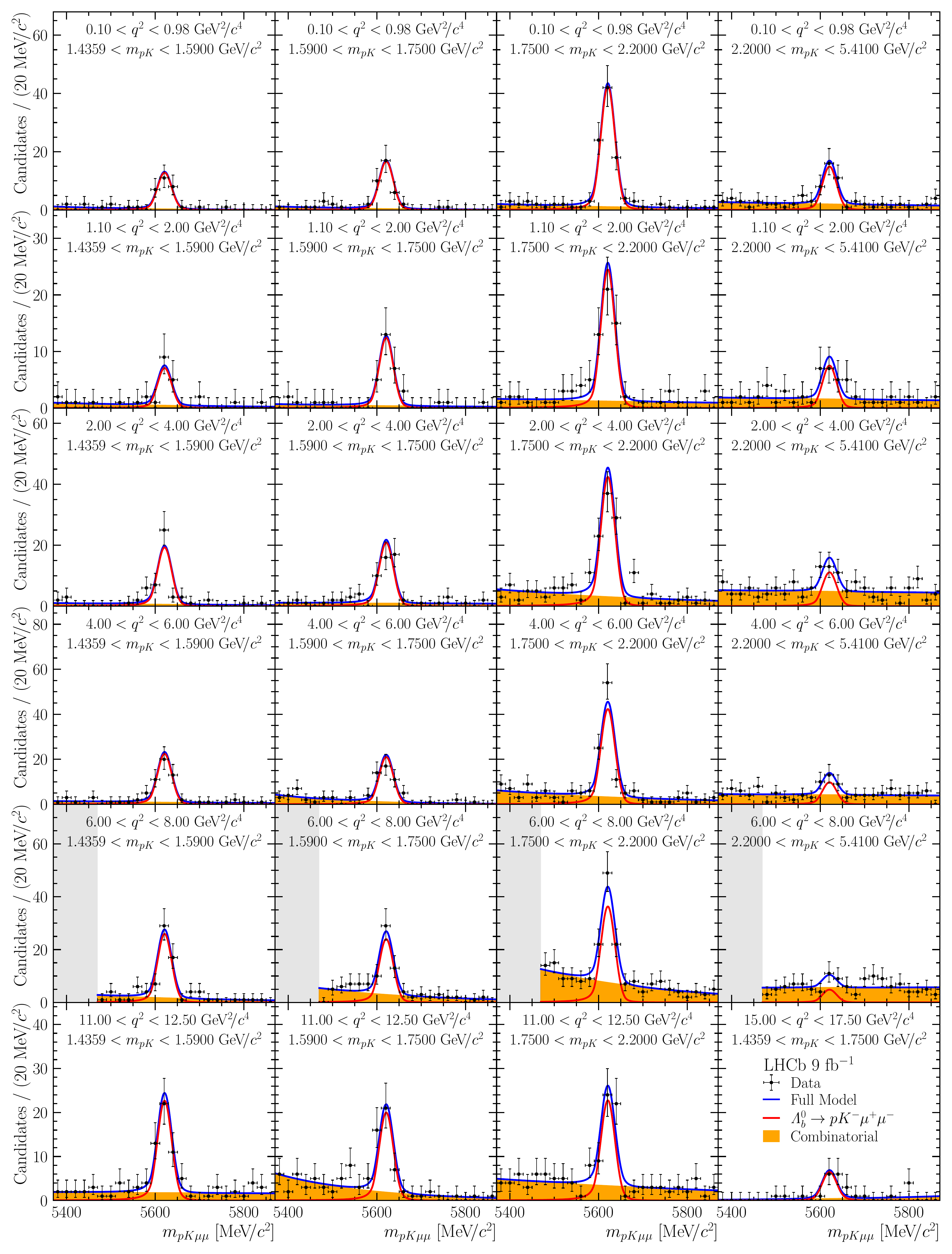}
    \caption{
    Mass distributions of selected \Lbpkmm decay candidates in bins of \mpk and \qsq.
    The data are overlaid with the result of the fit described in the text. 
    The shaded regions are populated by poorly reconstructed \Lbpkjpsi decays and are excluded from the analysis.      
    }
    \label{fig:massfit:rare}
\end{figure}

Figures~\ref{fig:massfit:jpsi} and \ref{fig:massfit:rare} show the \mpkmm distribution of the candidates in the different \qsq and \mpk bins. 
The distributions are overlaid with the result of the fit. 
In the \mbox{$6 < \qsq < 8$\gevgevcccc} bin, the range of the $m_{pK\mu\mu}$ distribution is reduced to remove background from poorly reconstructed \decay{\Lb}{\jpsi\proton\Km} decays that would otherwise pollute the low mass region.
Table~\ref{tab:appendix:yields} in the Appendix contains the yields of signal and background obtained from the fits.
The angular observables are determined for every bin of \qsq and \mpk that has a significant signal contribution.
For $2.20 < \mpk < 5.41\gevcc$, and the bin with $1.1 < \qsq < 2.0\gevgevcccc$ and $1.4359 < \mpk < 1.5900\gevcc$, the signal yields are so small that the angular observables cannot be reliably determined. 

\section{Efficiency model}
\label{sec:efficiency}

The efficiency in $\vec{\Phi}$ is parameterised by 
\begin{align}
    \varepsilon(\vec{\Phi}) = \sum\limits_{hijkl} e_{hijkl} P_{h}(m'_{pK})P_{i}(m_{\mu\mu}')P_{j}(\cos\theta_p)P_{k}(\cos\theta_\mu) \cos (l\phi)~,
\end{align}
where $P_n(x)$ are Legendre polynomials of order $n$, the $e_{hijkl}$ are a set of coefficients, and $m'_{pK}$ and $m'_{\mu\mu}$ are a mapping of the dihadron and dimuon masses to a square with range $[-1,+1]$. 
This model makes no assumptions about the factorisation of the different observables. 
Legendre polynomials and $\cos (l\phi)$ dependencies are used since these form an orthogonal basis.
As a result, the coefficients of the efficiency model are determined using the method of moments applied to the phase-space simulation samples.
The efficiency is parameterised using polynomials of up-to and including: order five for the transformed masses, $m'_{\mu\mu}$ and $m'_{pK}$; order four for $\cos\theta_p$; and order four for $\cos\theta_\mu$. 
The angle $\phi$ is described by functions up-to $\cos (3\phi)$.
The normalisation of the efficiency is arbitrary, and on average candidates are assigned values of 1.0. 
If $\varepsilon(\vec{\Phi}) < 0.15$, candidates are assigned an efficiency of 0.15 to avoid introducing large weights when determining the observables. 
This only affects a small number of candidates that all sit in the sidebands of the data. 
The impact of the minimum efficiency requirement is considered as a source of systematic uncertainty. 

The shape of $\varepsilon(\vec{\Phi})$ arises from kinematic and geometrical requirements in the selection and reconstruction of the proton, kaon and muons. 
The shape in $\cos\theta_\mu$ is symmetric as the detection asymmetry between positively and negatively charged muons is negligible~\cite{LHCb-PAPER-2013-033}, therefore odd order contributions are not considered in the model. 
Conversely, the shape in $\cos\theta_p$ is asymmetric due to the difference between the proton and kaon mass, leading to a momentum imbalance between the particles in the detector, and due to different momentum-dependent efficiencies of the PID requirements. 
For $\cos\theta_p$, both odd and even order contributions are used. 
The efficiency tends to be largest at intermediate \qsq and lower at the \qsq extremes.  
At small \qsq, the muons typically have smaller momentum and \pt, and are either not reconstructed, or do not meet the requirements of the hardware stage of the trigger.
At large \qsq, the hadrons are almost at rest in the \Lb rest frame and have small IP, and do not meet the requirements of the software trigger or offline selection.

The efficiency model is validated by inspecting the agreement between one-dimensional projections of the model and the simulation samples.
The agreement in the five-dimensional space is also checked using a multivariate classifier trained to separate the simulation sample and pseudoexperiments generated from the efficiency model. 
Several different combinations of polynomial orders are found to give a similar description of the simulated sample. 
This set of models is used to assign a systematic uncertainty on $\varepsilon(\vec{\Phi})$.

\section{Statistical and systematic uncertainties}
\label{sec:systematics}

The statistical uncertainty on the branching fraction measurements and the angular observables are determined using the variances defined in Sec.~\ref{sec:method}.
The appropriateness of the variance, as a measure of the uncertainty, is verified with pseudoexperiments. 
An alternative estimate, obtained by bootstrapping the data set~\cite{efron:1979} and repeating the determination of the different observables, is also considered. 
The two determinations give comparable results for bins that are well populated, but the variance is found to have better coverage in poorly populated bins and to cover correctly for the observables presented in this paper. 

Several sources of systematic uncertainties are considered.
For the differential branching fraction, the dominant contribution arises from the knowledge of the branching fraction of the \Lbpkjpsi decay.
The largest systematic uncertainty on the angular coefficients stems from the knowledge of the efficiency model.
Table~\ref{tab:syst} provides a summary of the different sources of systematic uncertainty. 
The total systematic uncertainty on the observables is determined by summing the individual sources in quadrature. 

\begin{table}[!tb]
    \centering
    \caption{
    Summary of different sources of systematic uncertainty on the differential branching fraction and angular coefficients, relative to the statistical uncertainty on the measurements.
    The values correspond to the median over all of the measured observables and bins. 
    The absolute range is given in parentheses.
    }
    \label{tab:syst}
    \begin{tabular}{lcccc}
        \toprule
        Source of systematic uncertainty & $\deriv^2\mathcal{B}/\deriv\qsq\deriv\mpksq$ & $\Kobs{i}$ \\
        \midrule
        $\mathcal{B}(\Lbpkjpsi)$ & 1.03 (0.35--1.58) & --- \\
        \midrule
        Efficiency model (sample size)      & 0.09 (0.03--0.25) & 0.07 (0.01--0.40) \\
        Efficiency model (polynomial order) & 0.10 (0.03--0.47) & 0.08 (0.01--1.15) \\
        Efficiency model (minimum value)    & 0.01 (0.00--0.22) & 0.00 (0.00--0.99) \\
        Efficiency model (\Lc veto)         & 0.01 (0.00--0.16) & 0.00 (0.00--0.27) \\
        \midrule
        Mass model (signal)      & 0.02 (0.00--0.18) & 0.01 (0.00--0.17) \\
        Mass model (background)  & 0.05 (0.00--0.95) & 0.01 (0.00--0.20) \\
        Mass model (sample size) & 0.01 (0.00--0.09) & 0.00 (0.00--0.05) \\
        \midrule
        Peaking backgrounds & 0.02 (0.00--0.09) & 0.01 (0.00--0.16) \\
        \midrule
        Resolution & 0.00 (0.00--0.02) & 0.00 (0.00--0.18) \\
        \midrule
        Simulation corrections (PID)           & 0.02 (0.00--0.08) & 0.01 (0.00--0.36) \\
        Simulation corrections (hadronisation) & 0.00 (0.00--0.02) & 0.00 (0.00--0.01) \\
        Simulation corrections (kinematic)     & 0.02 (0.00--0.06) & 0.01 (0.00--0.08) \\
        \bottomrule
    \end{tabular}
\end{table}

Four sources of systematic uncertainty are evaluated for the efficiency model.
The first source is due to the limited size of the simulation sample. 
To assess the systematic impact of this limitation, the simulated sample is bootstrapped 100 times and the determination of the efficiency model is repeated.
The observables are recalculated for each alternative efficiency model and the widths of the resulting distributions of the observables are taken as systematic uncertainties. 
The second source is due to the choice of the truncation order of the polynomials used in the efficiency model. 
The size of this effect is quantified by determining different efficiency models by either increasing or decreasing the polynomial order of the different dimensions by one, considering only the models that provide a similarly good description of the data to the baseline model.
The observables are then determined using these alternative efficiency models.
The largest deviation from the default value among the models is assigned as the systematic uncertainty on each observable.
Only an increase and decrease of one is considered because very high orders of the polynomial can lead to local regions with small or negative efficiencies in $\varepsilon(\vec{\Phi})$. 
The third source is due to the choice of the minimum efficiency. 
The impact of this choice is estimated by repeating the analysis on a single pseudoexperiment with a large sample size with the minimum efficiency set to 0.01.
The difference in the value of observables obtained using the default minimum and the new minimum efficiency is taken as systematic uncertainty.
The \Lc veto removes a very narrow region of phase space, which is difficult to model with polynomial functions.
Because the efficiency of the \Lc veto is 99.8\% on signal simulation samples, and no candidates fall in this region in data, this veto is not applied when determining the efficiency shape.
The fourth source of uncertainty accounts for this choice and is also determined using a single large pseudoexperiment.
The difference in absolute value between the observables calculated from the sample when neglecting or including the \Lc veto is taken as systematic uncertainty.

Three sources of systematic uncertainty are evaluated for the mass models.
The first source is due to the statistical uncertainty on the signal line shape. 
This uncertainty is assessed by varying the line shape parameters within their uncertainties, repeating the fits to the data and determining new values for the observables. 
The width of the resulting distributions of the observables are taken as systematic uncertainties.
The second and third sources are due to the choice of signal and background model. 
The systematic effect of the model choice is estimated using pseudoexperiments generated according to an alternative signal or background model. 
The observables are then determined using the baseline signal and background models and the resulting bias is assigned as a source of systematic uncertainty. 
For the signal, an alternative shape based on a modified asymmetric Apollonios function is used~\cite{Santos:2013gra}. 
For the background, a polynomial dependence is used with parameters determined from a sample selected with a looser multivariate classifier requirement. 
The uncertainty due to different signal and background line shapes is largest for $1.75 < \mpk < 2.20\gevcc$.

Contributions from misidentified backgrounds with the same topology as the signal are small and neglected in the analysis.
To estimate a systematic uncertainty due to omitting these backgrounds, a pseudoexperiment is generated with a large sample size according to the baseline signal and background models. 
Additional background contributions from misidentified particles are introduced at the expected level. 
The observables are then determined using the baseline signal and background models and the resulting bias is assigned as a systematic uncertainty. 
As little is known of the structure of the \decay{\Bs}{\Km\Kp\mumu} and \decay{\Bz}{\Kp\pim\mumu} decays, aside from in the region around the $\phi$ and $\Kstar(892)^0$ resonances, the additional candidates are selected from the data after applying alternative particle identification requirements. 

The resolution of the measured particle momentum has little impact on the measured angular observables and is neglected in the analysis.  
To assess the impact of this choice, a pseudoexperiment with a large sample size is generated in which $\theta_\mu$, $\theta_p$ and $\phi$ are smeared according to the resolutions on the angles determined from simulation.
The observables are then determined using the baseline model and the bias on the observables is assigned as a systematic uncertainty.

The determination of the efficiency is based on simulation samples that are calibrated to better describe the data.
The overall value of the calibrating weight is dominated by the correction of the \Lb production kinematics.
The kinematic correction is designed to correctly reproduce the \pt dependence of the \Lb production fraction relative to \Bz and \Bp mesons measured in Refs.~\cite{LHCB-PAPER-2011-018} and \cite{LHCb-PAPER-2018-050}, as well as the kinematic distributions of \Bz and \Bu mesons in the data. 
A systematic uncertainty on the correction to the \Lb kinematics is assessed in two ways.
First, the measurements in Refs.~\cite{LHCB-PAPER-2011-018} and \cite{LHCb-PAPER-2018-050} are varied within their uncertainties and new efficiency models are determined for each variation.
The observables are determined in data using the resulting alternate efficiency models.
The width of the resulting distributions of values of the different observables is assigned as a systematic uncertainty. 
Second, the correction weights are also extended to include the detector occupancy. 
The analysis is then repeated and the change in the values of the observables is assigned as a systematic uncertainty.
A systematic uncertainty on the hadron identification is also estimated by evaluating the efficiency model using information sampled from an alternative set of calibration samples.
Overall, the systematic uncertainties related to the calibration are small.

\section{Results and discussion}
\label{sec:results} 

\begin{figure}[!tb]
    \centering
    \includegraphics[width=0.49\textwidth]{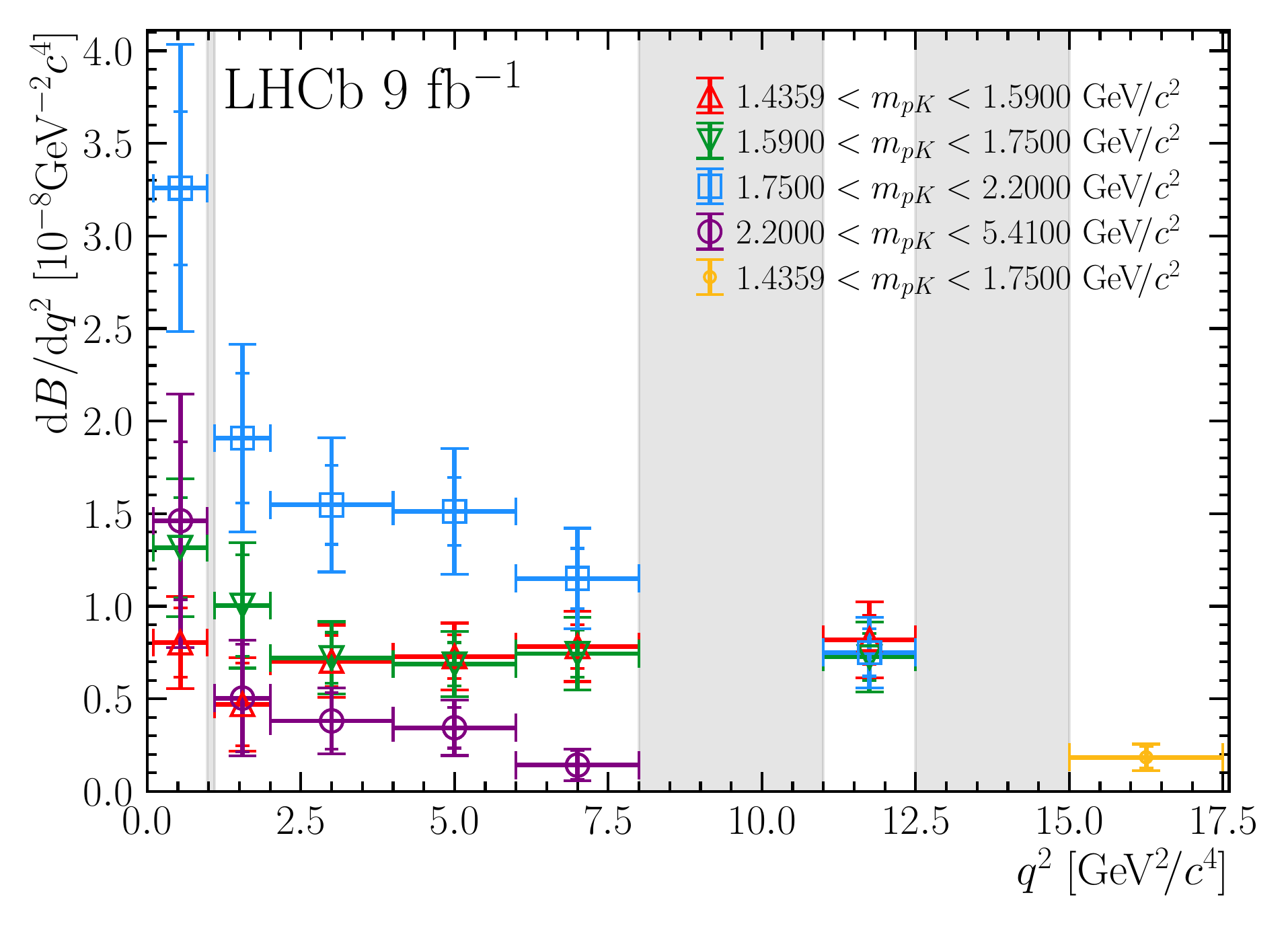}%
    \includegraphics[width=0.47\textwidth]{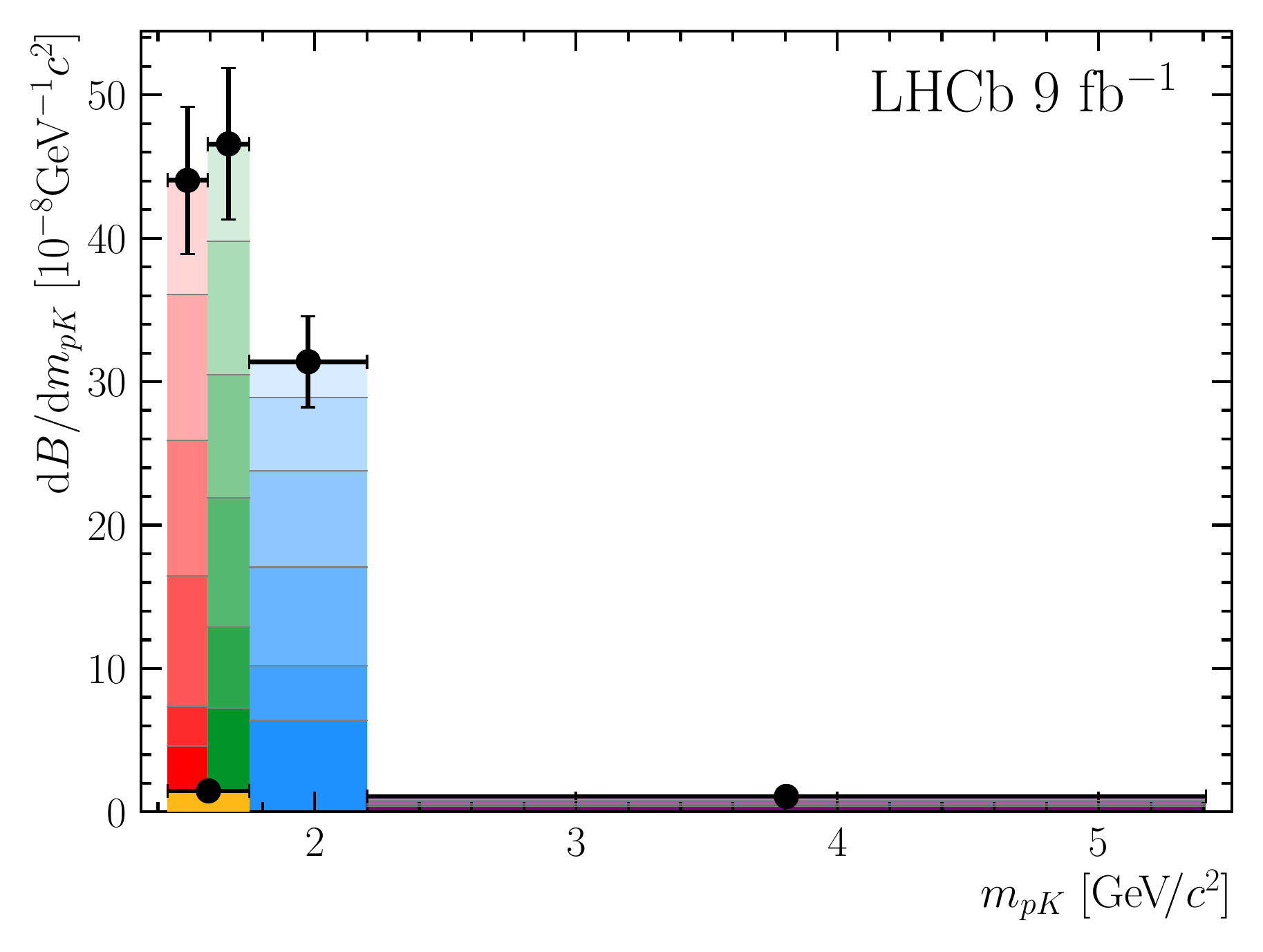}
    \caption{
    Differential branching fraction as a function of (left) \qsq and (right) \mpk.
    The stacked contributions with different shading in the right figure indicate the contributions from the different \qsq bins. 
    The darker hue corresponds to smaller values of \qsq. 
    }
    \label{fig:branchingfraction}
\end{figure}

Table~\ref{tab:branchingfraction} gives the differential branching fraction for every bin. 
The precision in most bins is limited by the knowledge of the \Lbpkjpsi branching fraction.
Figure~\ref{fig:branchingfraction} shows the results in bins of \qsq and \mpk.
The branching fraction as a function of \qsq in the first \mpk bin is compatible with the results presented in Ref.~\cite{LHCb-PAPER-2022-050}.
The branching fraction obtained by summing the contributions from different \qsq and \mpk bins over the ranges $0.1<\qsq<6.0\gevgevcccc$ and $\mpk < 2.6\gevcc$ is also compatible with the measurement presented in Ref.~\cite{LHCb-PAPER-2019-040}.
The variation of the differential branching fraction with \qsq obtained in this paper does not agree with predictions based on a quark model~\cite{Mott:2011cx} in any of the \mpk bins. 
The quark model predictions typically yield much smaller branching fractions at low-\qsq than seen in the data. 
A direct interpretation of the differences between the differential branching fraction in bins of \mpk and the resonance spectra in \Lbpkjpsi~\cite{LHCb-PAPER-2015-029} and \Lbpkg~\cite{LHCb-PAPER-2023-036} decays is difficult, due to the unknown interference pattern between the states in this analysis. 
The results are, however, qualitatively similar, with a variety of different resonances contributing to the total rate. 
The pattern is also consistent with the available phase space in the different decays. 

\begin{table}[!bt]
    \centering
    \caption{Differential branching fraction, $\deriv^2\BF/\deriv\qsq \deriv\mpksq$, in units of $10^{-8}\gev^{-4}c^8$ in bins of \qsq and \mpk.
    The first uncertainty is statistical, the second systematic, and the third due to the uncertainty on the \Lbpkjpsi branching fraction.
    The bin ranges are given in \gevcc for \mpk and in \gevgevcccc for \qsq.
    }
    \label{tab:branchingfraction}
    \scalebox{0.72}{
    \makebox[\textwidth]{
    \begin{tabular}{c|cccc}
    \toprule
    \backslashbox{\qsq}{\mpk} & $[1.4359,1.5900]$ & $[1.59,1.75]$ & $[1.75,2.20]$ & $[2.20,5.41]$ \\
    \midrule
    $[0.10,0.98]$ &  $5.22\pm1.21\pm0.43\pm0.98$&  $8.22\pm1.69\pm0.38\pm1.54$&  $7.24\pm0.92\pm0.52\pm1.36$&  $0.46\pm0.13\pm0.14\pm0.09$\\
    $[1.1,2.0]$ &  $3.05\pm1.45\pm0.51\pm0.57$&  $6.27\pm1.71\pm0.40\pm1.18$&  $4.24\pm0.78\pm0.16\pm0.80$&  $0.16\pm0.09\pm0.02\pm0.03$\\
    $[2.0,4.0]$ &  $4.56\pm0.90\pm0.26\pm0.86$&  $4.50\pm0.86\pm0.21\pm0.84$&  $3.44\pm0.47\pm0.08\pm0.64$&  $0.12\pm0.05\pm0.02\pm0.02$\\
    $[4.0,6.0]$ &  $4.72\pm0.76\pm0.15\pm0.89$&  $4.29\pm0.73\pm0.20\pm0.81$&  $3.36\pm0.41\pm0.07\pm0.63$&  $0.11\pm0.03\pm0.02\pm0.02$\\
    $[6.0,8.0]$ &  $5.08\pm0.76\pm0.12\pm0.95$&  $4.65\pm0.79\pm0.34\pm0.87$&  $2.56\pm0.36\pm0.05\pm0.48$&  $0.04\pm0.02\pm0.01\pm0.01$\\
    $[11,12.5]$ &  $5.32\pm0.86\pm0.20\pm1.00$&  $4.53\pm0.80\pm0.16\pm0.85$&  $1.67\pm0.28\pm0.03\pm0.31$& --- \\
    $[15.0,17.5]$ & \multicolumn{2}{c}{ $0.59\pm0.19\pm0.07\pm0.11$} & --- & --- \\
    \bottomrule
    \end{tabular}
    }
    }
\end{table}

\begin{figure}[!htb]
     \centering
     \includegraphics[width=0.49\textwidth]{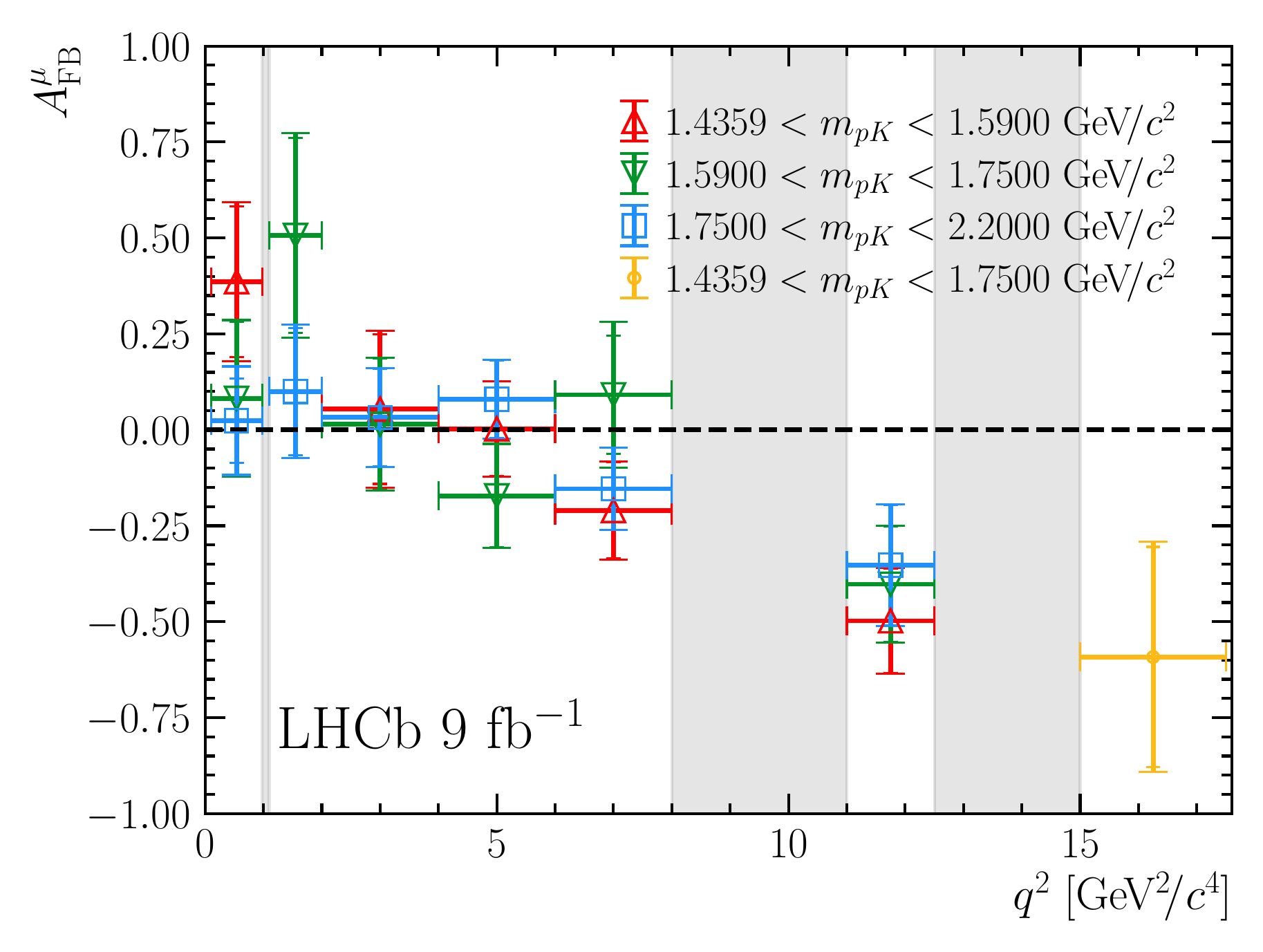}
     \includegraphics[width=0.49\textwidth]{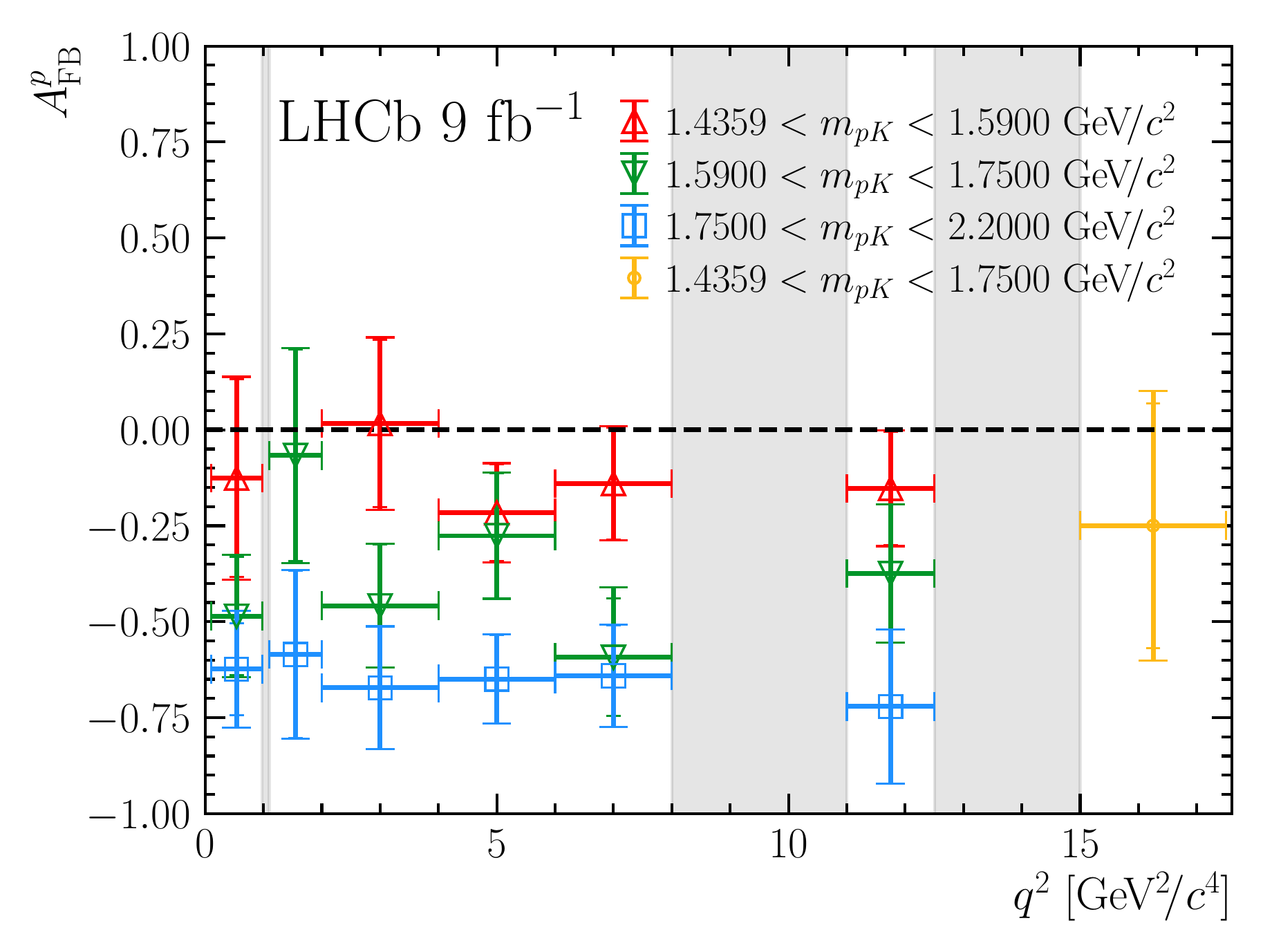}
     \caption{
     Values of the (left) dimuon forward-backward asymmetry and (right) dihadron forward-backward asymmetry in bins of \qsq and \mpk.
    }
     \label{fig:results:afb}
\end{figure}

Figures~\ref{fig:appendix:kobs:1}--\ref{fig:appendix:kobs:6} in the Appendix show the values of the complete set of angular observables.
Tabulated values of all of the observables are available as attached supplemental information.
Figure~\ref{fig:results:afb} shows the forward-backward asymmetries of the lepton and hadron systems computed from the observables. 
The lepton- and hadron-side forward-backward asymmetries correspond to
\begin{align}
A_{\rm FB}^{\mu} = \tfrac{3}{2}\Kobs{2} \quad\text{and}\quad A_{\rm FB}^{p} = \tfrac{3}{2} \Kobs{4} - \tfrac{\sqrt{21}}{8} \Kobs{10} + \tfrac{\sqrt{33}}{16}\Kobs{16}\,,
\end{align}
respectively~\cite{Beck:2022spd}. 
The lepton-side asymmetry is sensitive to interference between vector and axial-vector contributions to the decay. 
This asymmetry shows the same pattern observed in \decay{\Bz}{\Kstarz\mumu} decays with a characteristic sign-change between low and high \qsq~\cite{LHCb-PAPER-2020-002}. 
Note that the sign of the lepton-side asymmetry in this paper differs due to the angular basis used. 
A large hadron-side asymmetry is seen in many of the \qsq and \mpk bins, especially for $1.75 < \mpk < 2.20\gevcc$.
The hadron-side asymmetry is sensitive to the interference between states with different quantum numbers.

\section{Summary}
\label{sec:summary}

An analysis of the rate and angular distribution of \Lbpkmm decays, using data collected with \lhcb detector between 2011--2018, has been presented.
The analysis results in a first measurement of the differential branching fraction of the \Lbpkmm decay across its entire phase space, in bins of the dihadron mass and \qsq.  
The decay rate is dominated by contributions from resonances at low dihadron masses.
This paper also provides a first measurement of a complete set of angular observables in \Lbpkmm decays for \Lz states with spin less than $\tfrac{5}{2}$. 
These measurements are only provided in bins with sufficient signal yield. 
The angular coefficients indicate the presence of interference between states with different quantum numbers. They also show the pattern of interference between vector and axial-vector contributions that is characteristic of this type of rare FCNC decay. 
The pattern of measurements appears consistent with SM expectations. 
However, a detailed interpretation of the results requires a more complete understanding of the hadronic system and the different contributing states.

%% file: acknowledgements.tex
\section*{Acknowledgements}
%
%
\noindent We express our gratitude to our colleagues in the CERN
accelerator departments for the excellent performance of the LHC. We
thank the technical and administrative staff at the LHCb
institutes.
We acknowledge support from CERN and from the national agencies:
CAPES, CNPq, FAPERJ and FINEP (Brazil); 
MOST and NSFC (China); 
CNRS/IN2P3 (France); 
BMBF, DFG and MPG (Germany); 
INFN (Italy); 
NWO (Netherlands); 
MNiSW and NCN (Poland); 
MCID/IFA (Romania); 
MICIU and AEI (Spain);
SNSF and SER (Switzerland); 
NASU (Ukraine); 
STFC (United Kingdom); 
DOE NP and NSF (USA).
We acknowledge the computing resources that are provided by CERN, IN2P3
(France), KIT and DESY (Germany), INFN (Italy), SURF (Netherlands),
PIC (Spain), GridPP (United Kingdom), 
CSCS (Switzerland), IFIN-HH (Romania), CBPF (Brazil),
and Polish WLCG (Poland).
We are indebted to the communities behind the multiple open-source
software packages on which we depend.
Individual groups or members have received support from
ARC and ARDC (Australia);
Key Research Program of Frontier Sciences of CAS, CAS PIFI, CAS CCEPP, 
Fundamental Research Funds for the Central Universities, 
and Sci. \& Tech. Program of Guangzhou (China);
Minciencias (Colombia);
EPLANET, Marie Sk\l{}odowska-Curie Actions, ERC and NextGenerationEU (European Union);
A*MIDEX, ANR, IPhU and Labex P2IO, and R\'{e}gion Auvergne-Rh\^{o}ne-Alpes (France);
AvH Foundation (Germany);
ICSC (Italy); 
Severo Ochoa and Mar\'ia de Maeztu Units of Excellence, GVA, XuntaGal, GENCAT, InTalent-Inditex and Prog. ~Atracci\'on Talento CM (Spain);
SRC (Sweden);
the Leverhulme Trust, the Royal Society
 and UKRI (United Kingdom).

%% file: appendix.tex
\clearpage 
\section*{Appendices}

\appendix

\section{Signal and background yields}

Table~\ref{tab:appendix:yields} provides the observed signal and background yield in each of the $\qsq$ and \mpk bins from the unbinned extended maximum-likelihood fit to $p\Km\mumu$ distributions. 
The uncertainties on the yields are calculated from the profile likelihood~\cite{James:1975dr}.

\begin{table}[!hbt]
    \centering
    \caption{
    Signal and background yields obtained from fitting $p\Km\mumu$ mass distributions in the data in the different bins of \qsq and \mpk. 
    }
    \label{tab:appendix:yields}
\bgroup
\def\arraystretch{1.2}
    \begin{tabular}{cc|cc}
    \toprule
    \qsq $[\!\gev^{2}/c^4]$ & \mpk $[\!\gevcc]$ & Signal & Background \\
    \midrule
    \multirow{4}{*}{0.10--0.98} 
    & 1.4359--1.5900 & $27^{+6\hphantom{1}}_{-5\hphantom{1}}$ & $11^{+4}_{-3}$ \\
    & 1.59--1.75 & $35^{+7\hphantom{1}}_{-6\hphantom{1}}$ & $13^{+5}_{-4}$ \\
    & 1.75--2.20 & $90^{+10}_{-10}$ & $29^{+7}_{-6}$ \\
    & 2.20--5.41 & $32^{+7\hphantom{1}}_{-6\hphantom{1}}$ & $51^{+8}_{-8}$ \\
    \midrule
    \multirow{4}{*}{1.1--2.0} 
    & 1.4359--1.5900 &  $15^{+5}_{-4}$ & $13^{+4}_{-4}$ \\
    & 1.59--1.75 & $27^{+6}_{-5}$ & $10^{+4}_{-3}$ \\
    & 1.75--2.20 & $52^{+8}_{-8}$ & $31^{+7}_{-6}$ \\
    & 2.20--5.41 & $16^{+6}_{-5}$ & $40^{+8}_{-7}$ \\
    \midrule
    \multirow{4}{*}{2.0--4.0} 
    & 1.4359--1.5900 &  $42^{+7\hphantom{1}}_{-7\hphantom{1}}$ & $15^{+5}_{-4}$ \\
    & 1.59--1.75 & $45^{+8\hphantom{1}}_{-7\hphantom{1}}$ & $24^{+6}_{-5}$ \\
    & 1.75--2.20 & $92^{+11}_{-11}$ & $\hphantom{1}81^{+11}_{-10}$ \\
    & 2.20--5.41 & $24^{+7\hphantom{1}}_{-7\hphantom{1}}$ & $121^{+12}_{-12}$ \\
    \midrule
    \multirow{4}{*}{4.0--6.0} 
    & 1.4359--1.5900 &  $48^{+8\hphantom{1}}_{-7\hphantom{1}}$ & $23^{+6}_{-5}$ \\
    & 1.59--1.75 & $45^{+8\hphantom{1}}_{-7\hphantom{1}}$ & $36^{+7}_{-6}$ \\
    & 1.75--2.20 & $91^{+11}_{-10}$ & $\hphantom{1}87^{+11}_{-10}$ \\
    & 2.20--5.41 & $21^{+7\hphantom{1}}_{-6\hphantom{1}}$ & $105^{+12}_{-11}$ \\
    \midrule
    \multirow{4}{*}{6.0--8.0} 
    & 1.4359--1.5900 &  $55^{+9\hphantom{1}}_{-8\hphantom{1}}$ & $31^{+7}_{-6}$ \\
    & 1.59--1.75 & $51^{+9\hphantom{1}}_{-8\hphantom{1}}$ & $54^{+9}_{-8}$ \\
    & 1.75--2.20 & $77^{+11}_{-10}$ & $137^{+14}_{-13}$ \\
    & 2.20--5.41 & $10^{+6\hphantom{1}}_{-5\hphantom{1}}$ & $113^{+12}_{-11}$ \\
    \midrule
    \multirow{3}{*}{11.0--12.5} 
    & 1.4359--1.5900 &  $49^{+8\hphantom{1}}_{-8\hphantom{1}}$ & $44^{+8}_{-7}$ \\
    & 1.59--1.75 & $43^{+8\hphantom{1}}_{-7\hphantom{1}}$ & $59^{+9}_{-8}$ \\
    & 1.75--2.20 & $49^{+9\hphantom{1}}_{-8\hphantom{1}}$ & $\hphantom{1}86^{+11}_{-10}$ \\
    \midrule
    15.0--17.0 & 1.4359--1.5900 &  $14^{+5\hphantom{1}}_{-4\hphantom{1}}$ & $12^{+4}_{-4}$ \\
    \bottomrule
    \end{tabular}
\egroup
\end{table}

\section{Angular observables}

\begin{figure}[!htb]
     \centering
     \includegraphics[width=0.48\textwidth]{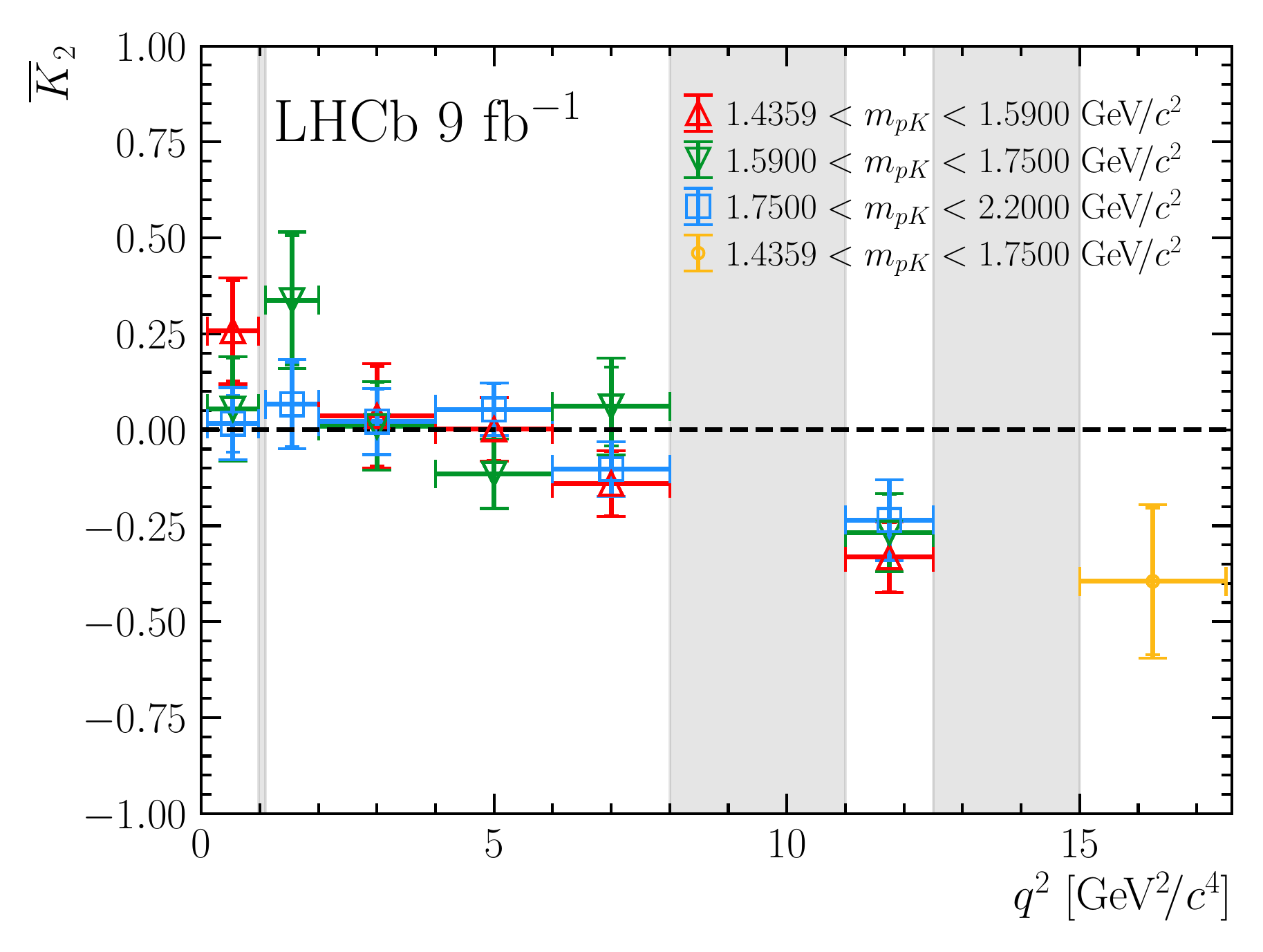}
     \includegraphics[width=0.48\textwidth]{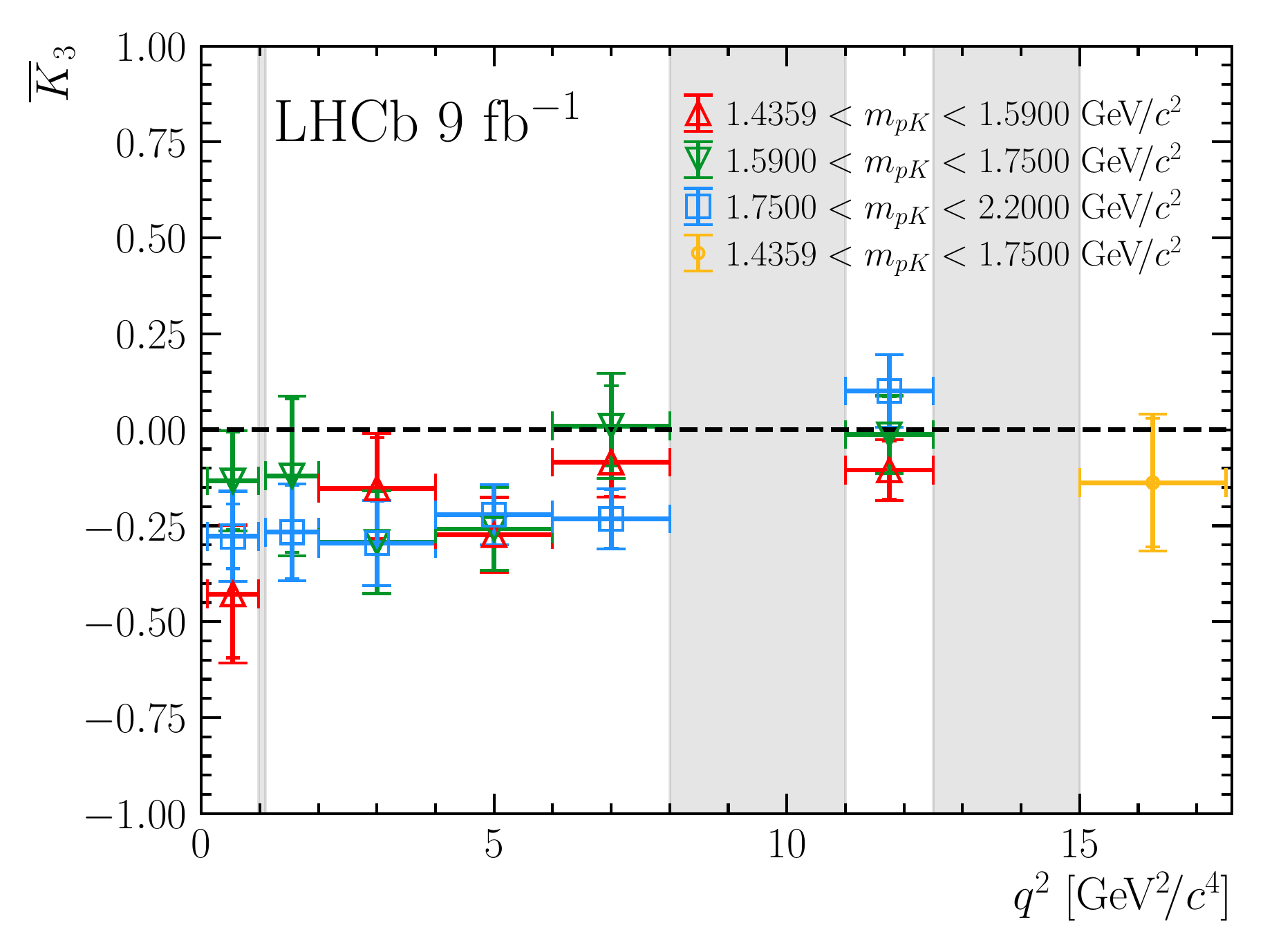} \\
     \includegraphics[width=0.48\textwidth]{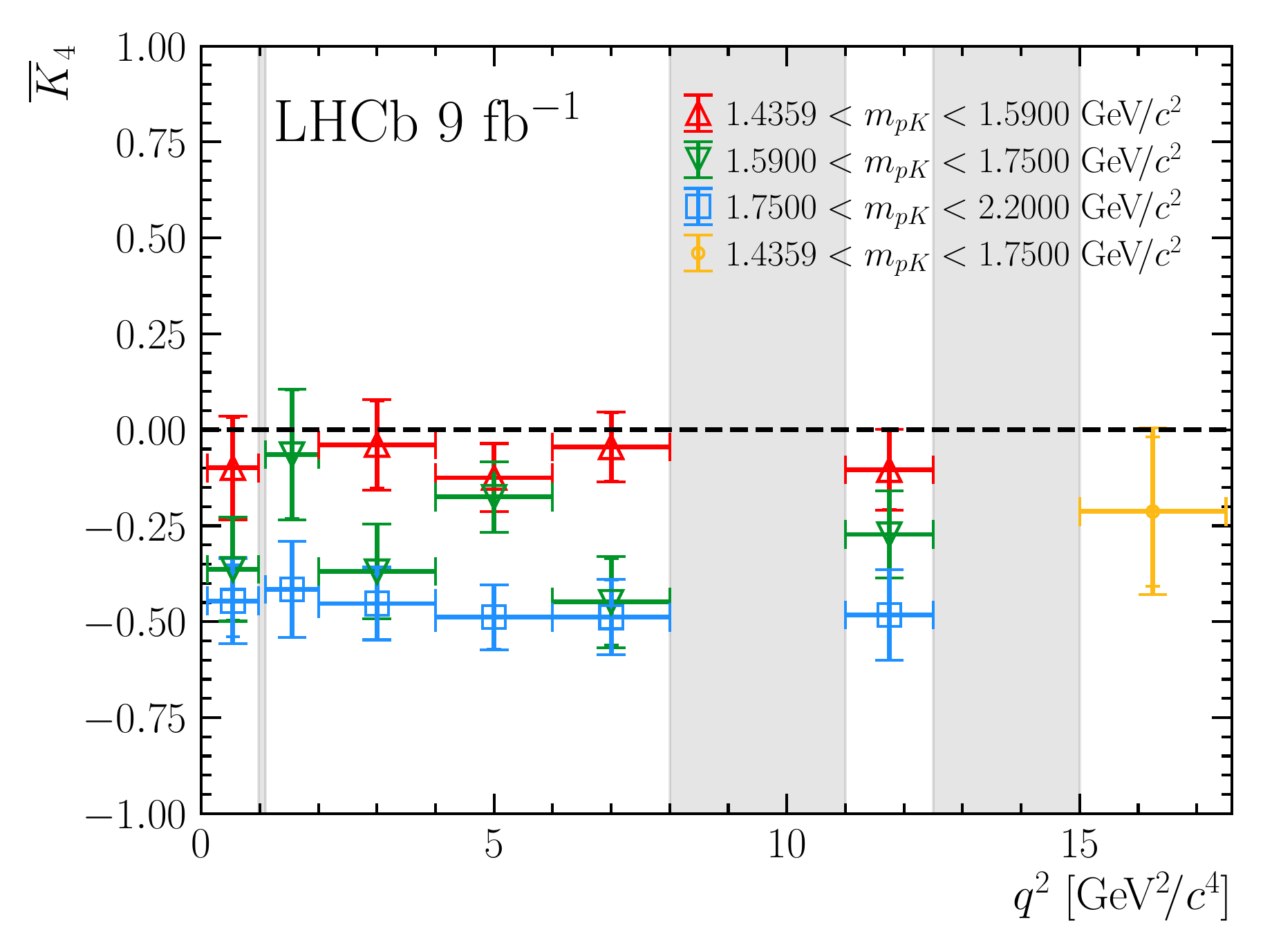}
     \includegraphics[width=0.48\textwidth]{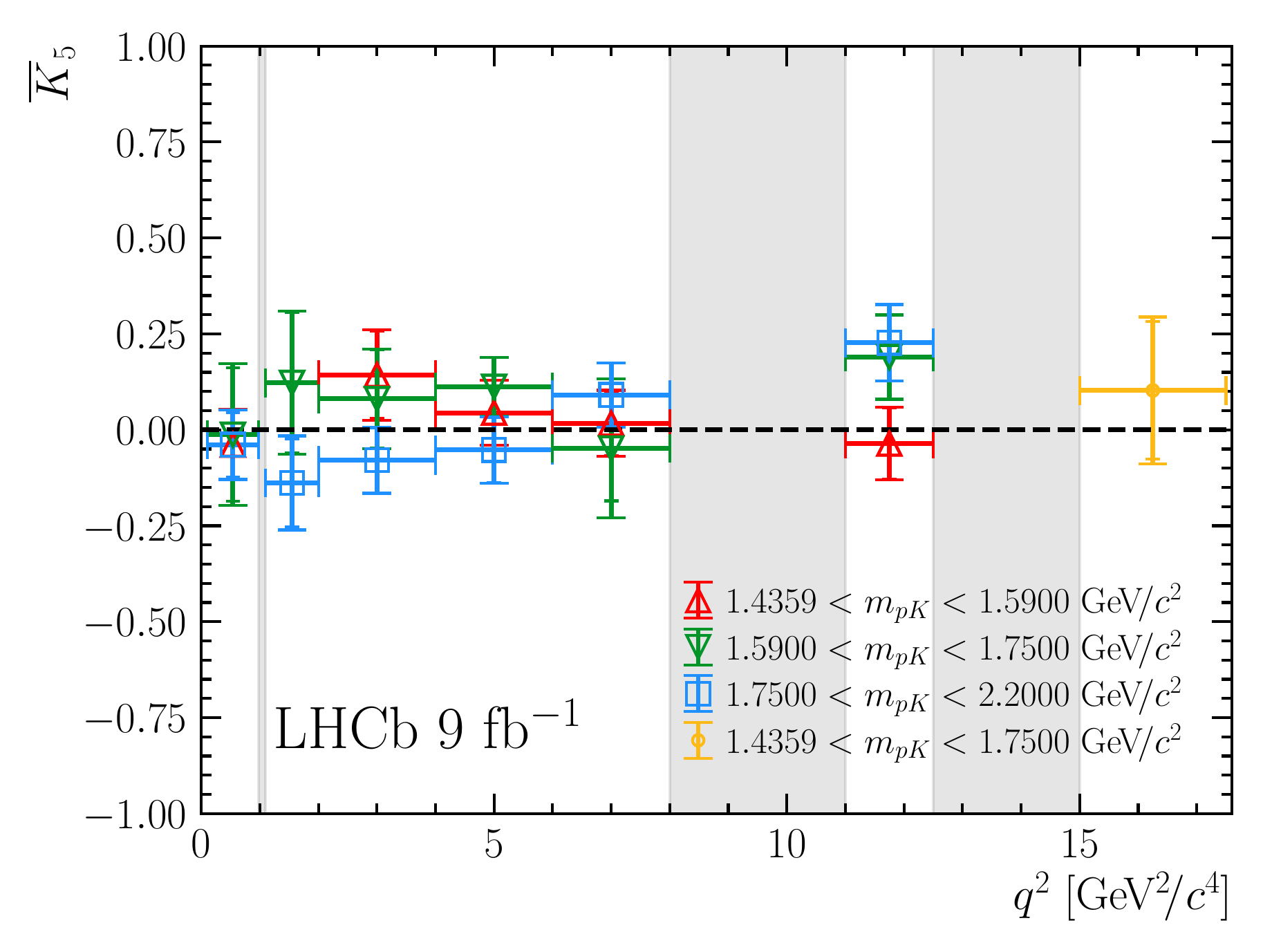} \\
     \includegraphics[width=0.48\textwidth]{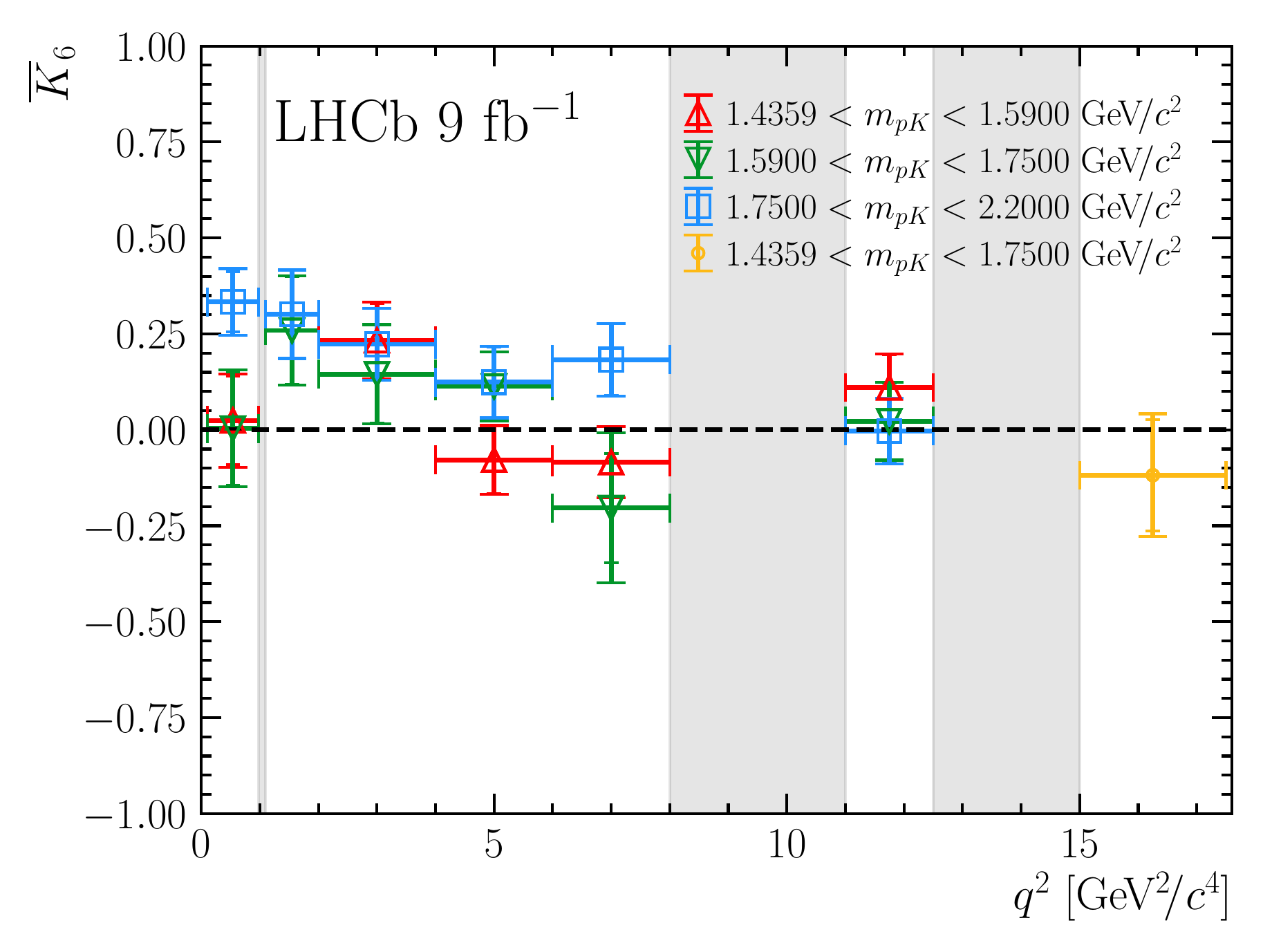}
     \includegraphics[width=0.48\textwidth]{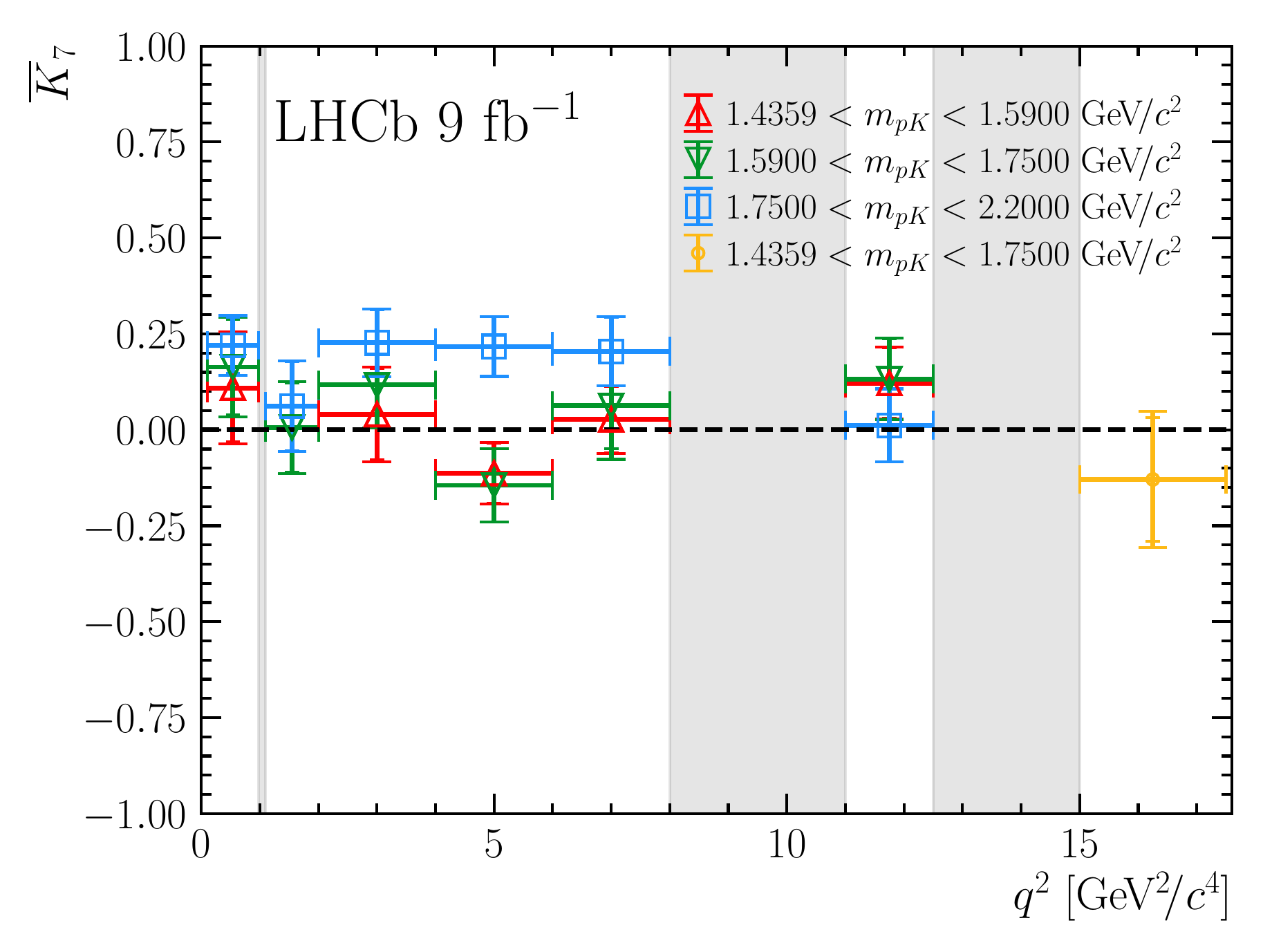}
     \caption{
     Values of \Kobs{2}--\Kobs{7} in bins of \qsq and \mpk.
    }
     \label{fig:appendix:kobs:1}
\end{figure}

\begin{figure}[!htb]
     \centering
     \includegraphics[width=0.48\textwidth]{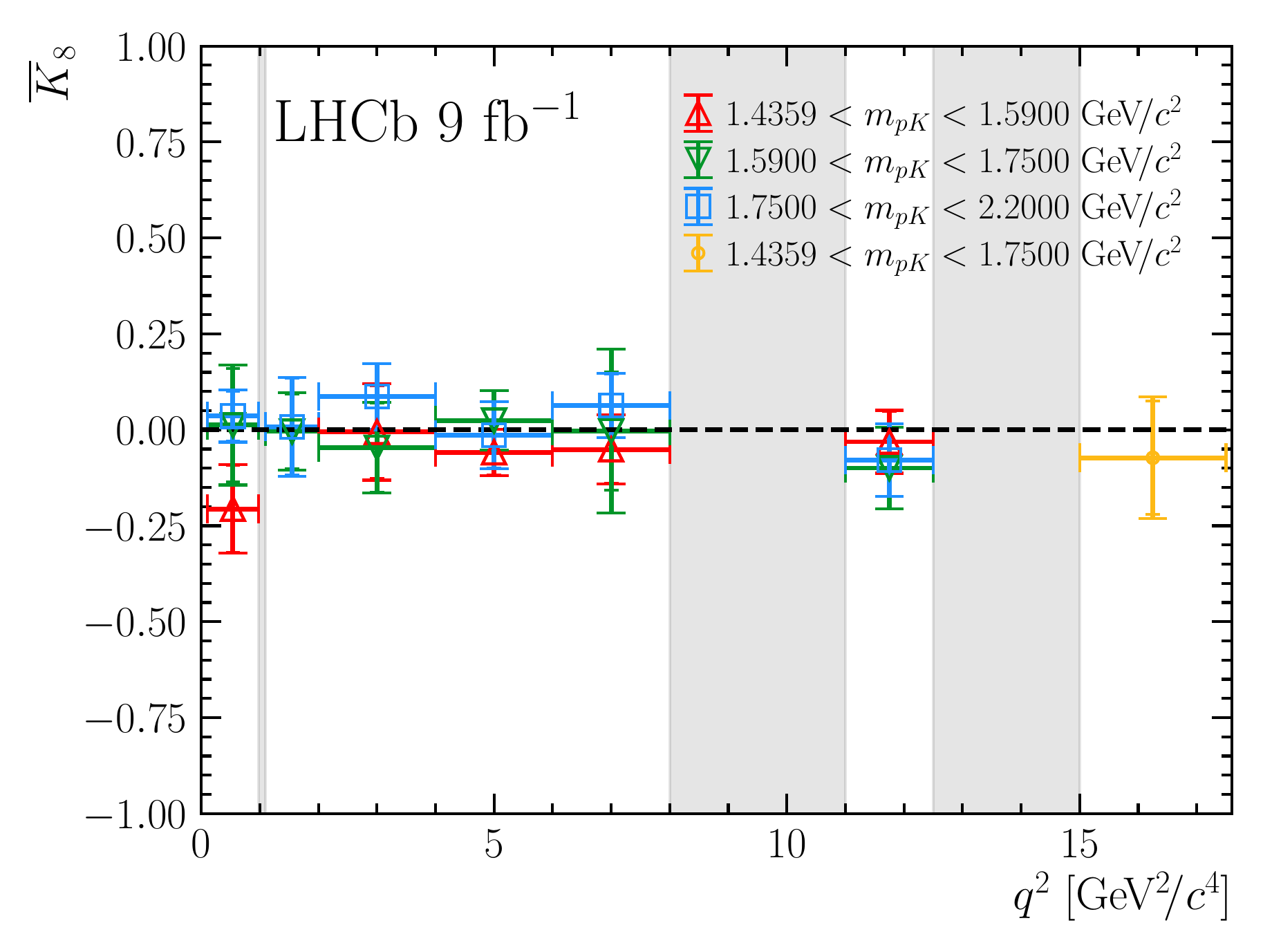} 
     \includegraphics[width=0.48\textwidth]{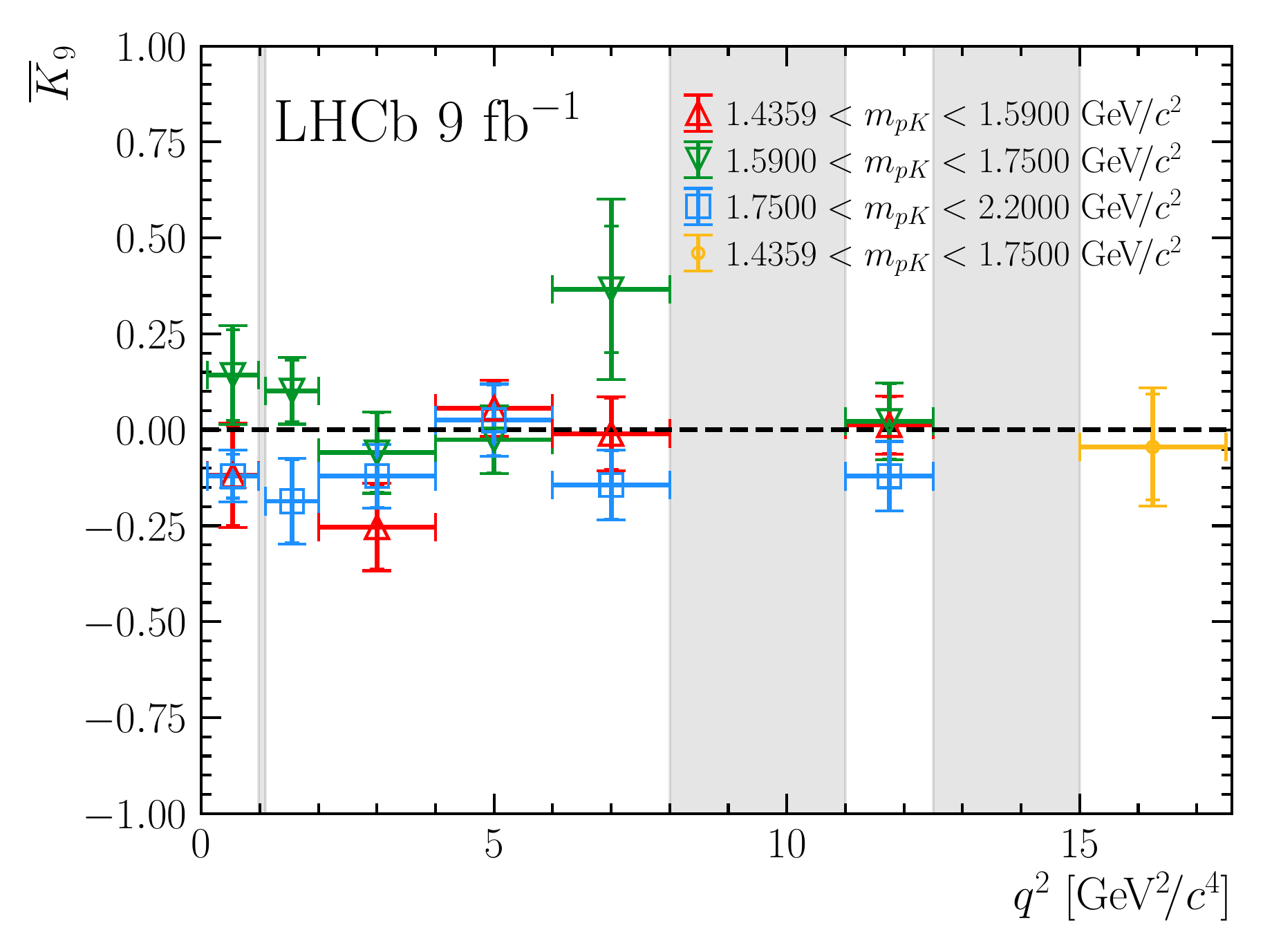} \\
     \includegraphics[width=0.48\textwidth]{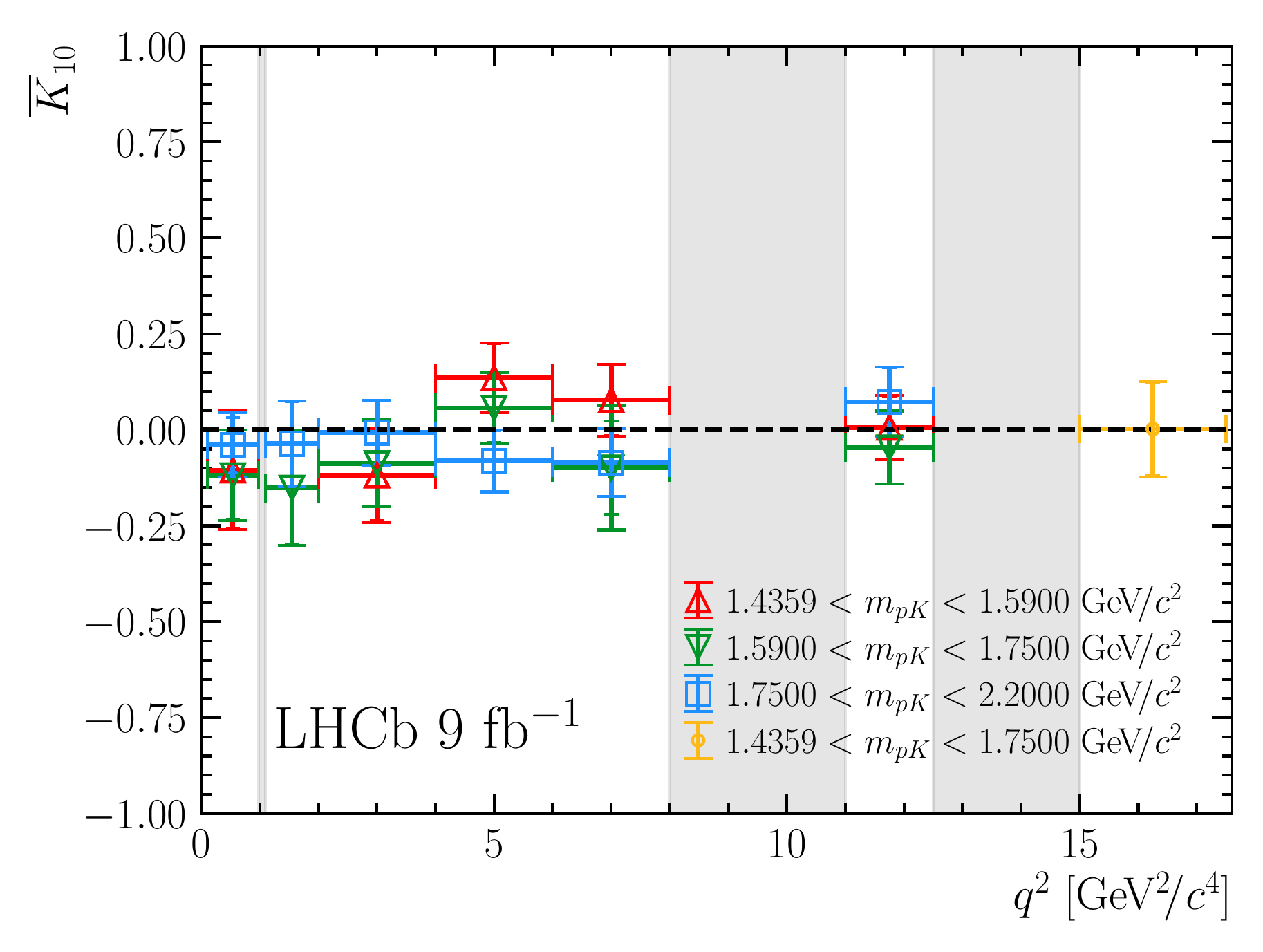}
     \includegraphics[width=0.48\textwidth]{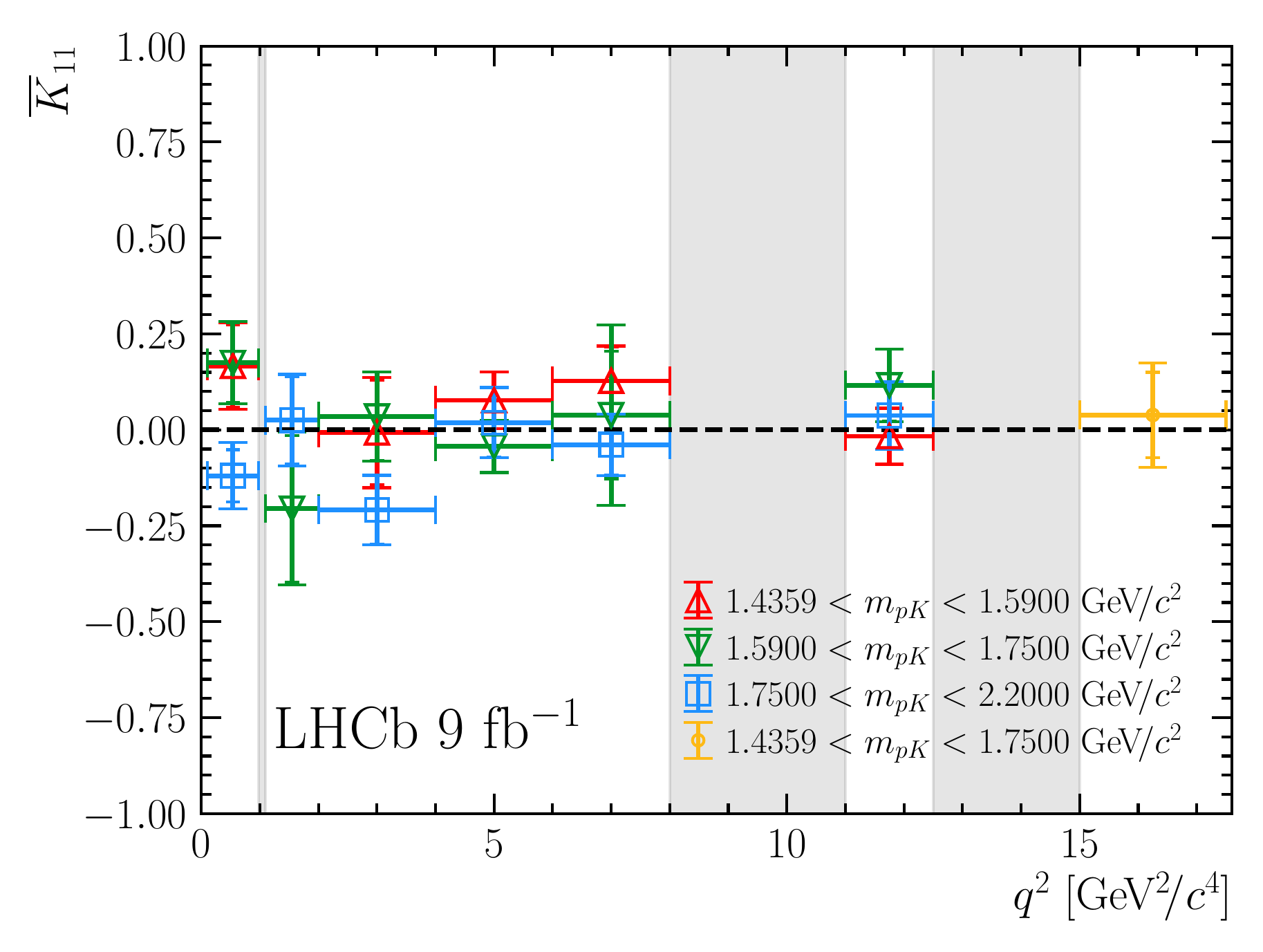} \\
     \includegraphics[width=0.48\textwidth]{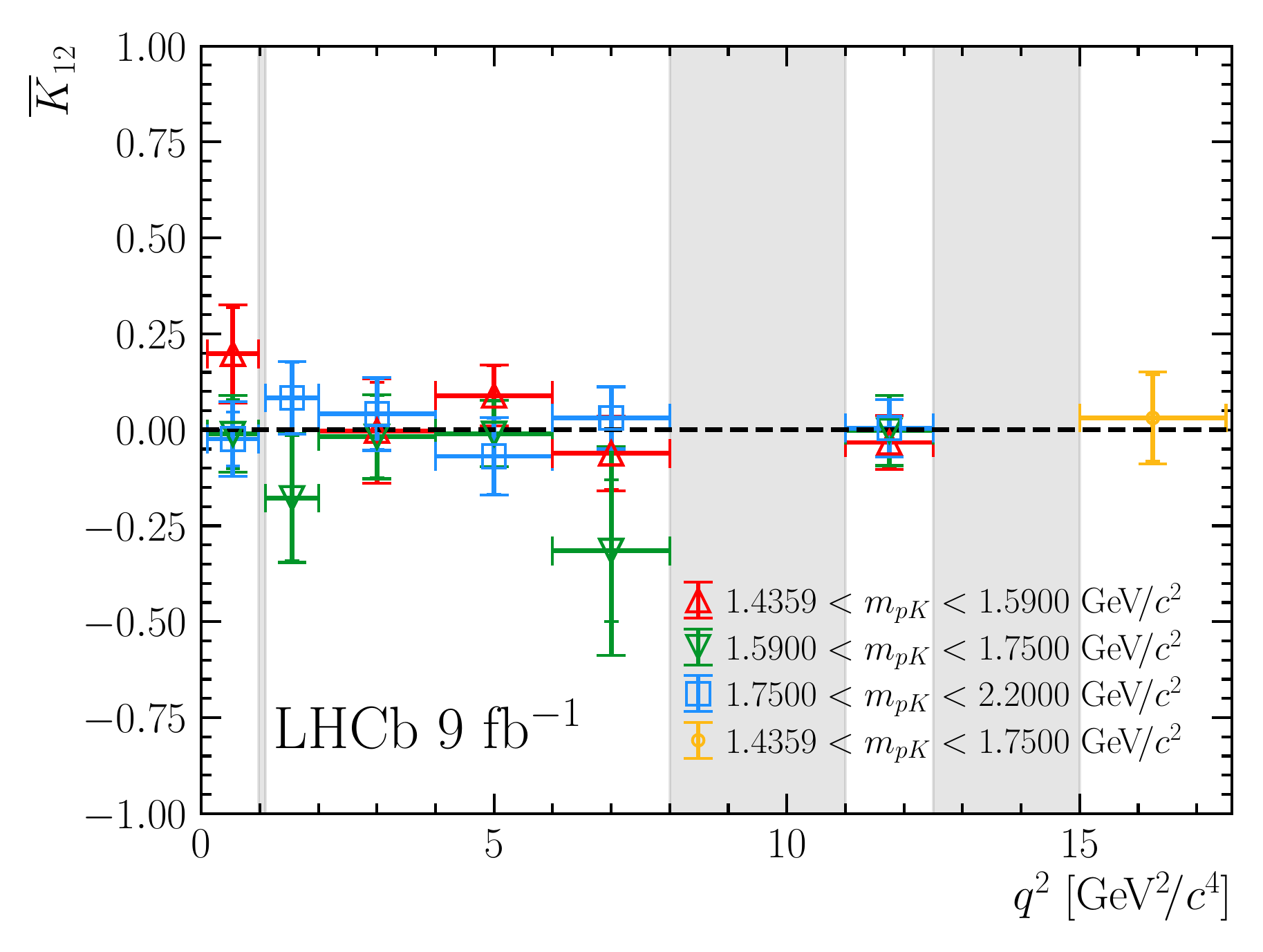}
     \includegraphics[width=0.48\textwidth]{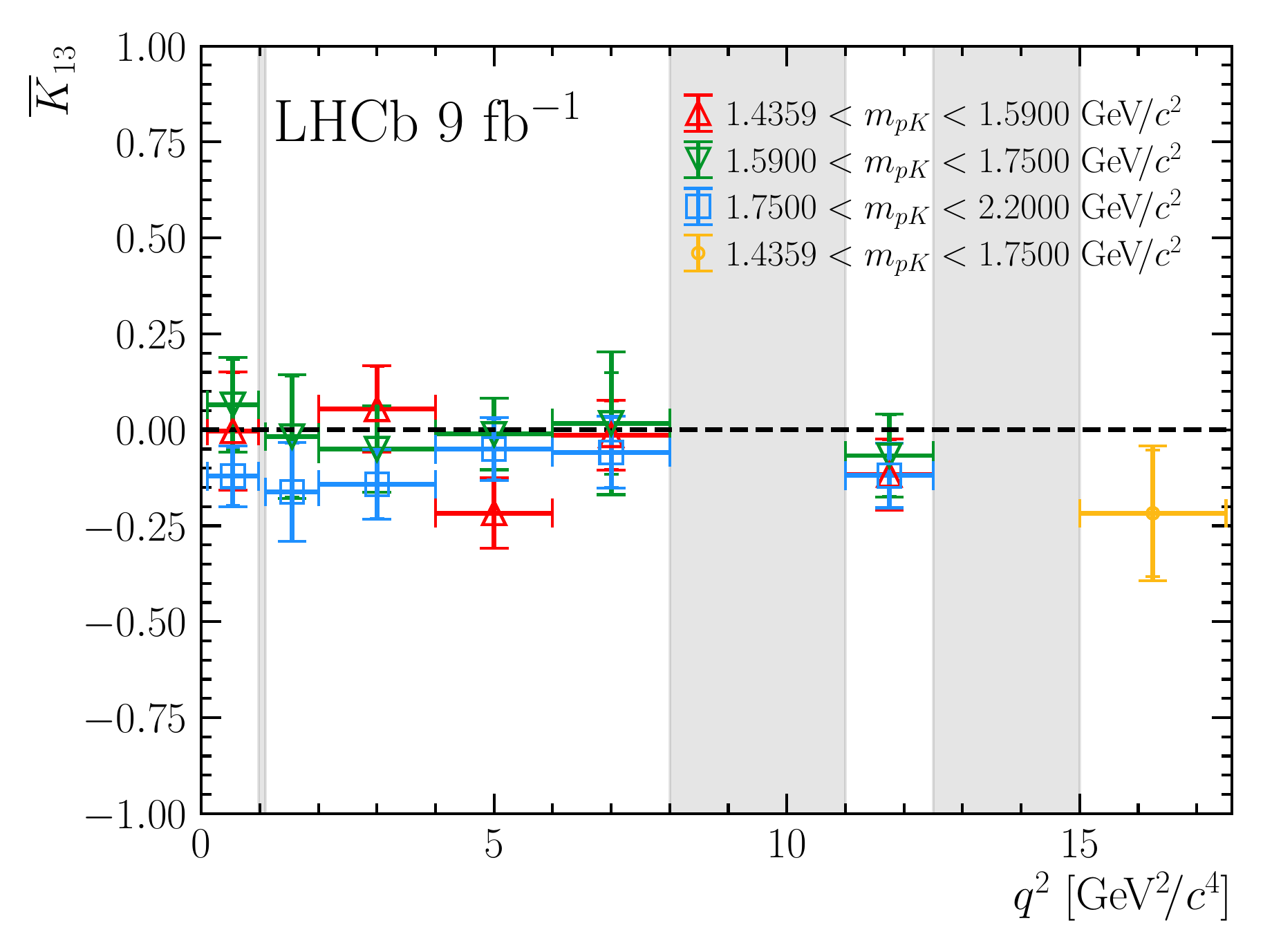} \\
     \includegraphics[width=0.48\textwidth]{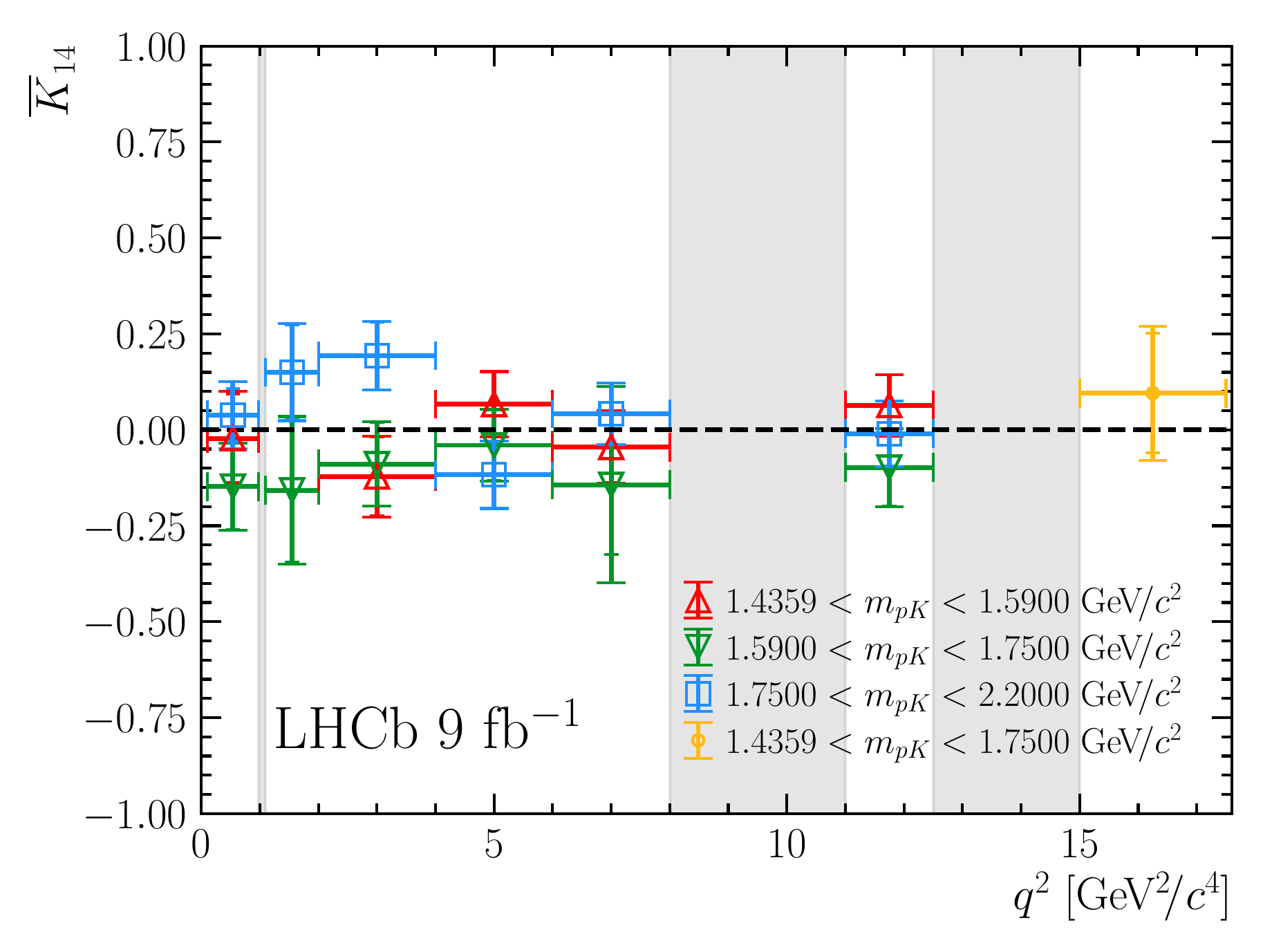}
     \includegraphics[width=0.48\textwidth]{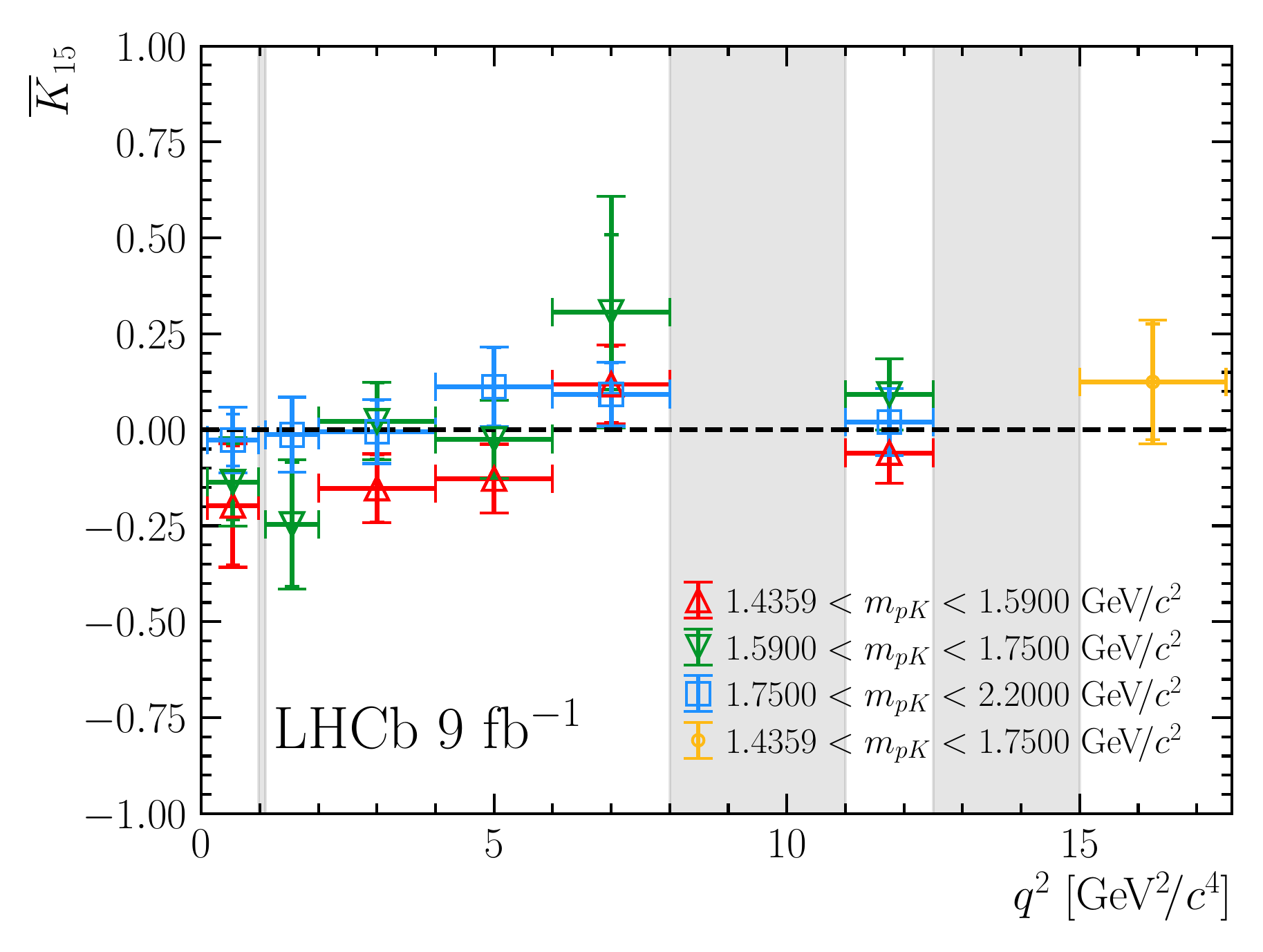}
     \caption{
     Values of \Kobs{8}--\Kobs{15} in bins of \qsq and \mpk.
    }
     \label{fig:appendix:kobs:2}
\end{figure}

\begin{figure}[!htb]
     \centering
     \includegraphics[width=0.48\textwidth]{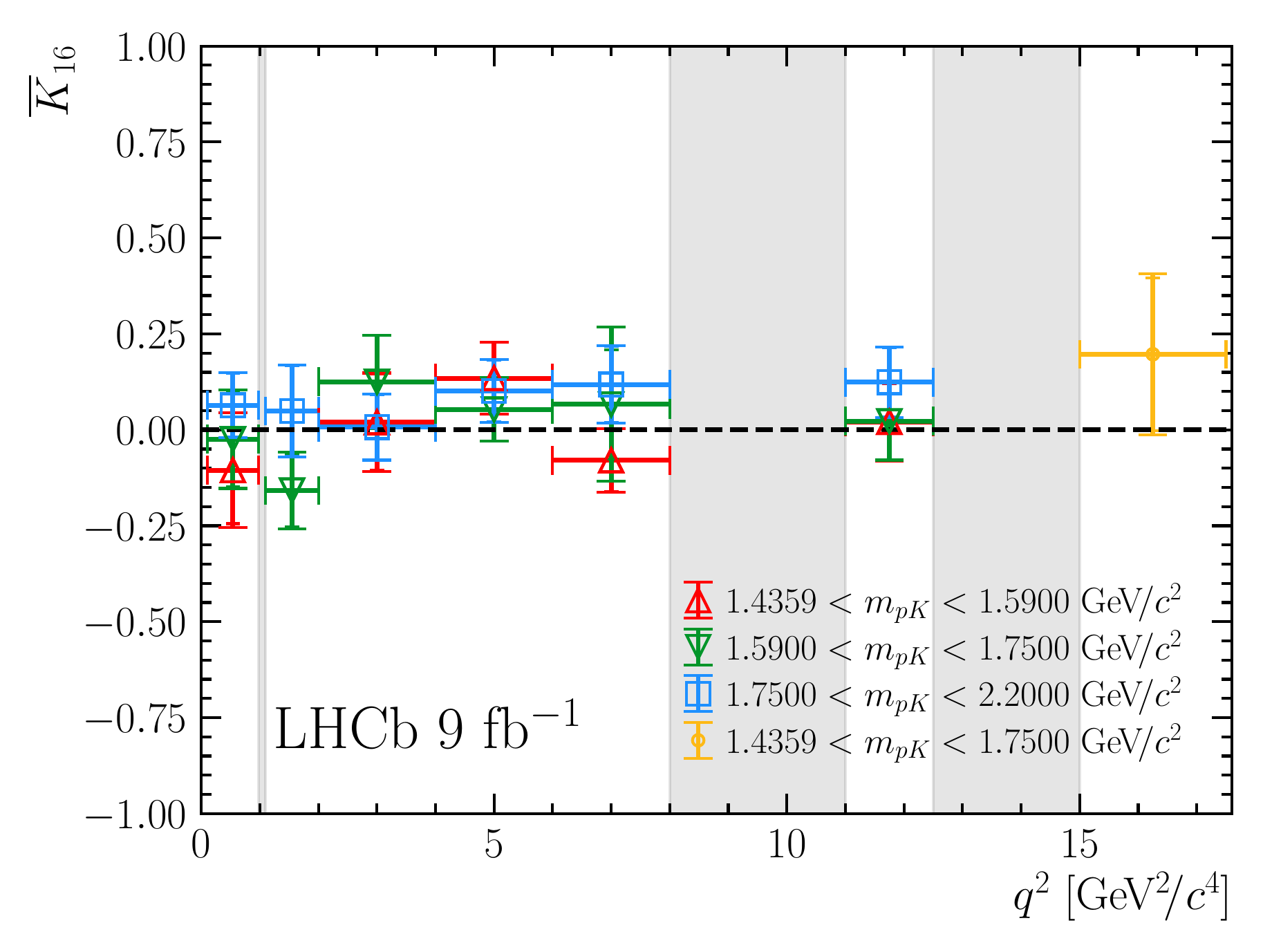}
     \includegraphics[width=0.48\textwidth]{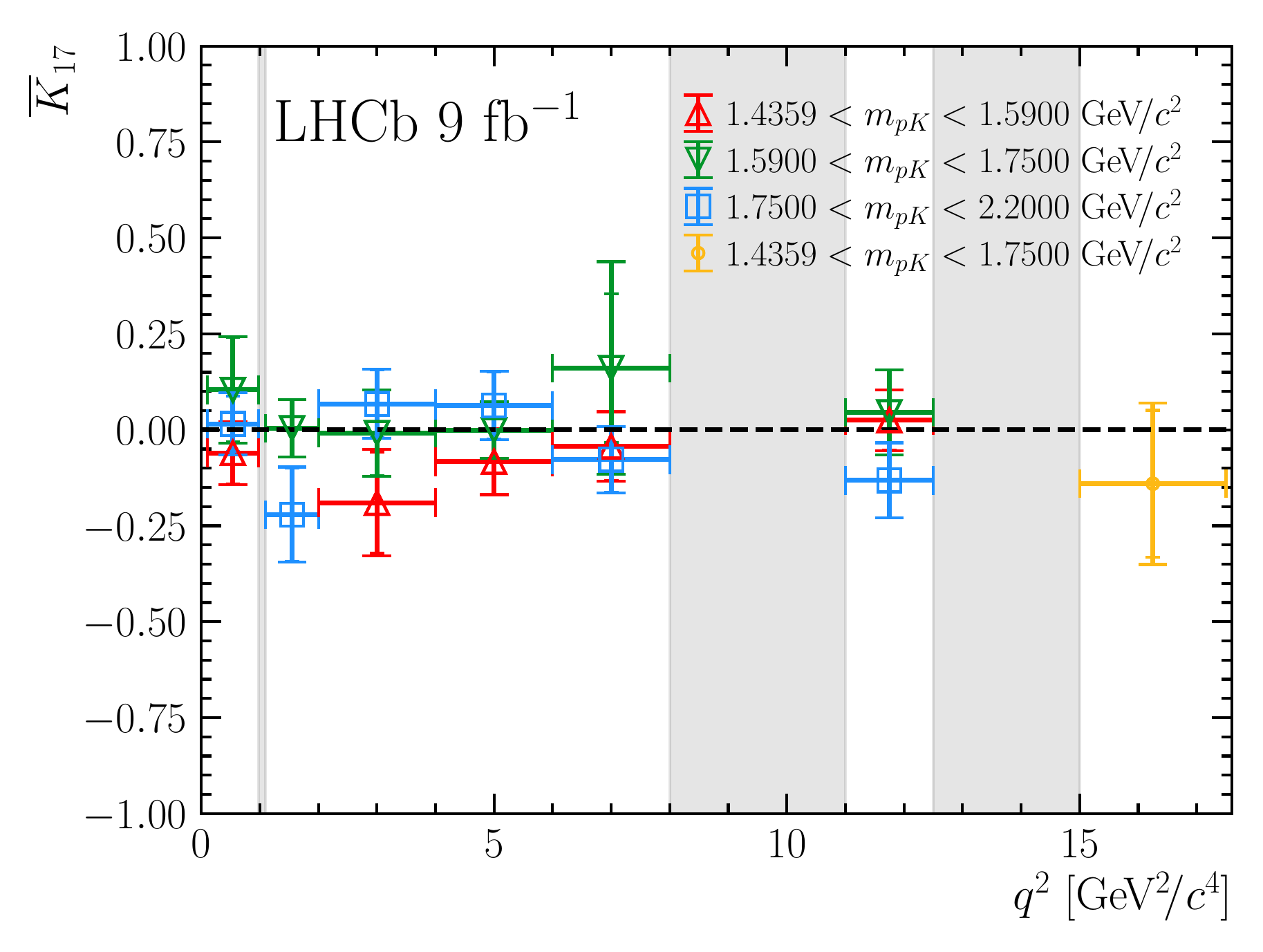} \\
     \includegraphics[width=0.48\textwidth]{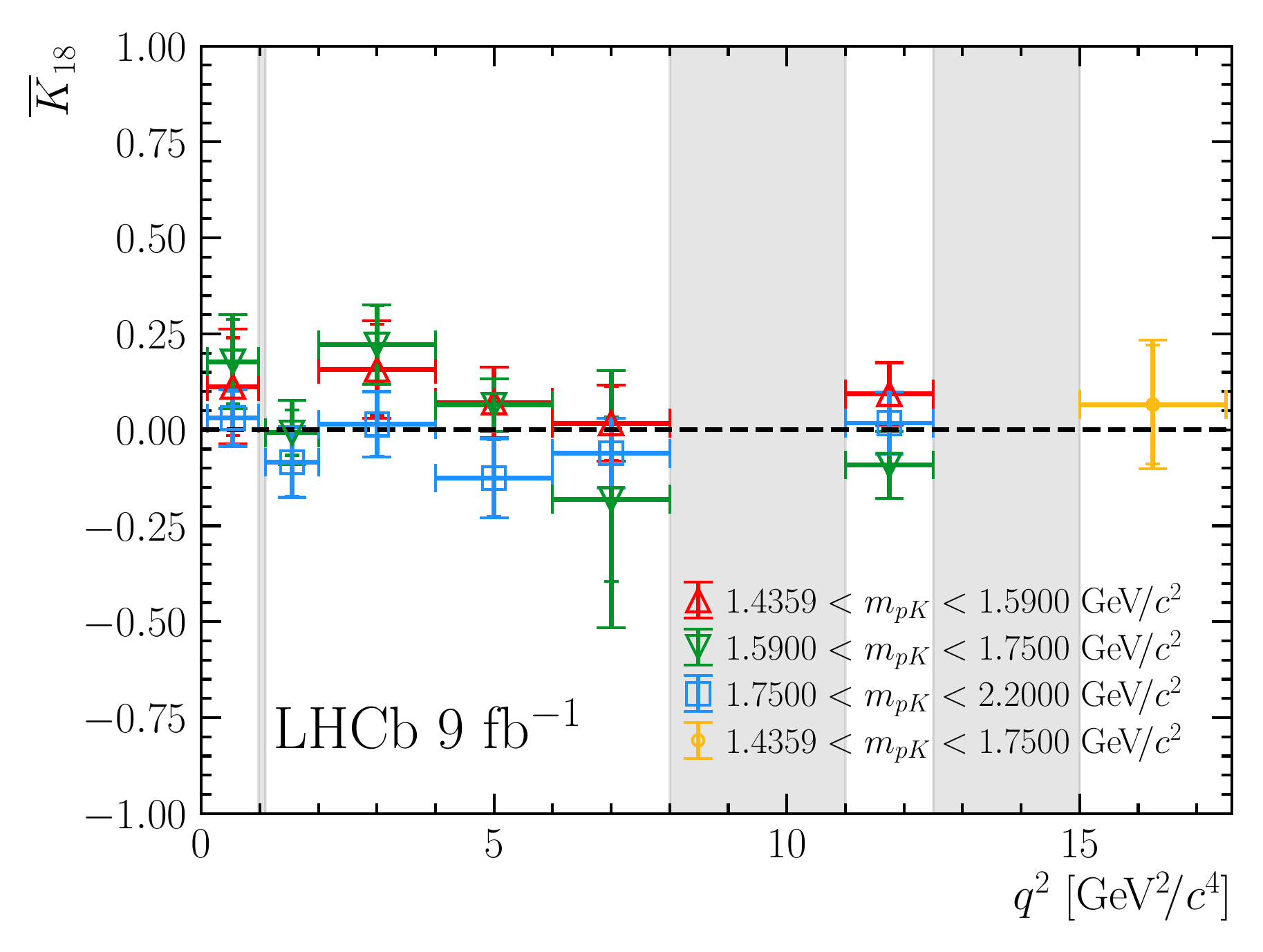}
     \includegraphics[width=0.48\textwidth]{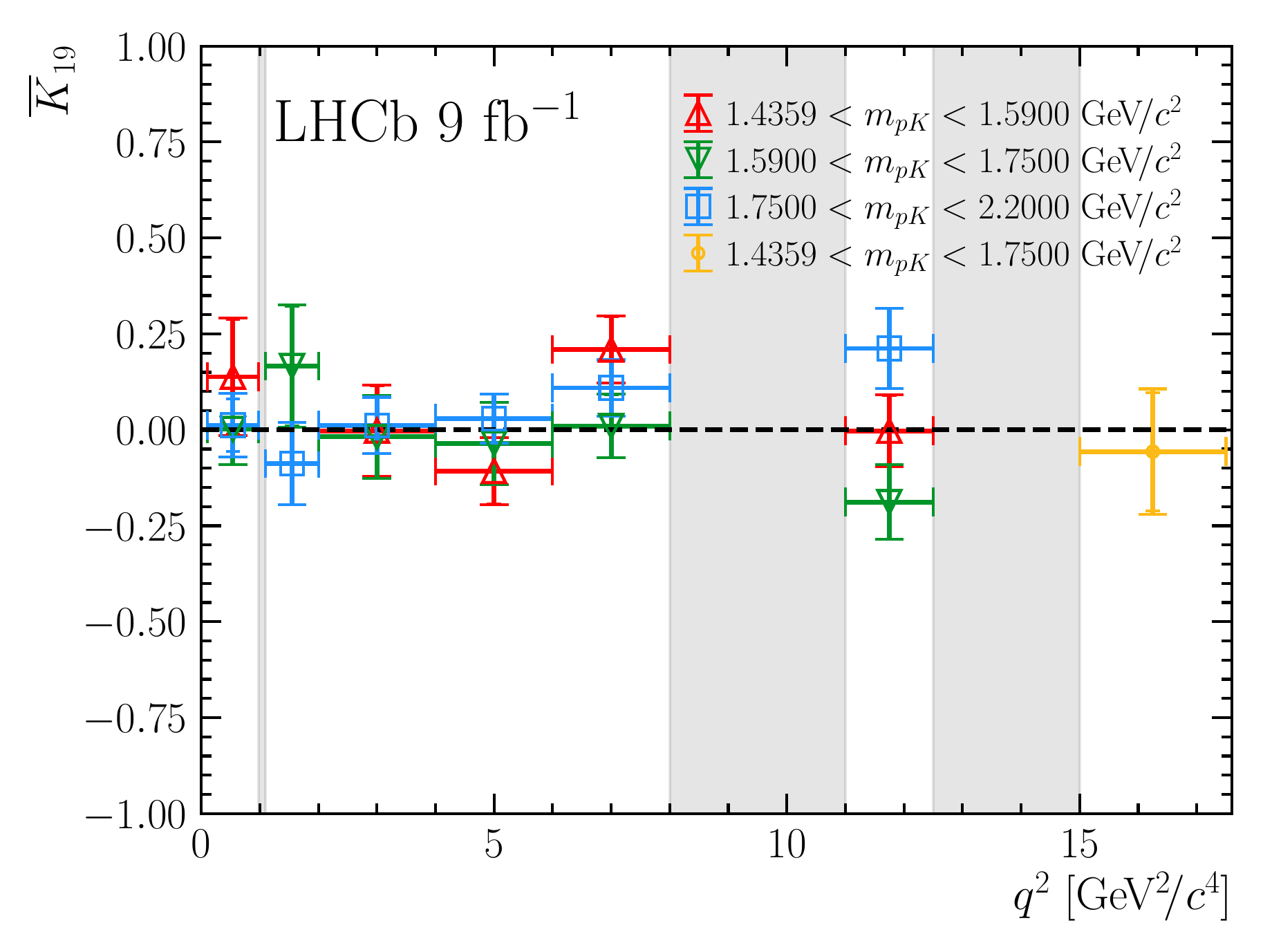} \\
     \includegraphics[width=0.48\textwidth]{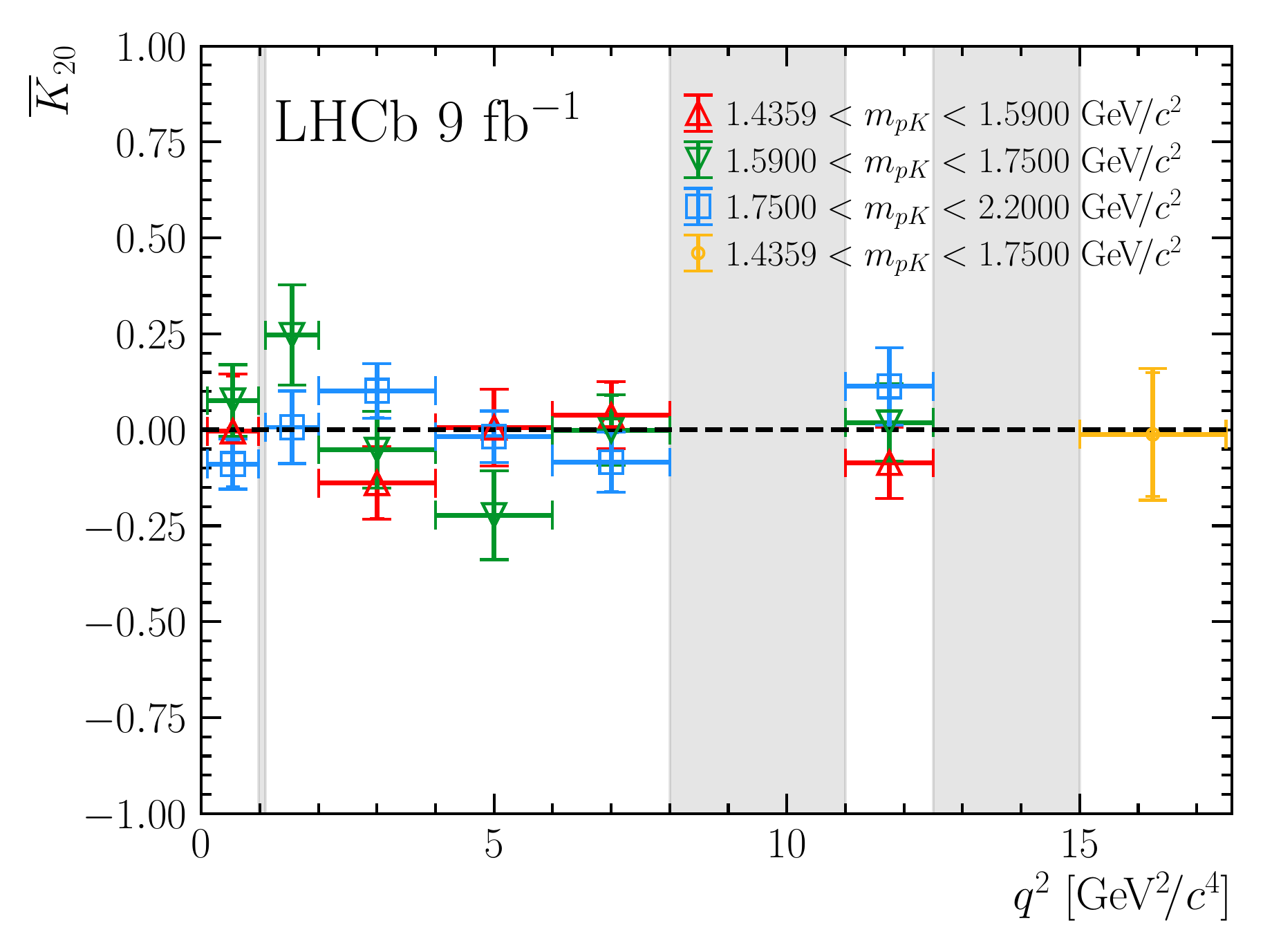}
     \includegraphics[width=0.48\textwidth]{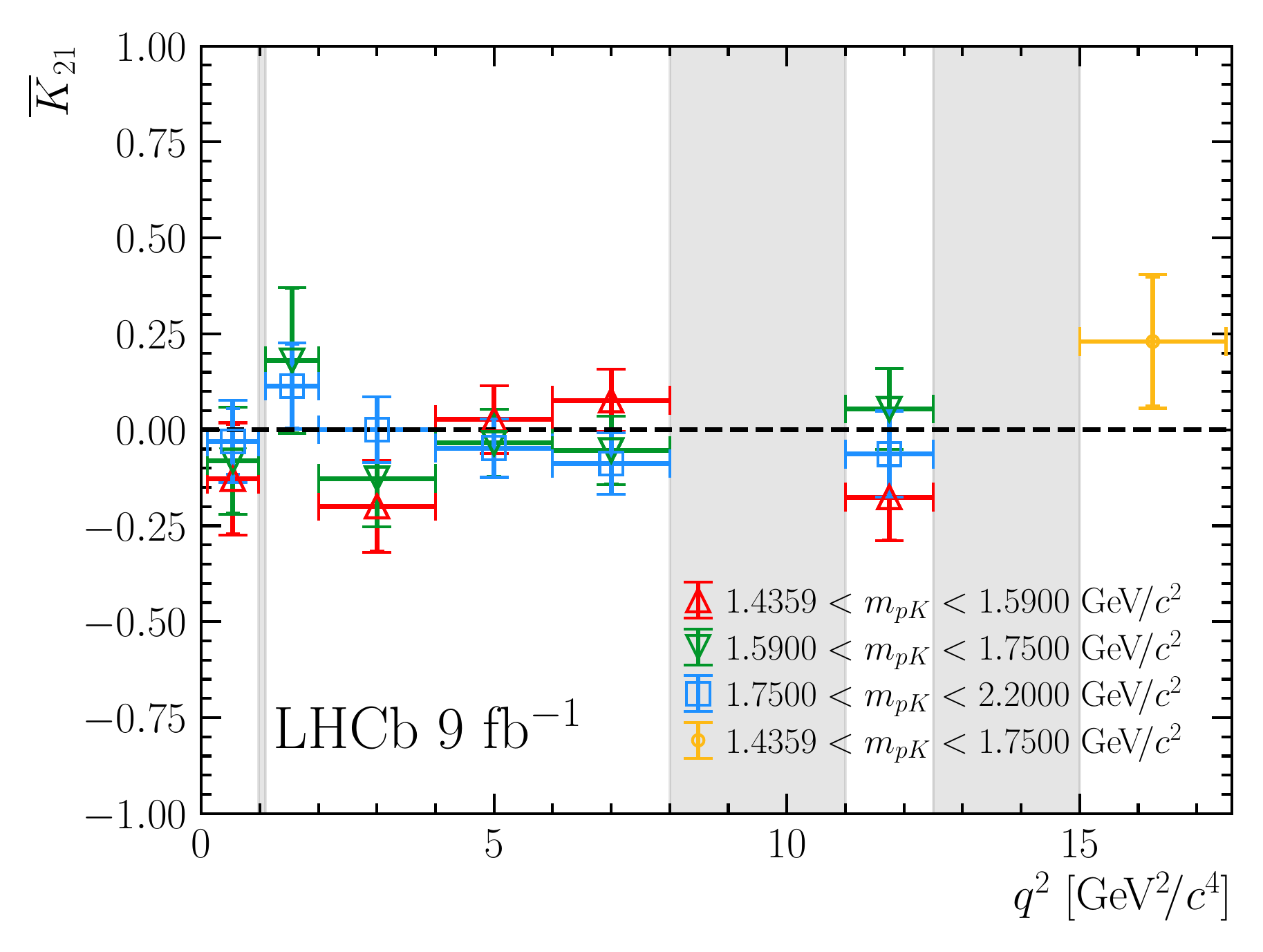} \\
     \includegraphics[width=0.48\textwidth]{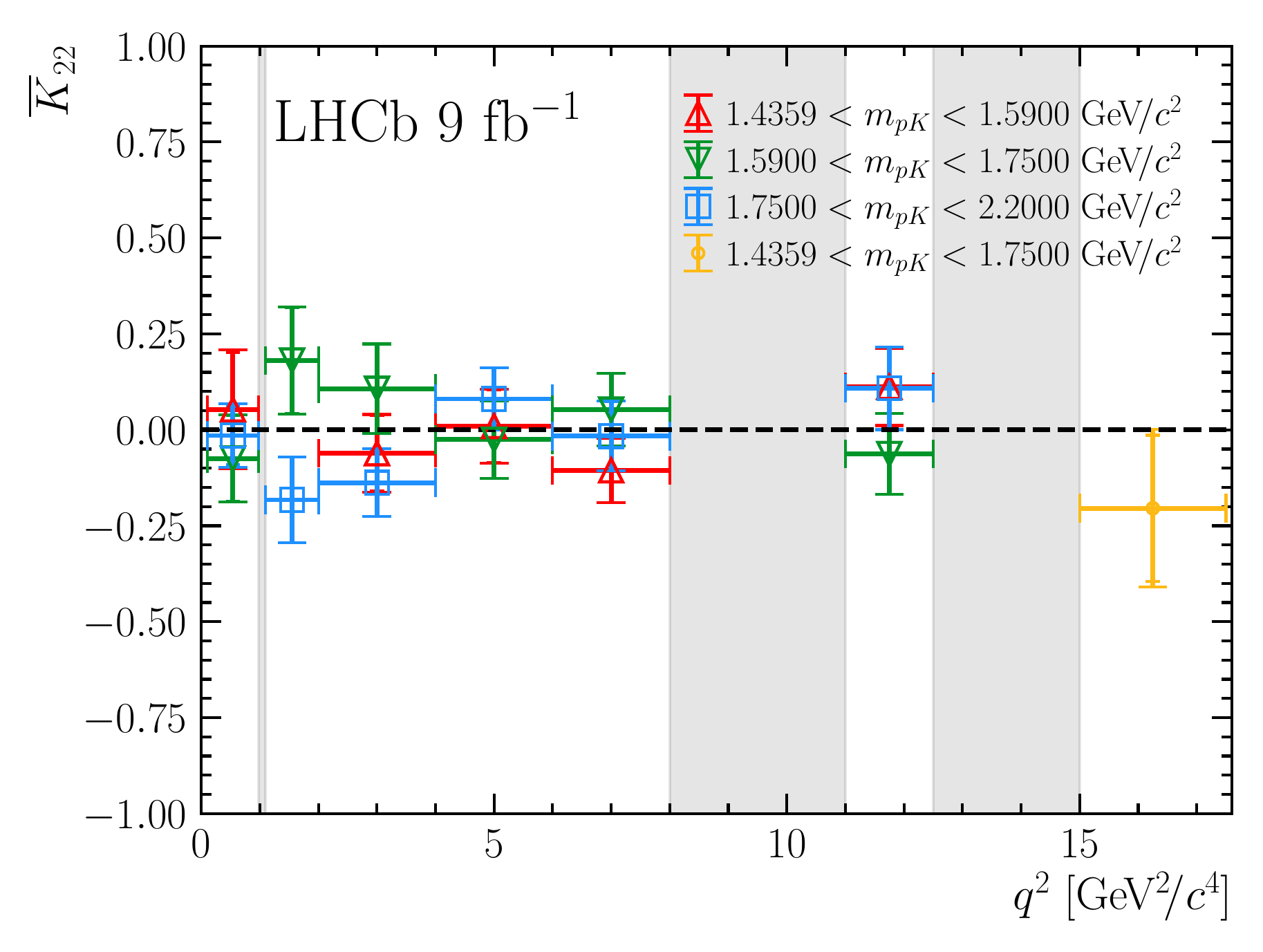}
     \includegraphics[width=0.48\textwidth]{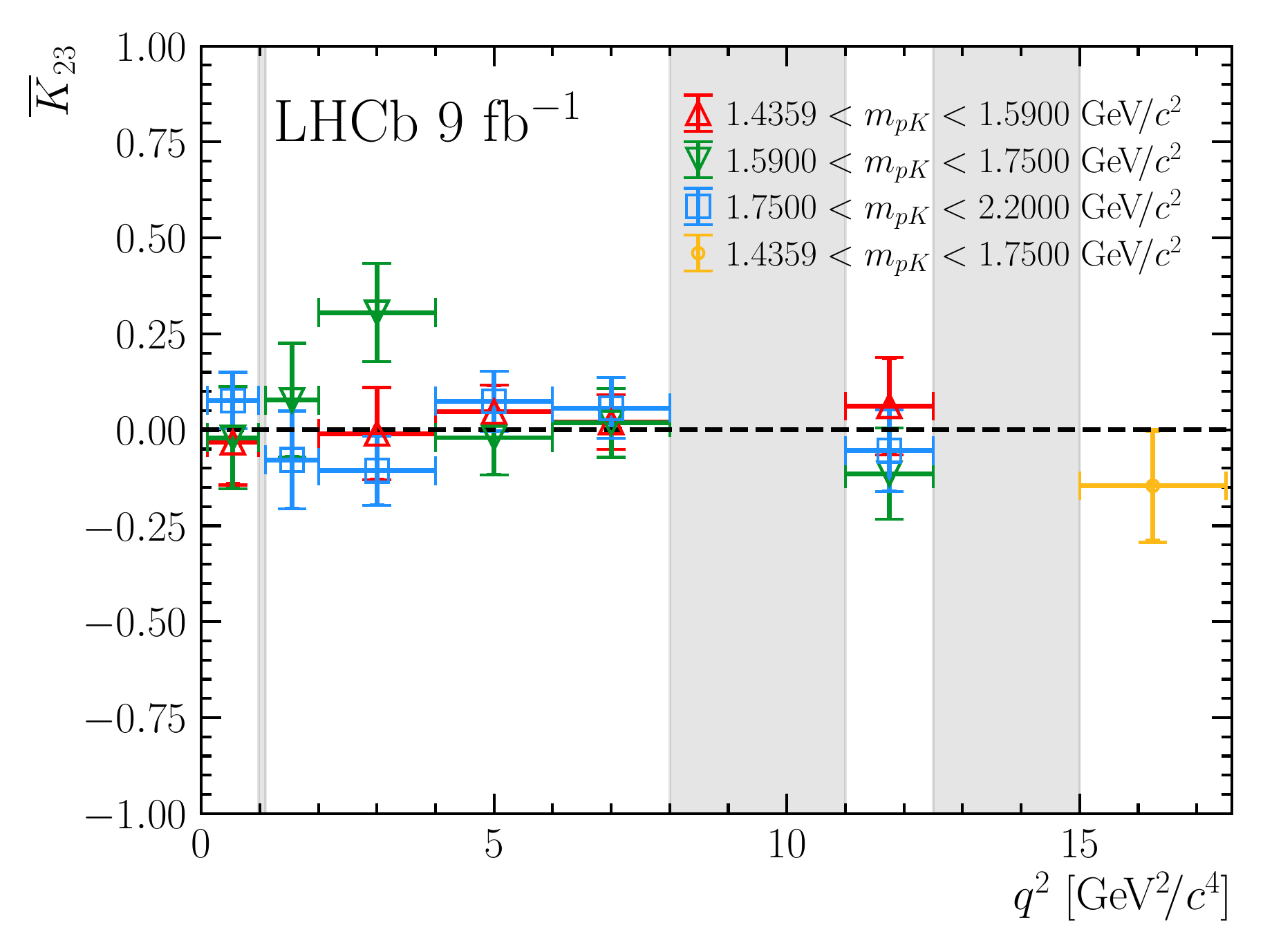}
     \caption{
     Values of \Kobs{16}--\Kobs{23} in bins of \qsq and \mpk.
    }
     \label{fig:appendix:kobs:3}
\end{figure}

\begin{figure}[!htb]
     \centering
     \includegraphics[width=0.48\textwidth]{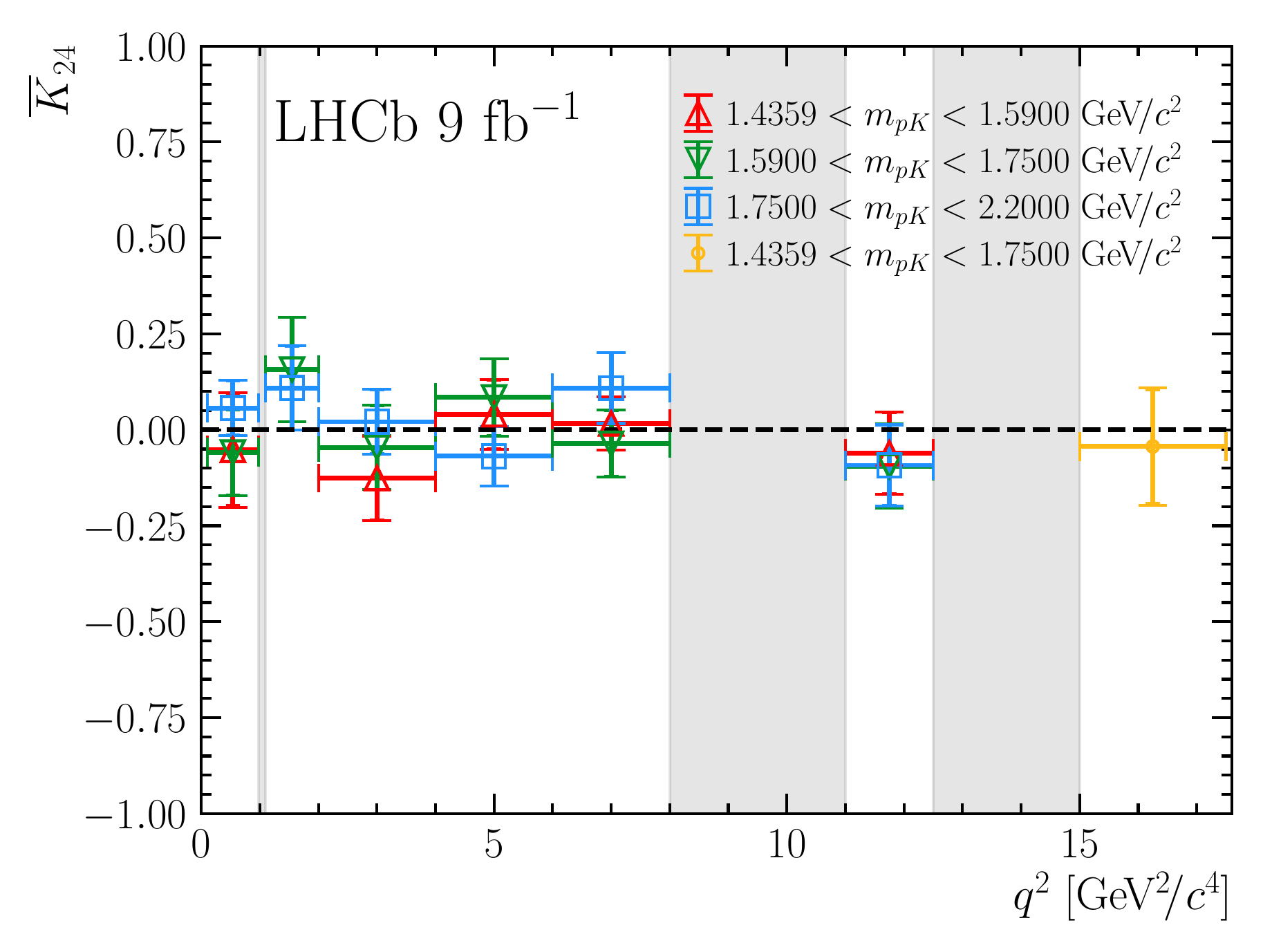}
     \includegraphics[width=0.48\textwidth]{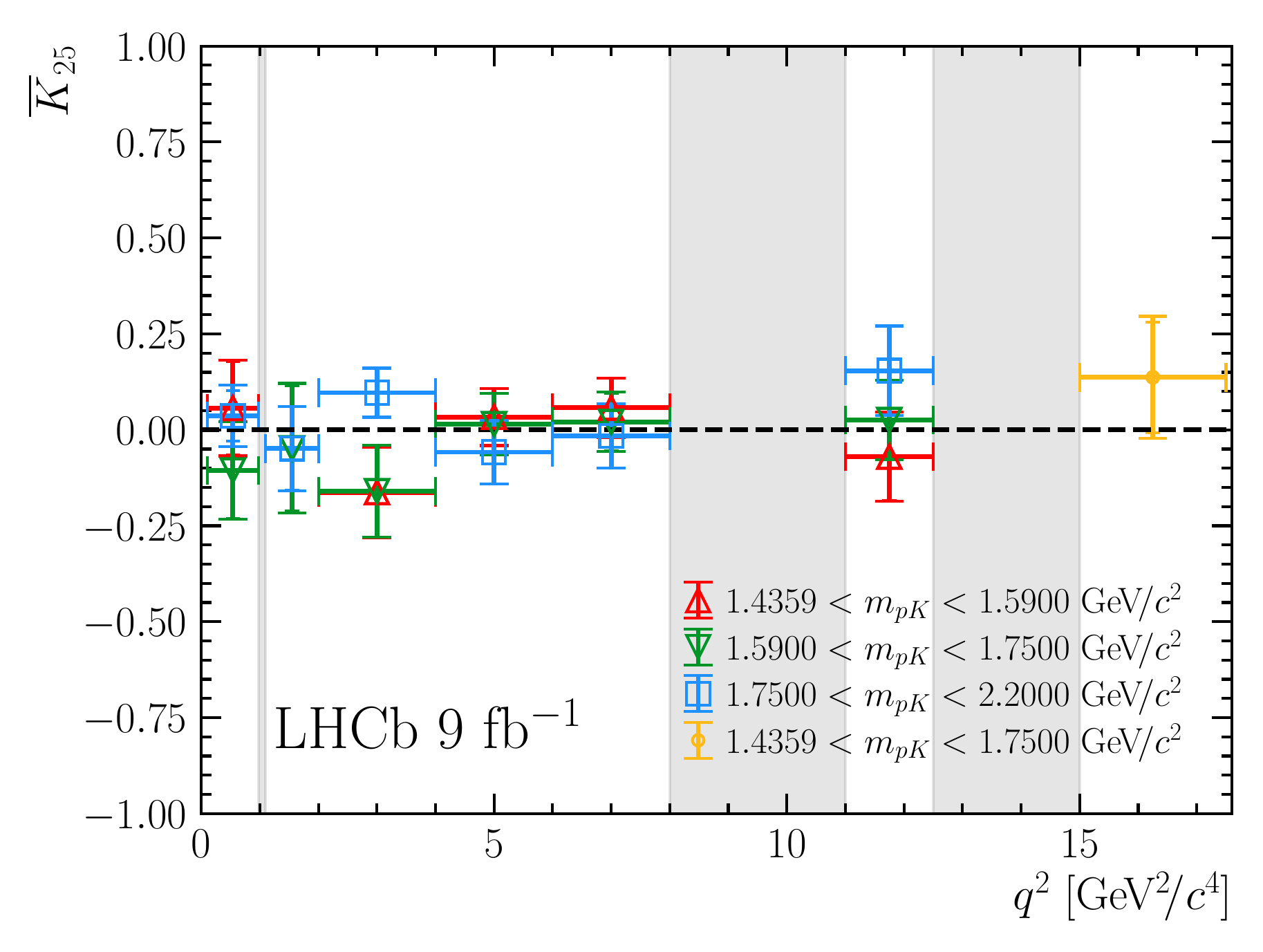} \\
     \includegraphics[width=0.48\textwidth]{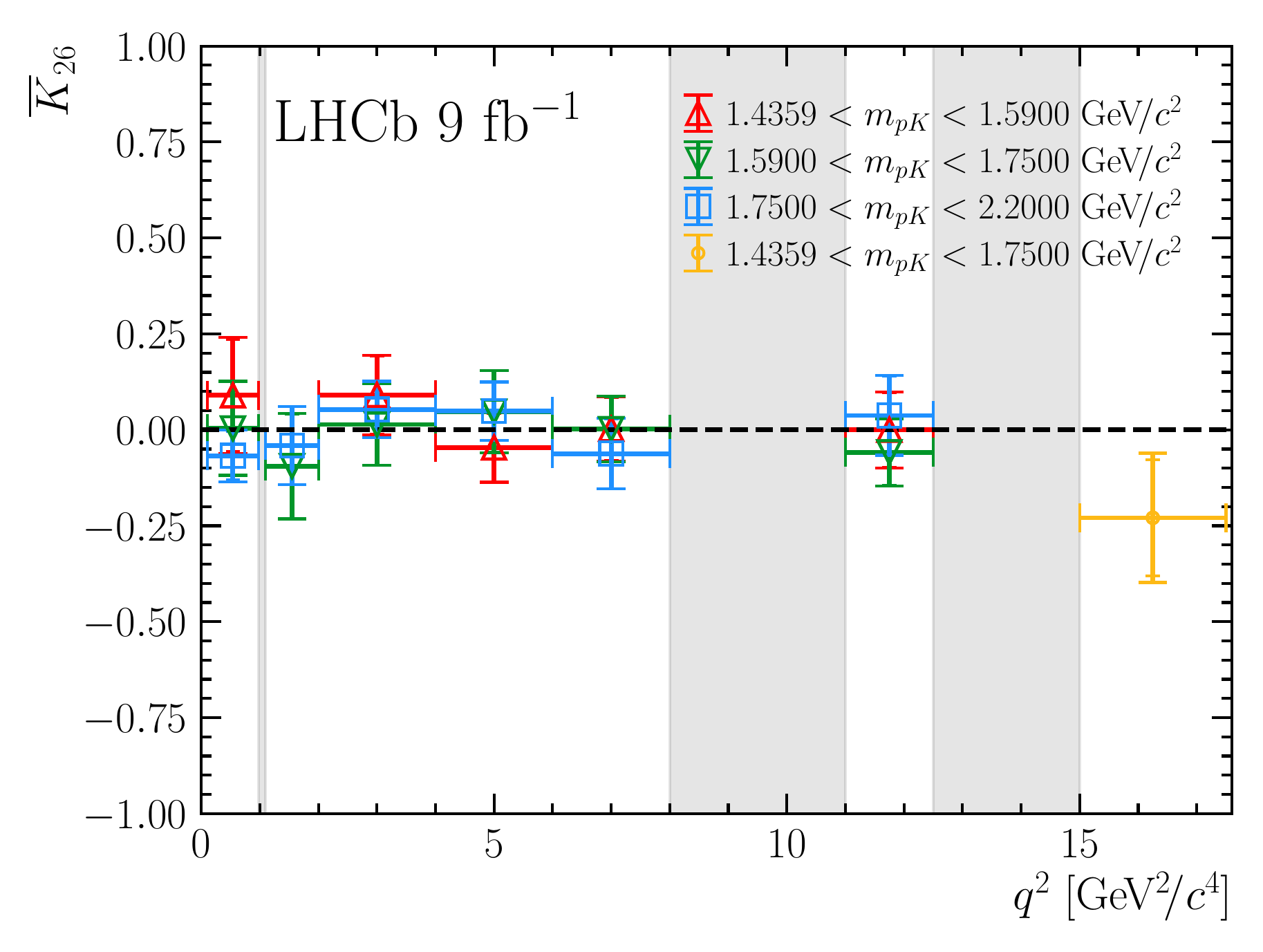}
     \includegraphics[width=0.48\textwidth]{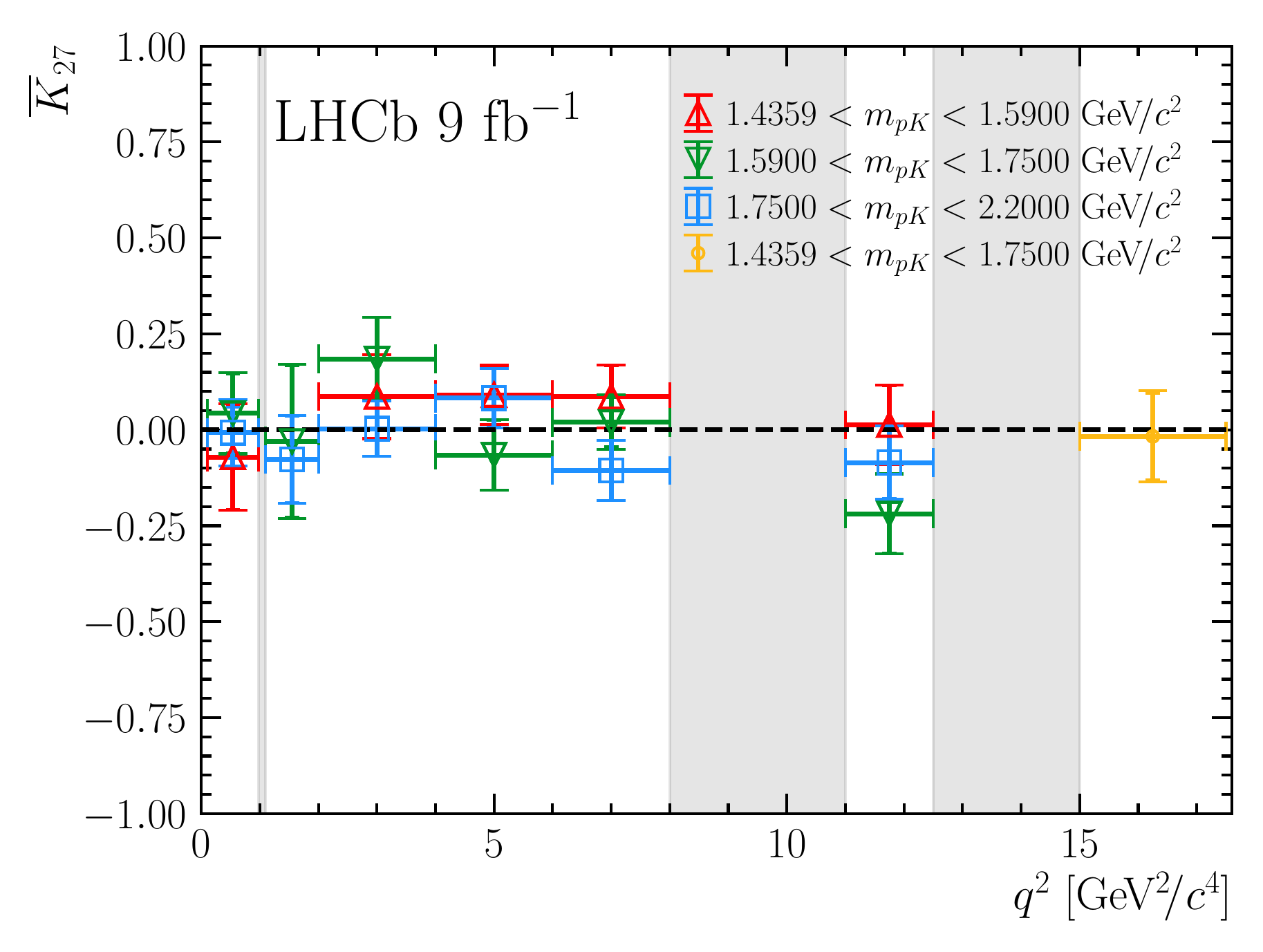} \\
     \includegraphics[width=0.48\textwidth]{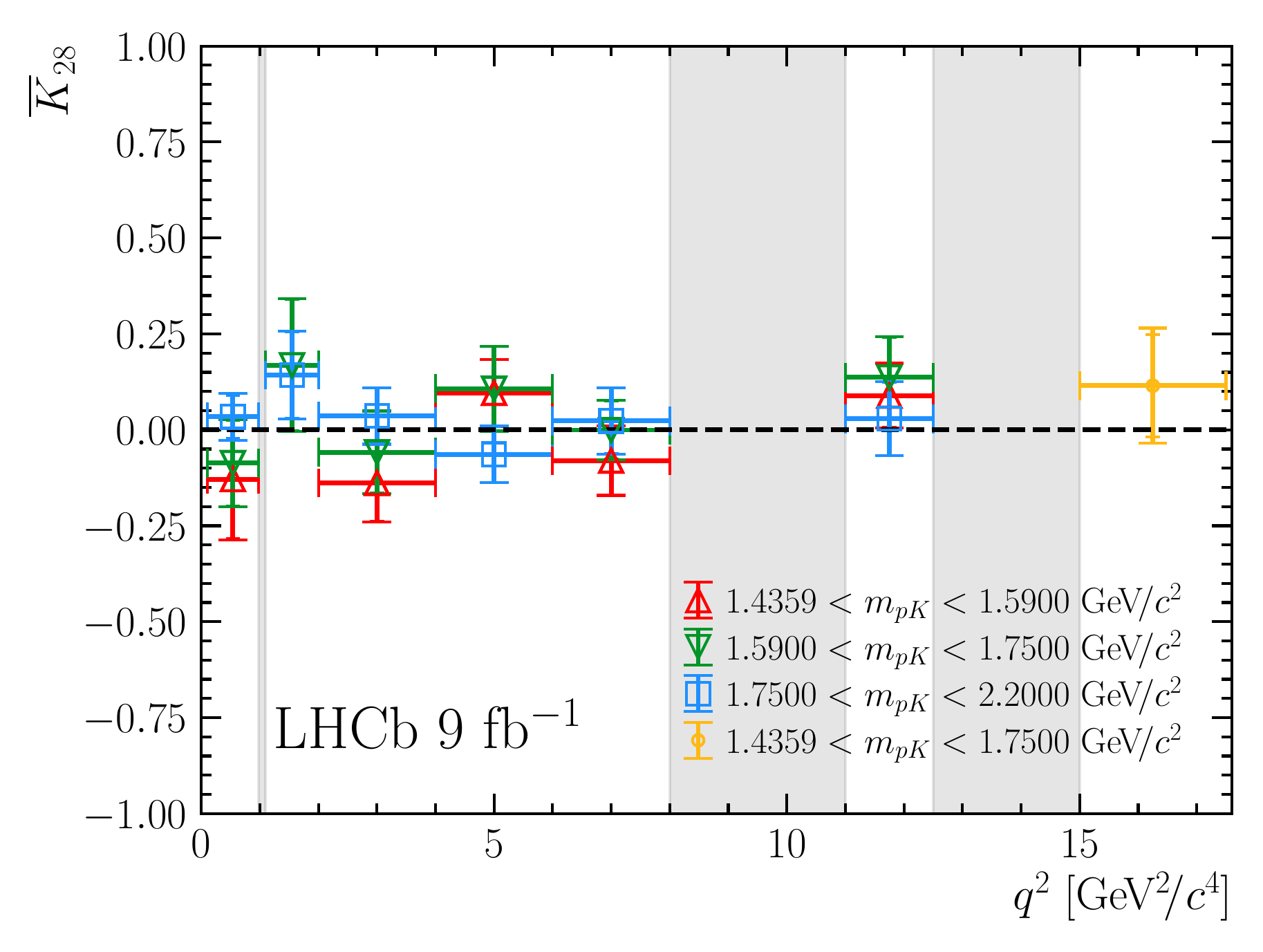}
     \includegraphics[width=0.48\textwidth]{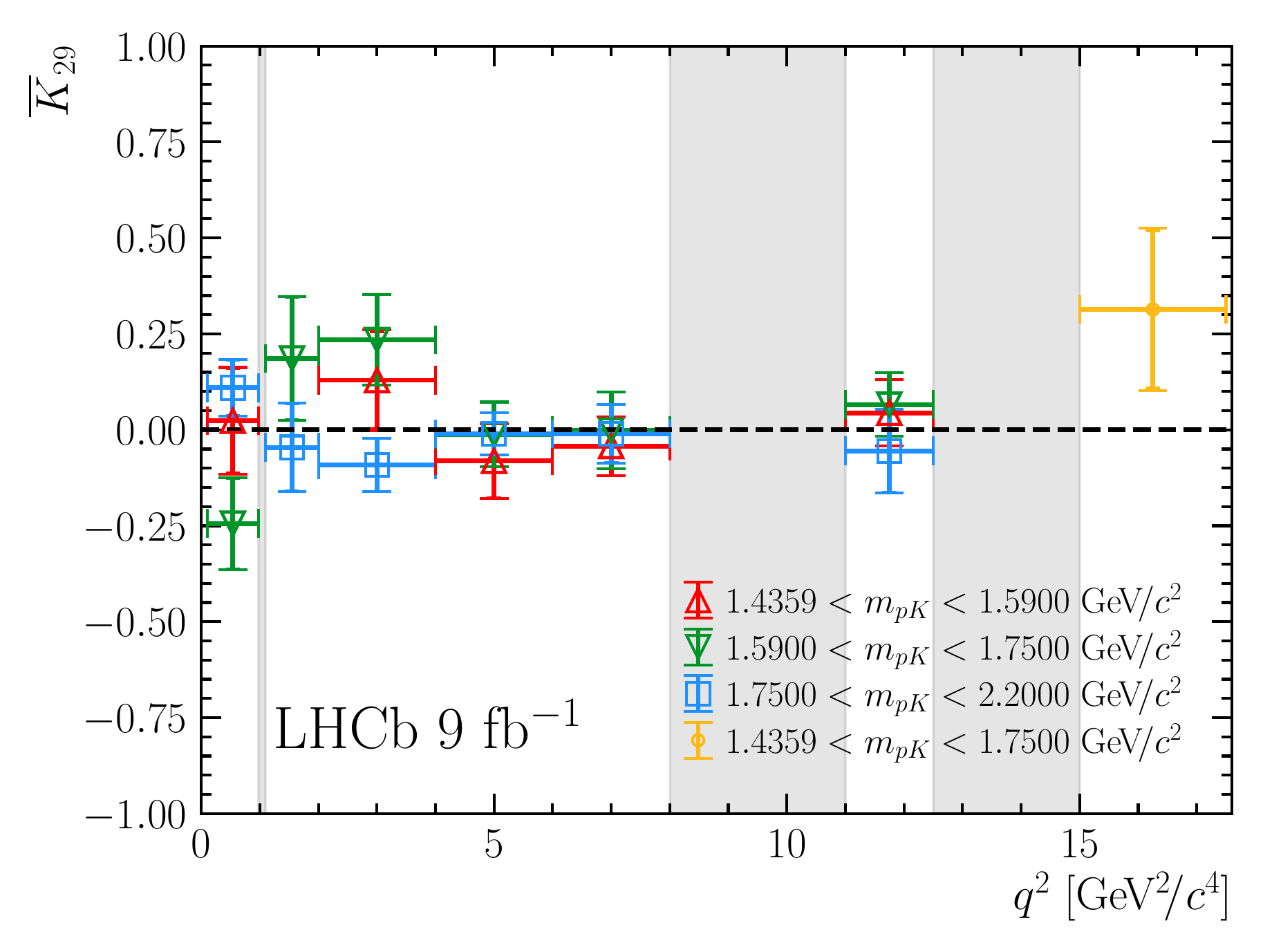} \\
     \includegraphics[width=0.48\textwidth]{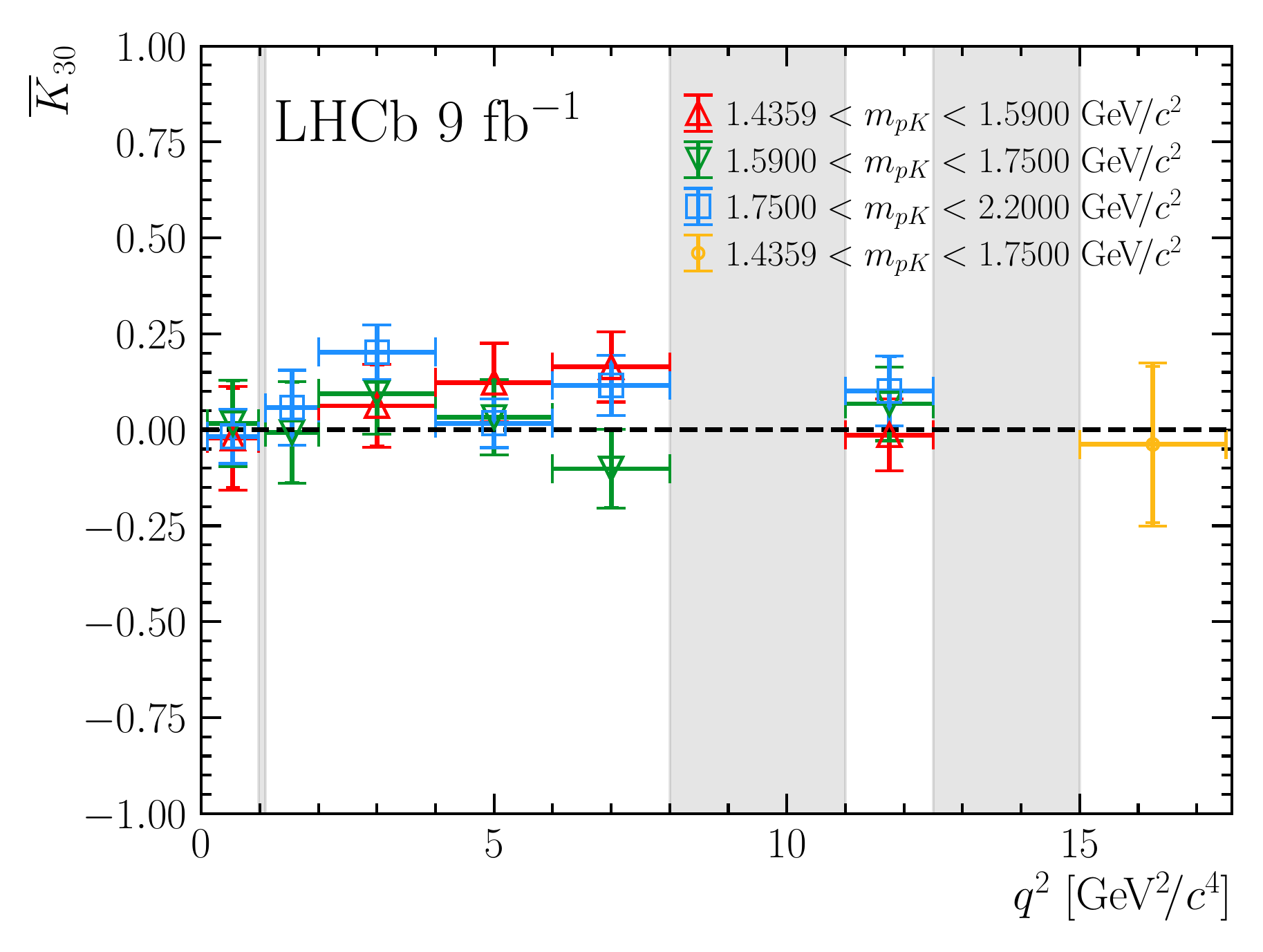}
     \includegraphics[width=0.48\textwidth]{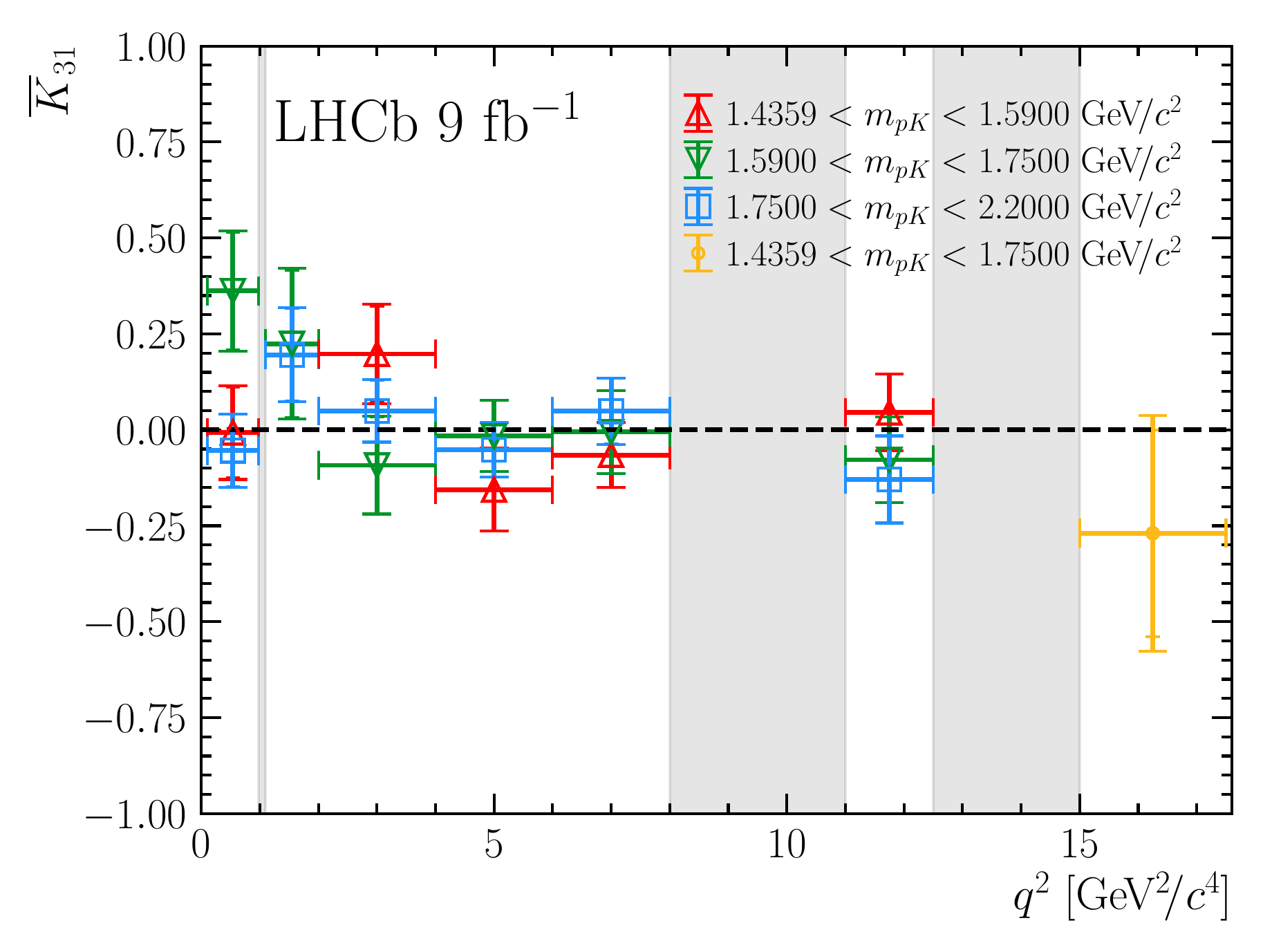}
     \caption{
     Values of \Kobs{24}--\Kobs{31} in bins of \qsq and \mpk.
    }
     \label{fig:appendix:kobs:4}
\end{figure}

\begin{figure}[!htb]
     \centering
     \includegraphics[width=0.48\textwidth]{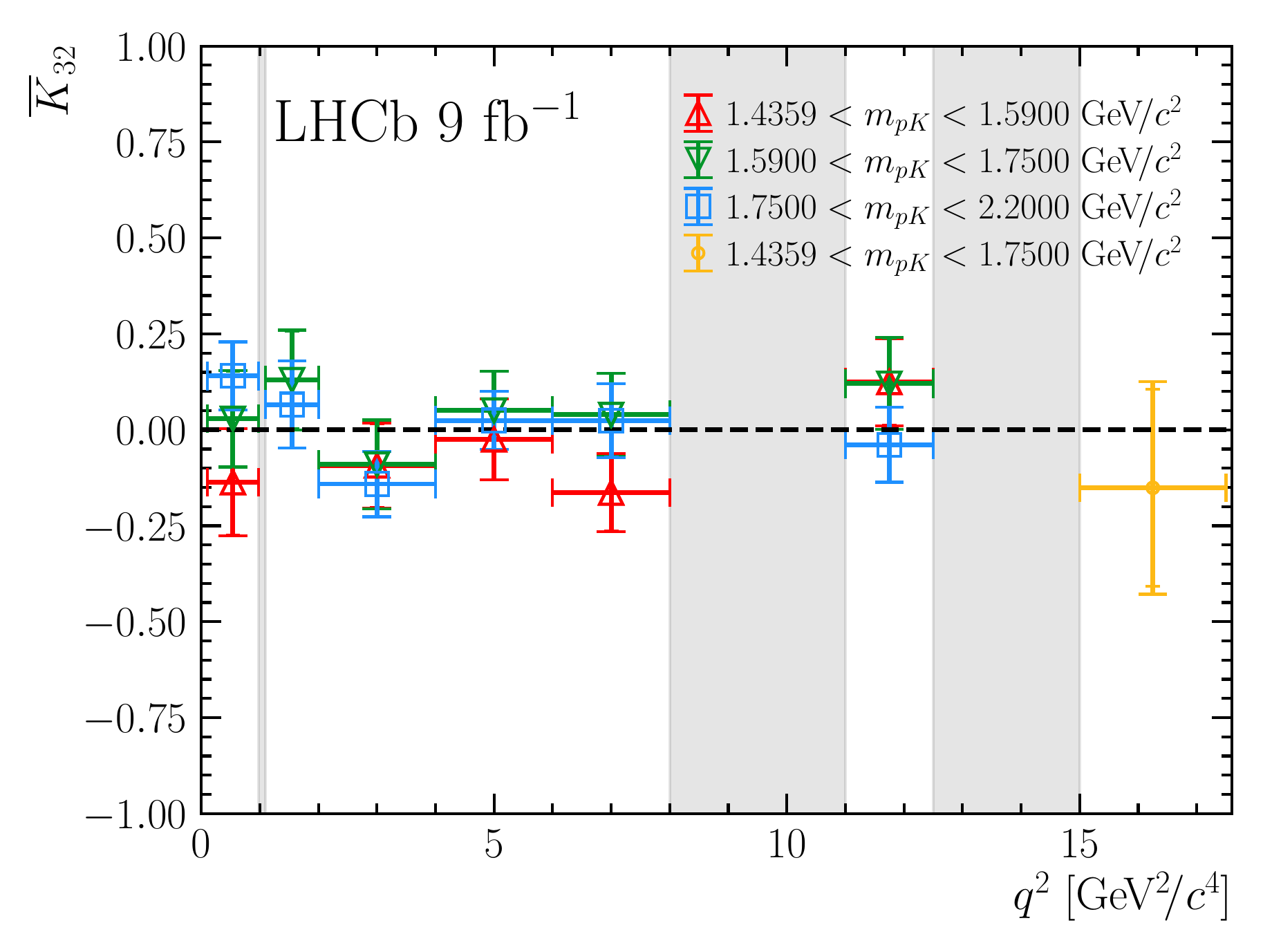}
     \includegraphics[width=0.48\textwidth]{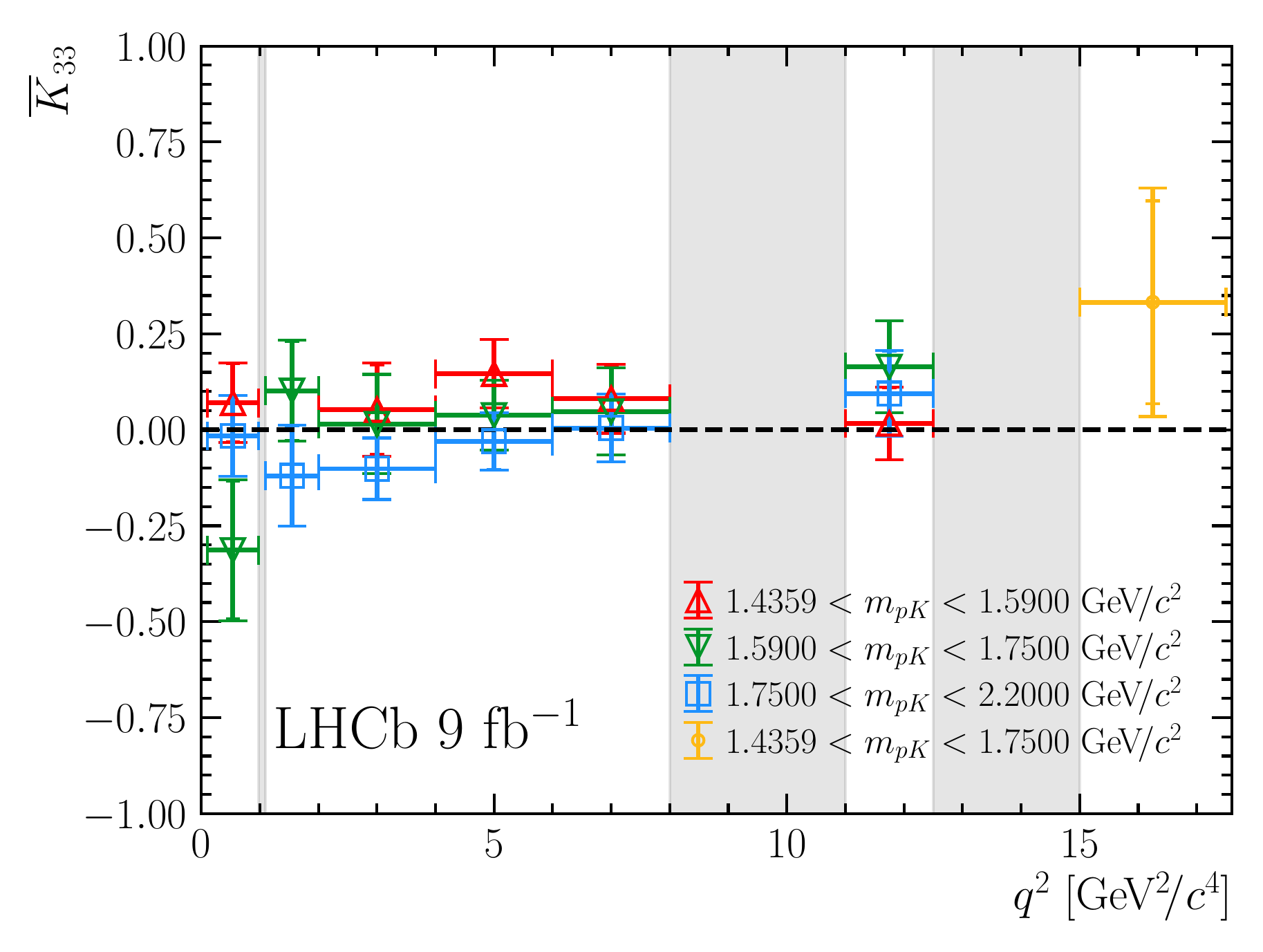} \\
     \includegraphics[width=0.48\textwidth]{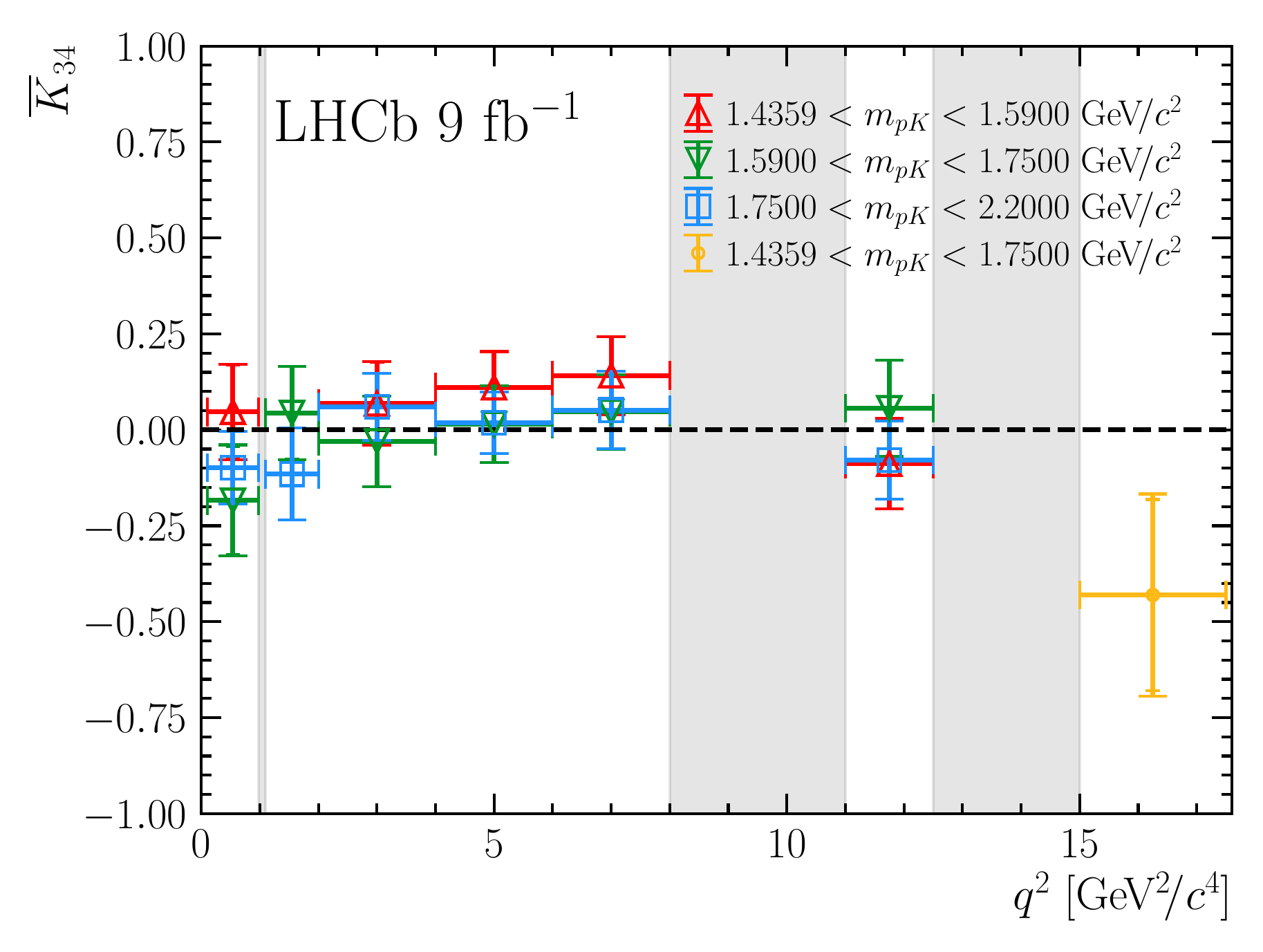}
     \includegraphics[width=0.48\textwidth]{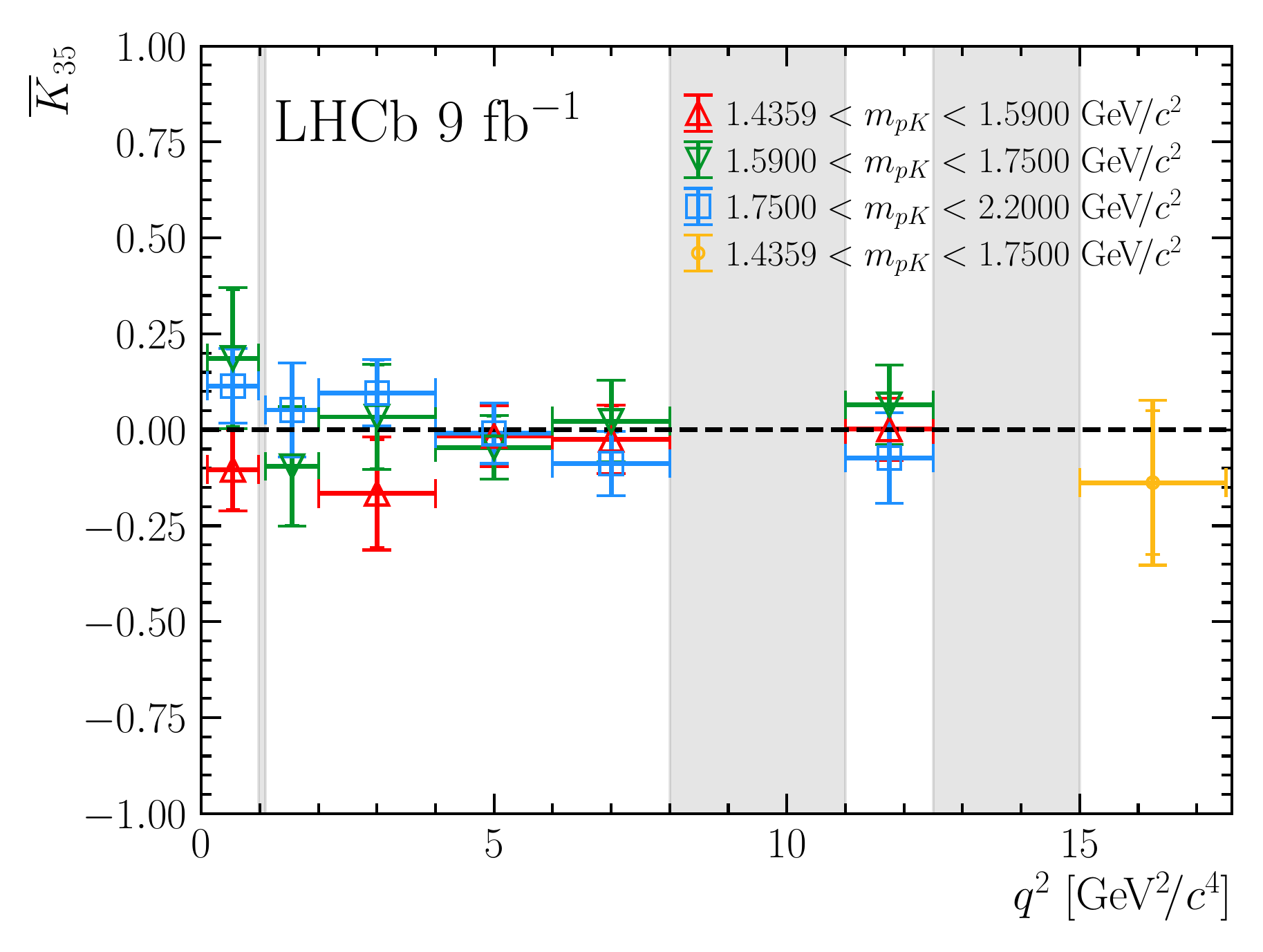} \\
     \includegraphics[width=0.48\textwidth]{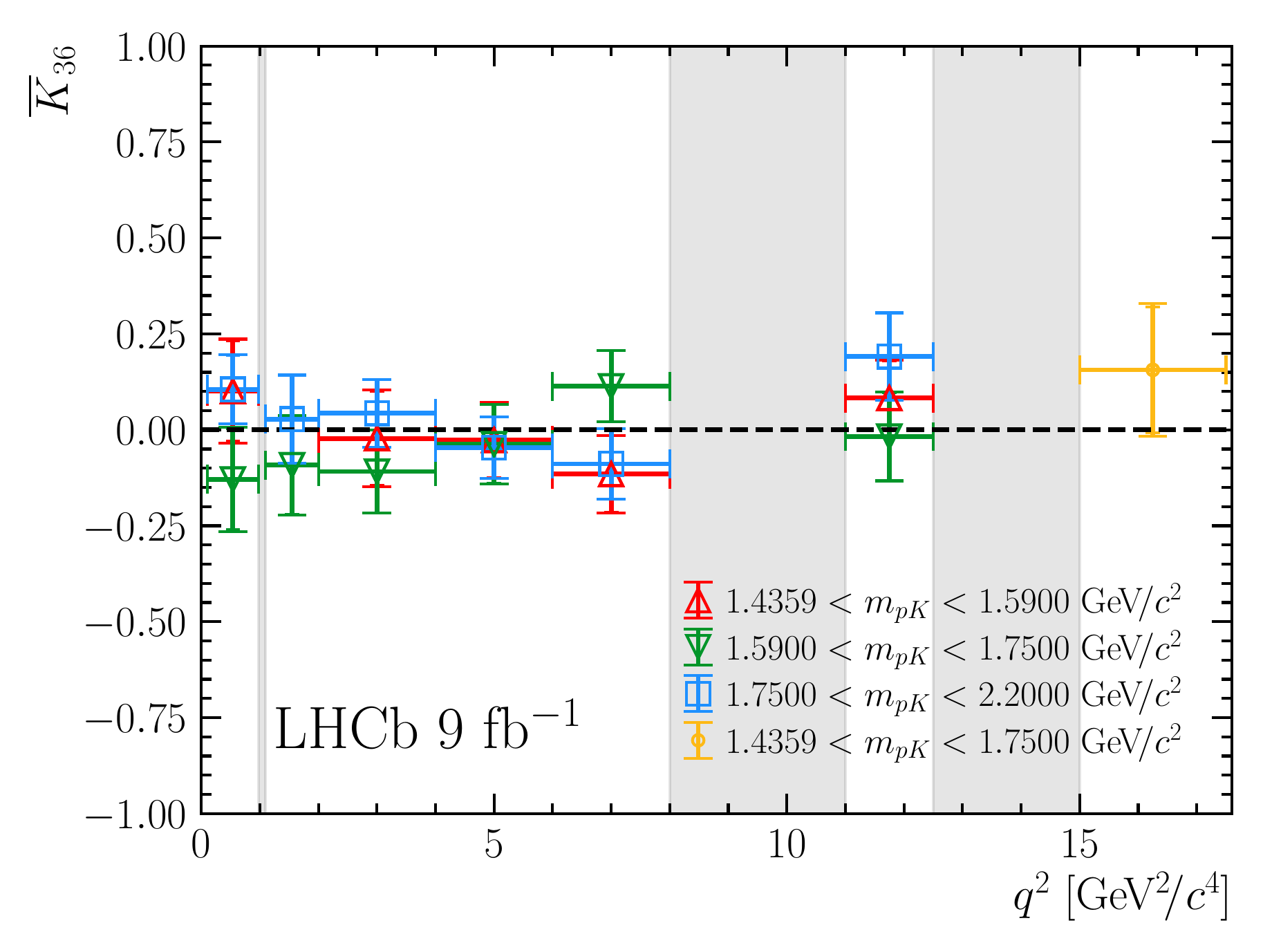}
     \includegraphics[width=0.48\textwidth]{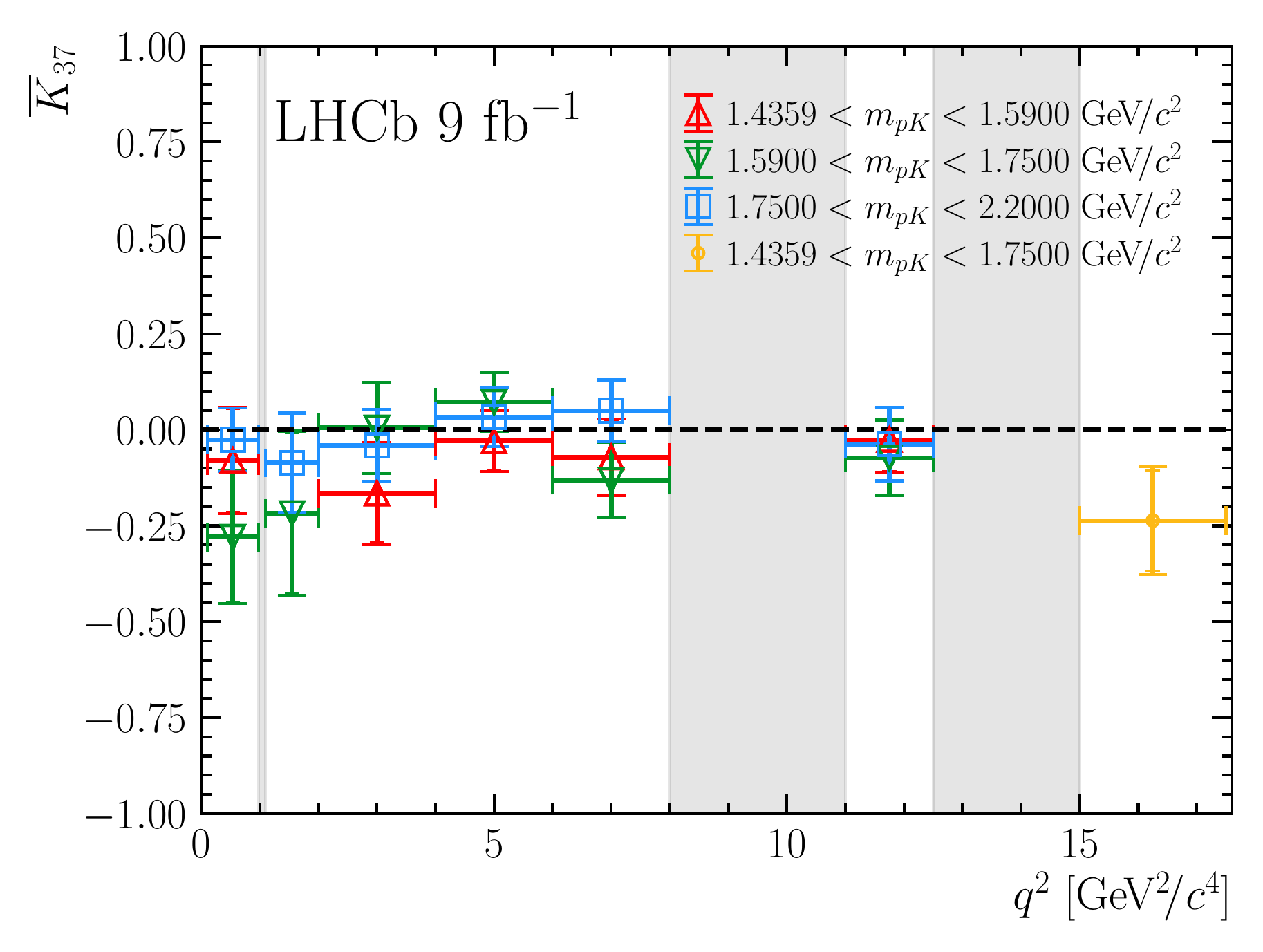} \\
     \includegraphics[width=0.48\textwidth]{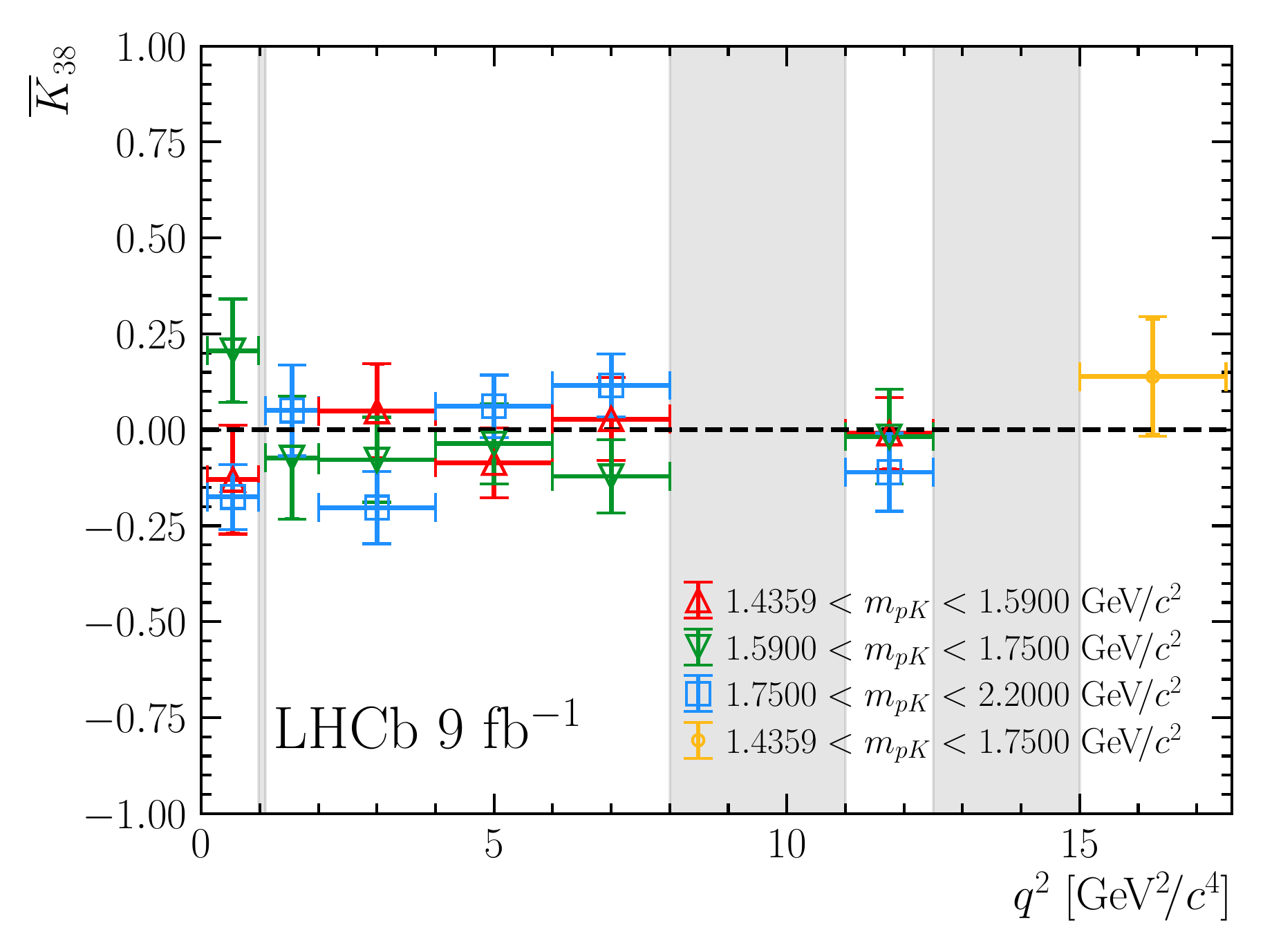}
     \includegraphics[width=0.48\textwidth]{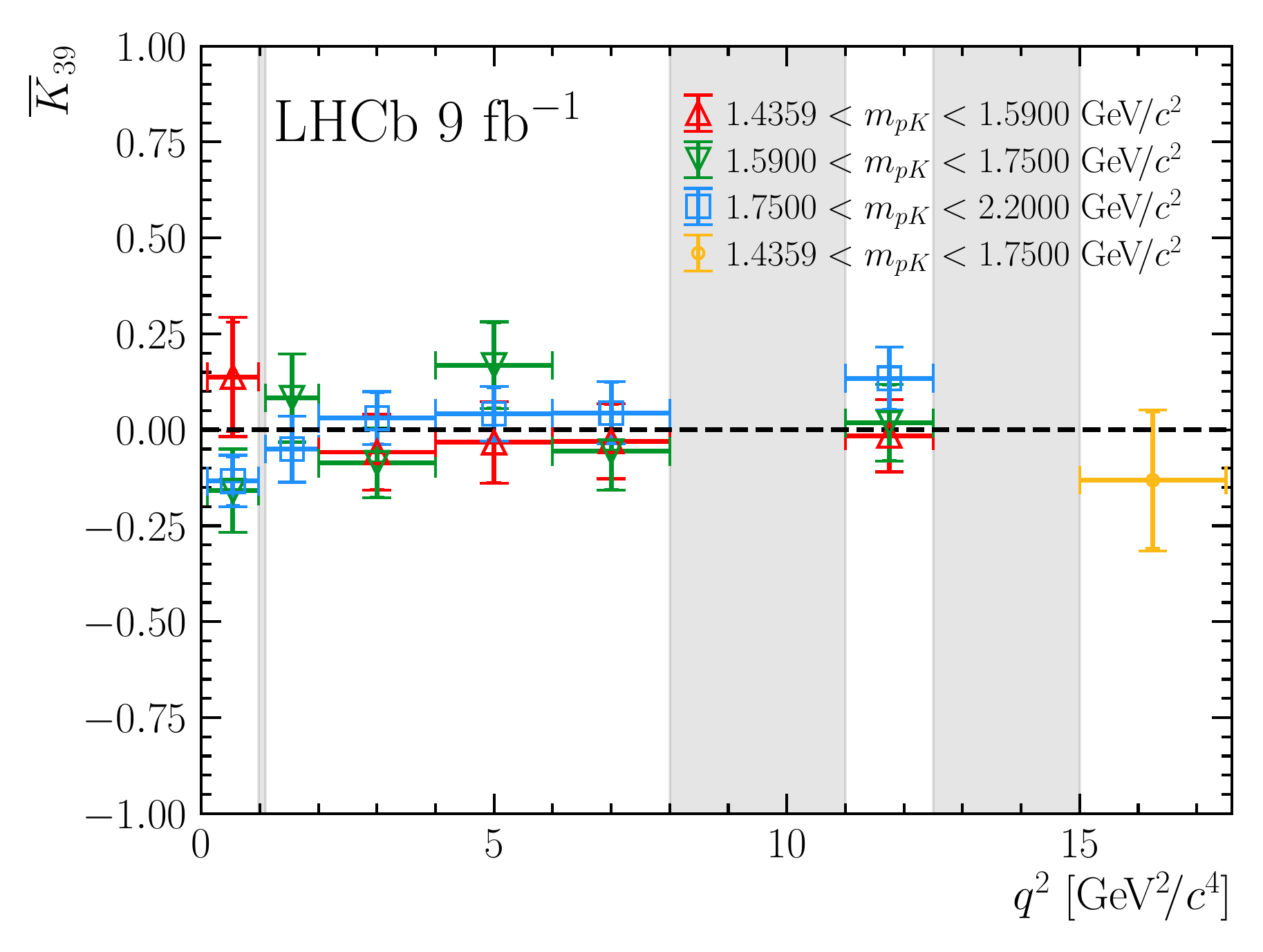}
     \caption{
     Values of \Kobs{32}--\Kobs{39} in bins of \qsq and \mpk.
    }
     \label{fig:appendix:kobs:5}
\end{figure}

\begin{figure}[!htb]
     \centering
     \includegraphics[width=0.48\textwidth]{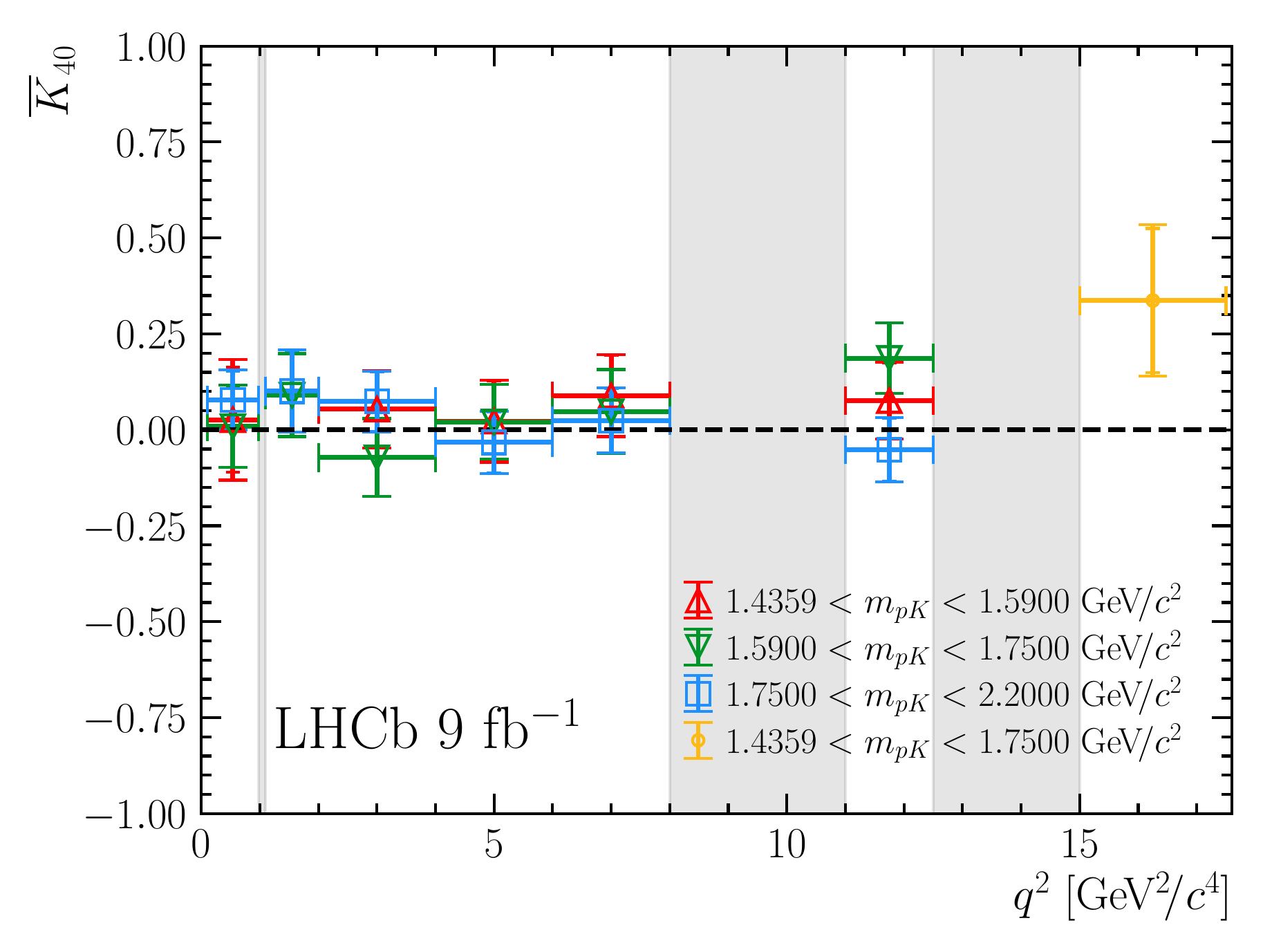}
     \includegraphics[width=0.48\textwidth]{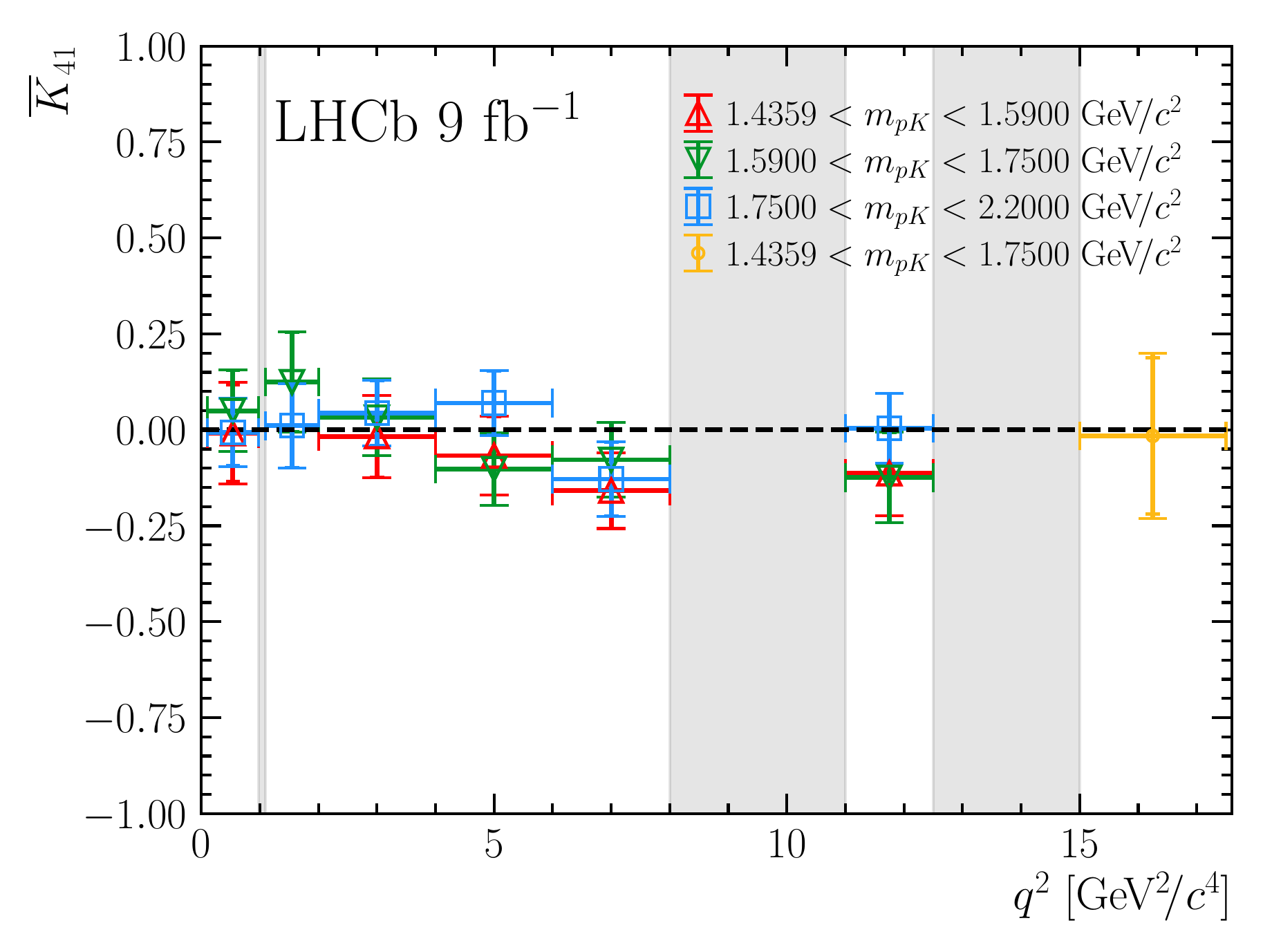} \\
     \includegraphics[width=0.48\textwidth]{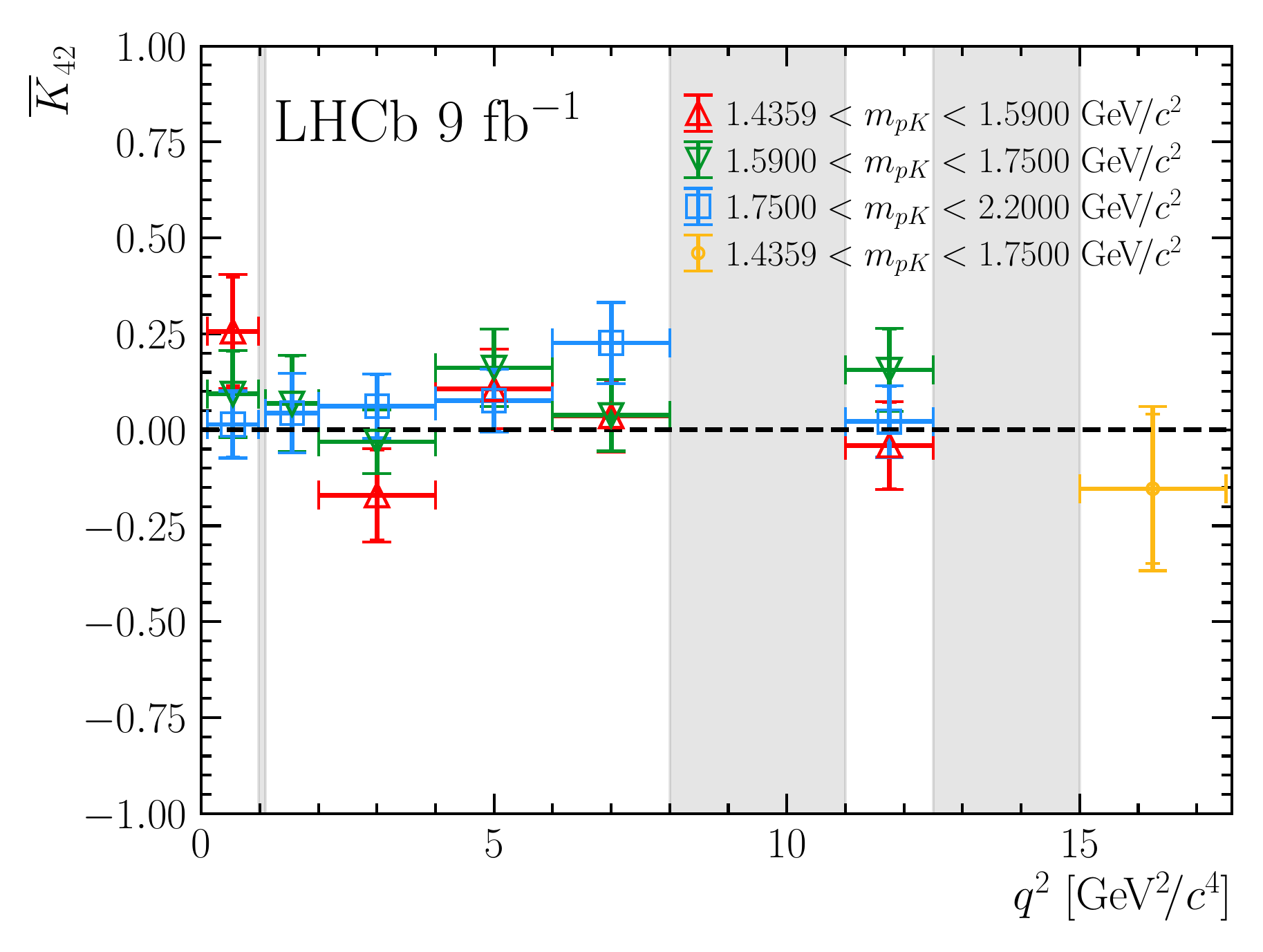}
     \includegraphics[width=0.48\textwidth]{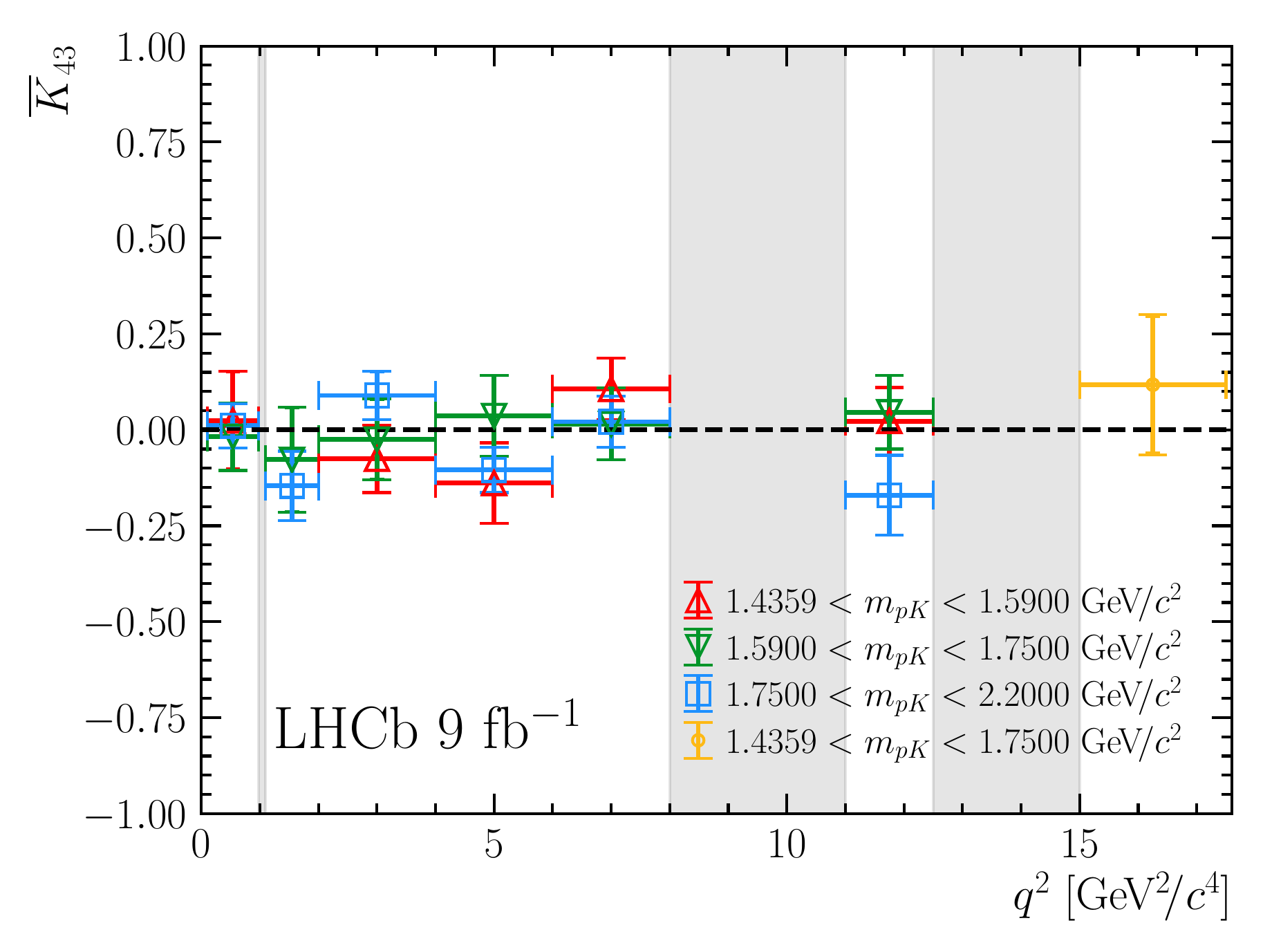} \\
     \includegraphics[width=0.48\textwidth]{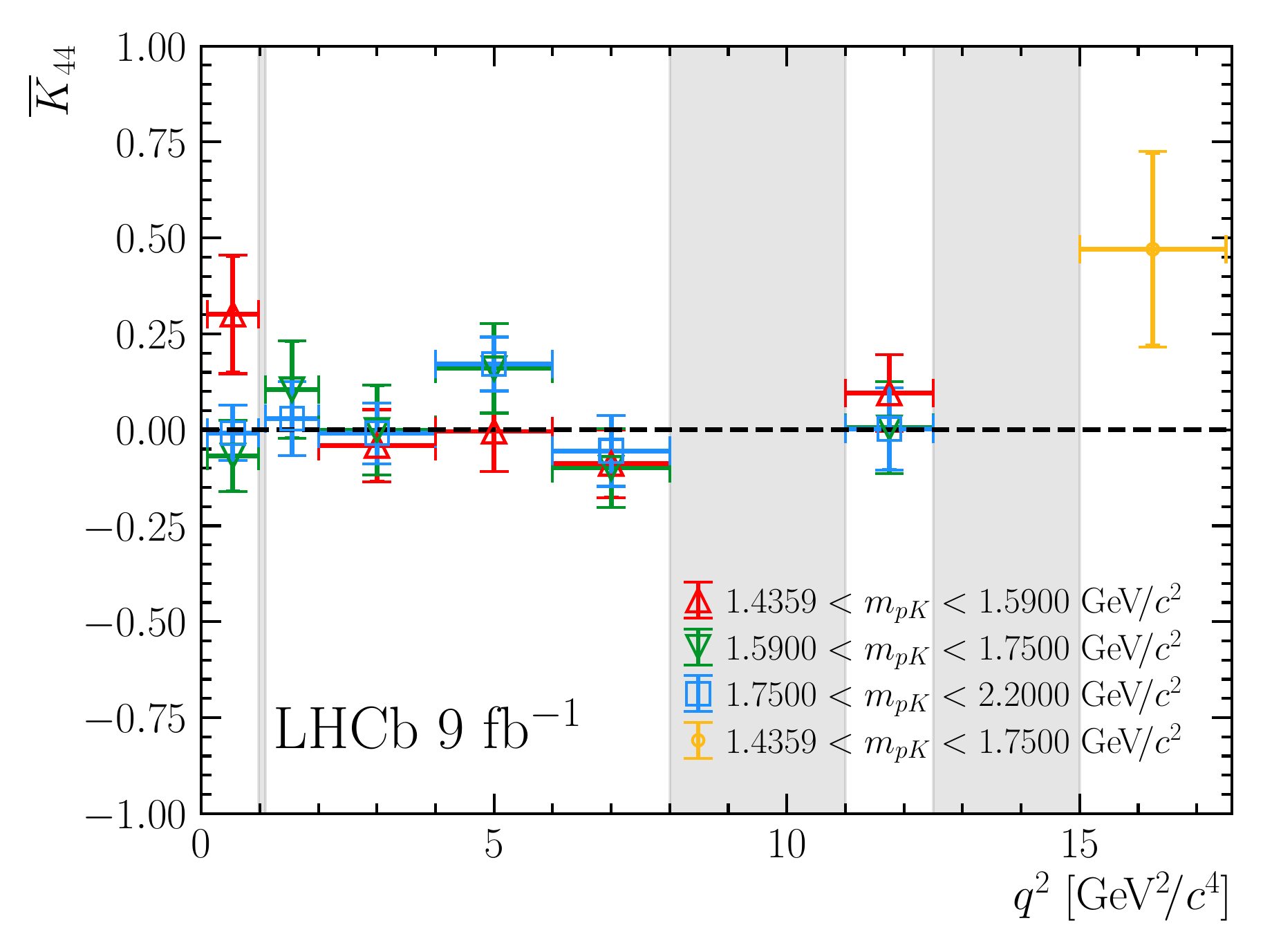}
     \includegraphics[width=0.48\textwidth]{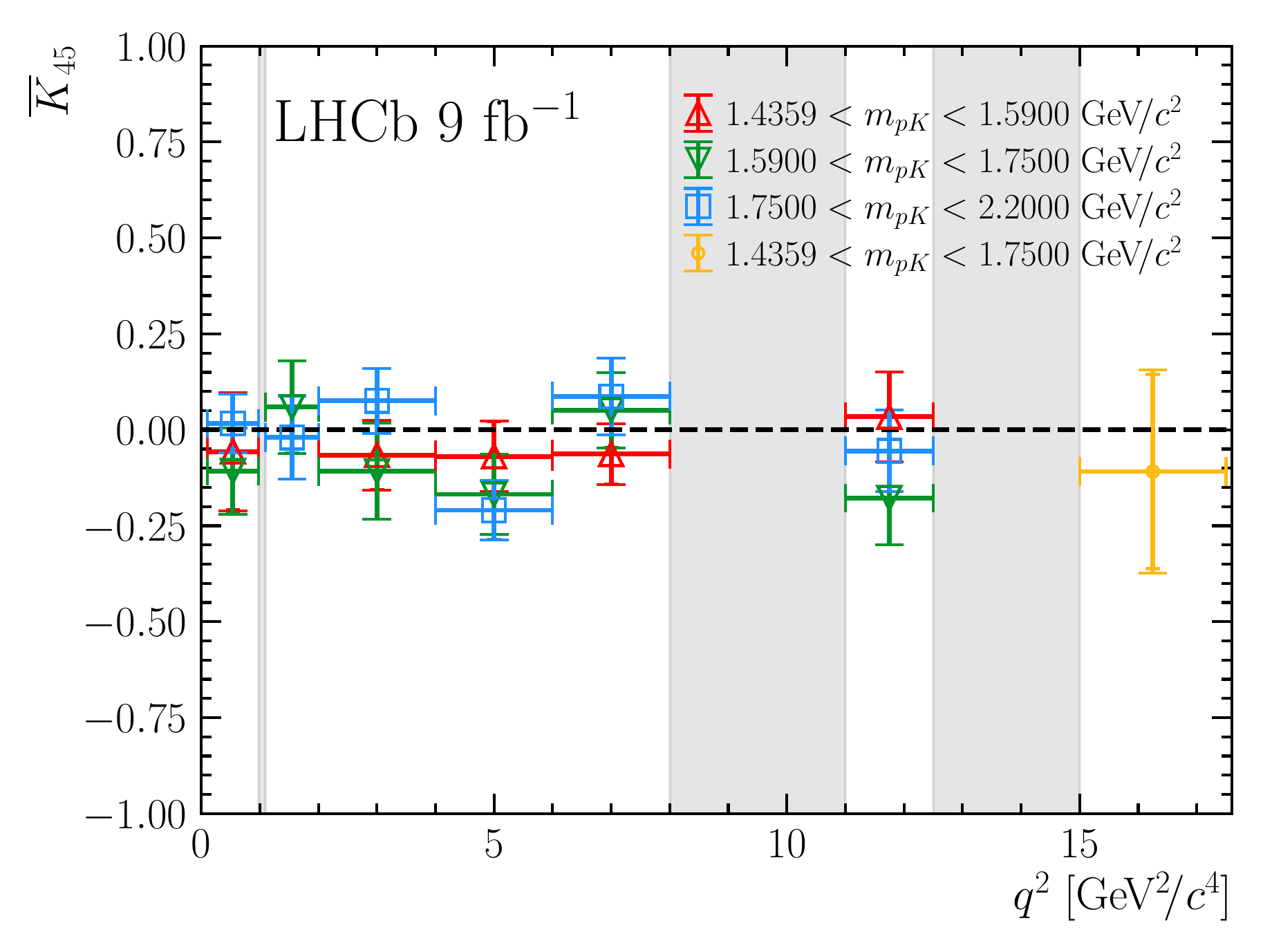} \\
     \includegraphics[width=0.48\textwidth]{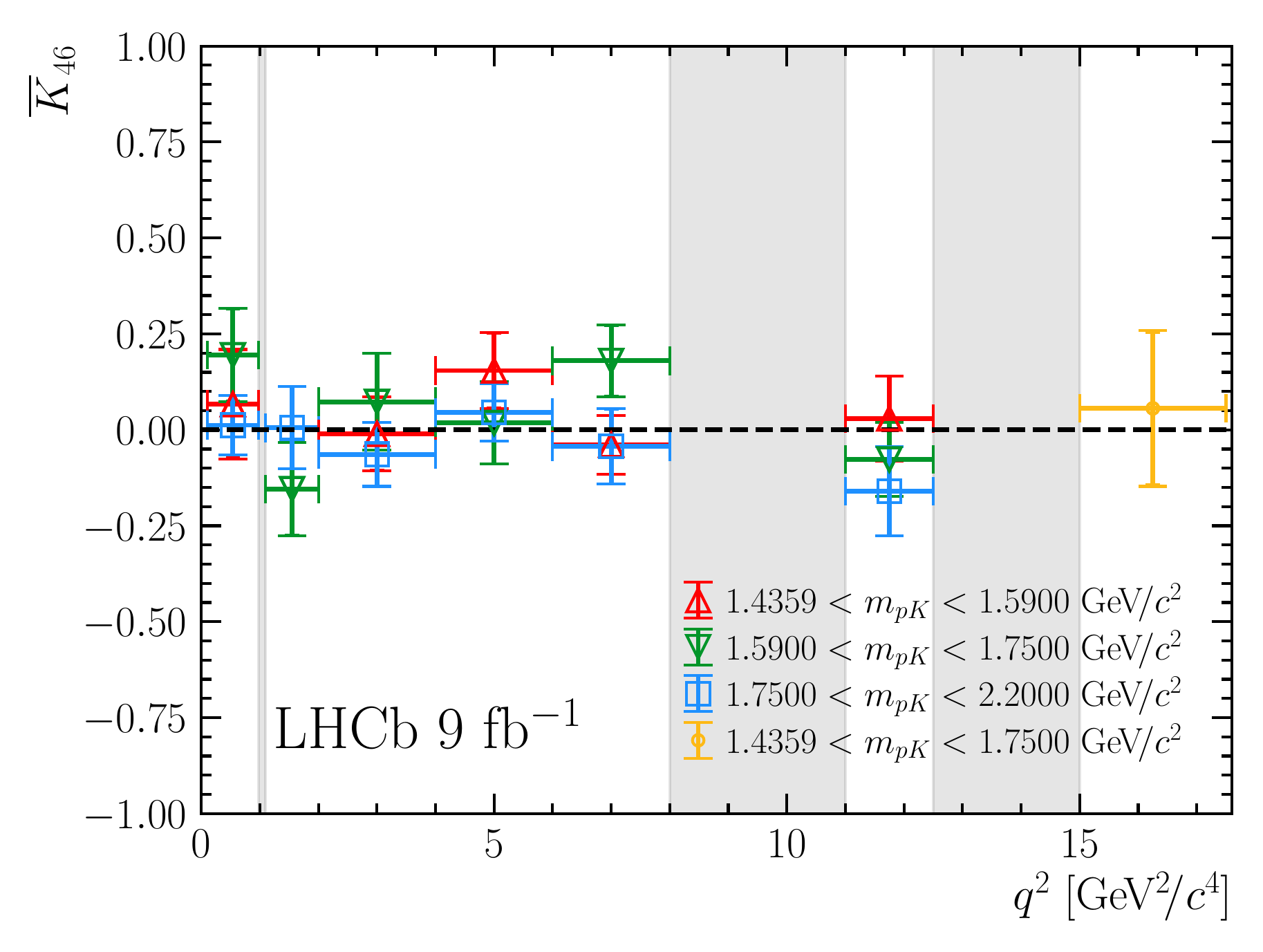}
     \caption{
     Values of \Kobs{40}--\Kobs{46} in bins of \qsq and \mpk.
    }
     \label{fig:appendix:kobs:6}
\end{figure}

The values of the \Kobs{2}--\Kobs{46} angular observables in the different \qsq and \mpk bins are provided in Figs.~\ref{fig:appendix:kobs:1}--\ref{fig:appendix:kobs:6}. 

%% file: Authorship_LHCb-PAPER-2024-024.tex
\centerline
{\large\bf LHCb collaboration}
\begin
{flushleft}
\small
R.~Aaij$^{37}$\lhcborcid{0000-0003-0533-1952},
A.S.W.~Abdelmotteleb$^{56}$\lhcborcid{0000-0001-7905-0542},
C.~Abellan~Beteta$^{50}$,
F.~Abudin{\'e}n$^{56}$\lhcborcid{0000-0002-6737-3528},
T.~Ackernley$^{60}$\lhcborcid{0000-0002-5951-3498},
A. A. ~Adefisoye$^{68}$\lhcborcid{0000-0003-2448-1550},
B.~Adeva$^{46}$\lhcborcid{0000-0001-9756-3712},
M.~Adinolfi$^{54}$\lhcborcid{0000-0002-1326-1264},
P.~Adlarson$^{81}$\lhcborcid{0000-0001-6280-3851},
C.~Agapopoulou$^{14}$\lhcborcid{0000-0002-2368-0147}
% ...

{\footnotesize \it

% ...
$^{14}$Universit{\'e} Paris-Saclay, CNRS/IN2P3, IJCLab, Orsay, France\\
% ...
$^{37}$Nikhef National Institute for Subatomic Physics, Amsterdam, Netherlands\\
% ...
$^{46}$Instituto Galego de F{\'\i}sica de Altas Enerx{\'\i}as (IGFAE), Universidade de Santiago de Compostela, Santiago de Compostela, Spain\\
% ...
$^{50}$Physik-Institut, Universit{\"a}t Z{\"u}rich, Z{\"u}rich, Switzerland\\
% ...
$^{54}$H.H. Wills Physics Laboratory, University of Bristol, Bristol, United Kingdom\\
% ...
$^{56}$Department of Physics, University of Warwick, Coventry, United Kingdom\\
% ...
$^{59}$School of Physics and Astronomy, University of Glasgow, Glasgow, United Kingdom\\
$^{60}$Oliver Lodge Laboratory, University of Liverpool, Liverpool, United Kingdom\\
% ...
$^{68}$Syracuse University, Syracuse, NY, United States\\
% ...
$^{81}$Department of Physics and Astronomy, Uppsala University, Uppsala, Sweden, associated to $^{59}$\\
% ...
}
\end{flushleft}